\documentclass[preprint, number]{elsarticle} \usepackage[utf8]{inputenc}
\usepackage[english]{babel}

\usepackage{doi} \usepackage{natbib}

\addto\captionsenglish{}

\usepackage{amsmath, amsfonts, mathrsfs}
\usepackage{graphicx}
\usepackage{subcaption}
\usepackage{bm}
\usepackage[margin=1.0in]{geometry}

\usepackage{amsthm}

\newcommand{\norm}[1]{\left\lVert#1\right\rVert}

\usepackage{hyperref}
\hypersetup{
    colorlinks=true,             }

\begin{document}
\title{A fourth-order accurate finite volume method for ideal MHD via upwind constrained transport}
\date{\today}

\author[pacm]{Kyle Gerard Felker\corref{cor1}}
\ead{kfelker@math.princeton.edu}

\author[pacm,astro]{James Stone}
\ead{jmstone@princeton.edu}
\cortext[cor1]{Corresponding author}
\address[pacm]{Program in Applied and Computational Mathematics, Princeton University, Princeton, NJ 08544, United States}
\address[astro]{Department of Astrophysical Sciences, Princeton University, Princeton, NJ 08544, United States}
\begin{abstract} We present a fourth-order accurate finite volume method for the solution of ideal magnetohydrodynamics (MHD). The numerical method combines high-order quadrature rules in the solution of semi-discrete formulations of hyperbolic conservation laws with the upwind constrained transport (UCT) framework to ensure that the divergence-free constraint of the magnetic field is satisfied. A novel implementation of UCT that uses the piecewise parabolic method (PPM) for the reconstruction of magnetic fields at cell corners in 2D is introduced.
The resulting scheme can be expressed as the extension of the second-order accurate constrained transport (CT) Godunov-type scheme that is currently used in the Athena astrophysics code. After validating the base algorithm on a series of hydrodynamics test problems, we present the results of multidimensional MHD test problems which demonstrate formal fourth-order convergence for smooth problems, robustness for discontinuous problems, and improved accuracy relative to the second-order scheme.
\end{abstract}

\begin{keyword}
magnetohydrodynamics \sep numerical methods \sep high-order finite volume method \sep constrained transport
\end{keyword}
\maketitle

\section{Introduction} \label{sec:intro}
Numerical solutions to the equations of ideal compressible magnetohydrodynamics (MHD) are widely used to study astrophysical phenomena including jets, accretion disks, winds, solar flares, and magnetospheres. Solutions to these problems may possess both complex smooth features and strong shocks. It is a challenge for numerical methods to resolve the complex turbulent dynamics and capture discontinuous features in a robust and computationally cost effective manner.

Among methods for gas dynamics, Godunov-type schemes based on the local solutions to Riemann problems are well-suited to both capturing shocks and resolving nonlinear waves; therefore, they have been extended to MHD in a variety of approaches \cite{GardinerStone2005, GardinerStone2008, Crockett2005, Fromang2006, Dedner2002, Toth2000, RyuMinatiJonesFrank1998, DaiWoodward1998b, BalsaraSpicer1999}. Second-order spatially and temporally accurate Godunov-type methods are among the most popular upwind schemes in computational astrophysics. However, higher-order (\(\mathcal{O}(\Delta x^r, \Delta t^p)\) where \( r, p \geq 3 \) here) schemes can improve smooth solutions at a much faster rate than second-order methods as the discrete resolution is improved. Further, high-order methods typically improve the spatial locality and reuse of memory references of the algorithm, and thus they reduce data transfers and operate more efficiently than low-order schemes \cite{Loffeld2017, Guzik2015}.  The on-node performance benefit increases for larger local domain box sizes, especially for higher dimensional problems on high-performance manycore architectures \cite{Loffeld2017, Olschanowsky2014}.

We are unaware of any finite volume methods for MHD of fourth-order or greater accuracy in active use in the computational astrophysics community. Discontinuous Galerkin (DG) methods for MHD have been formulated at arbitrarily high-order accuracy \cite{Mocz2014, Luo2008, LiShu2005}, however such schemes are rarely used at or beyond \(\mathcal{O}(\Delta x^4)\) accuracy, as it remains a challenge to prevent unphysical oscillations at shocks and discontinuities. Finite difference (FD) methods have been applied to MHD with third and fifth-order accuracy \cite{DelZanna2007, Mignone2010a}. High-order finite volume (FV) methods offer the accuracy of DG and FD methods with robust shock capturing. Several high-order FV methods, typically employing variants of weighted essentially non-oscillatory (WENO) reconstruction, have been coupled to techniques such as GLM-MHD to correct errors that arise when evolving the magnetic fields \cite{Susanto2013, NunezMunz2016-I}. 

In addition to the challenges of nonlinearity inherent in the Euler equations of compressible hydrodynamics,  numerical integration of MHD must also contend with the non-convexivity and non-strict hyperbolicity of the system of equations. Moreover, the equations of MHD are intrinsically multidimensional and thus cannot be consistently expressed in a dimensionally split approach, a technique commonly used to reduce the complexity of a system of equations to a series of 1D problems. The ideal MHD system does not trivially map to a conservation law formulation, since the
\begin{equation}
\nabla \cdot \mathbf{B} = 0 \, .\label{eq:div-b}
\end{equation}
It is well known that the divergence-free constraint must be satisfied to numerical precision to omit nonphysical solutions generated by the presence of numerical magnetic monopoles \cite{BrackbillBarnes1980}.

McCorquodale and Colella \cite{McCorquodaleColella2011} formulated a fourth-order method for hydrodynamics that is extensible to mapped, multiblock grids with adaptive mesh refinement (AMR) \cite{ColellaDorrHittingerMartin2009, ColellaDorrHittingerMartin2011, Guzik2015}. High-order accuracy is achieved for smooth problems using a combination of quadrature rules evaluated with finite difference approximations and \(\mathcal{O}(\Delta t^4)\) Runge-Kutta (RK) temporal integration. The piecewise parabolic method (PPM) is used to guarantee the robustness of the method for problems with strong discontinuities. We use this approach as the basis of the algorithm we present here.

The constrained transport (CT) technique was introduced by Evans and Hawley \cite{EvansHawley1988} and addresses the need to enforce the divergence-free constraint. Unlike alternative MHD algorithms such as divergence cleaning \cite{BalsaraKim2004}, Hodge projection, 8-wave schemes \cite{Powell1999}, or GLM-MHD \cite{Dedner2002, Mignone2010a}, the CT discretization evolves the magnetic field using the induction equation directly. While this approach can strictly enforce the solenoidal condition in Equation~\eqref{eq:div-b} to round-off error, there are difficulties in applying the technique in practice. Most crucially, the original CT formulation is limited to second-order accuracy \cite{Matsumoto2016}. When combined with upwinding of hydrodynamic fluxes, the CT discretization may lead to a scheme with dual independent representations of magnetic field quantities. The consistency of \(\mathbf{B}\) on a staggered mesh relative to the conserved variable mesh quantity \(\mathbf{B}\) is often guaranteed on a case-by-case basis for the particular underlying finite volume scheme \cite{GardinerStone2005, GardinerStone2008}.

The upwind constrained transport (UCT) framework of Londrillo and Del Zanna \cite{Londrillo2000, Londrillo2004} extends CT formalism to a class of hybrid methods for MHD that inherently 1) maintain the divergence-free condition and 2) are consistent with the underlying Godunov scheme. The authors detail and test several high-order examples of their framework; a third-order accurate scheme based on WENO reconstruction produces the best results for their MHD test problems \cite{Londrillo2004}.

In this paper, we combine the fourth-order finite volume method of McCorquodale and Colella for compressible hydrodynamics with a novel implementation of UCT that uses PPM for reconstructing field quantities across cell-faces \cite{McCorquodaleColella2011}. The paper is organized as follows. Section~\ref{sec:fv-hydro} reviews the high-order finite volume framework and specifies the details of the particular implementation used here for the hydrodynamics subsystem. Then, Section~\ref{sec:hydro-tests} validates the high-order hydrodynamics algorithm via three classes of test problems. The upwind constrained transport framework of Londrillo and Del Zanna is reviewed in Section~\ref{sec:uct}, and the novel implementation using PPM coupled to the fourth-order finite volume hydrodynamics subsystem is introduced. Section~\ref{sec:mhd-tests} evaluates the overall algorithm using MHD test problems and compares the high-order scheme to a second-order scheme. Section~\ref{sec:conclusion} gives concluding remarks and discusses future work.

\section{Hydrodynamics subsystem: fourth-order finite volume methods for hyperbolic conservation laws} \label{sec:fv-hydro}
The equations governing compressible hydrodynamics can be formulated as a system of hyperbolic conservation laws
\begin{equation}
\frac{\partial \mathbf{U}}{\partial t} + \vec{\nabla} \cdot \vec{\bm{\mathrm{F}}}(\bm{\mathrm{U}}) = 0 \, , \label{eq:conservation-law}
\end{equation}
with conserved variable vector \(\mathbf{U}= [\rho, \rho\mathbf{v}, E] \) and nonlinear flux function \( \vec{\bm{\mathrm{F}}}(\bm{\mathrm{U}}) \), where \(\rho\) is the density, \(\mathbf{v}\) is the gas velocity, and \(E\) is the total energy per unit mass. In this section and the next, we consider the solution of the system using Godunov-type methods based on the solution of Riemann problems. The resulting finite volume method for the hydrodynamics subsystem will form the basis of the overall MHD scheme when combined with the constrained transport discretization of Section~\ref{sec:uct}. This terminology is equivalent to the ``underlying scheme'' in broader constrained transport literature and the ``base scheme'' in the nomenclature of T\'{o}th \cite{Toth2000}.

Colella, et al. \cite{ColellaDorrHittingerMartin2009, ColellaDorrHittingerMartin2011} devised an approach to constructing high-order finite volume methods for the solution of Equation~\eqref{eq:conservation-law} on mapped grids. An essential component of this class of methods is the use of high-order quadrature rules evaluated on cell interfaces, in particular for the numerical fluxes. It was first introduced by Barad and Colella \cite{BaradColella2005} in the context of a finite volume solver for the Poisson equation. The framework was extended from scalar hyperbolic PDEs to systems of nonlinear hyperbolic conservation laws by McCorquodale and Colella in \cite{McCorquodaleColella2011}. Guzik et al. in \cite{Guzik2015} combine the advances of \cite{ColellaDorrHittingerMartin2011} and \cite{McCorquodaleColella2011} to introduce a fourth-order accurate finite volume method for nonlinear systems of equations on mapped, adaptively refined grids. The method's capabilities are demonstrated on canonical hydrodynamics problems involving strong shocks.

We focus our analysis on the algorithmic components that are essential to the high-order finite volume methods regardless of the coordinate system and mesh. Therefore, throughout this work we assume a uniform Cartesian mesh composed of cubic control volumes with grid spacing \(h \), and we consider a physical domain equivalent to this abstract \emph{computational space} in the mapped grid literature. An advantage of the mapped grid formalism is that the particular finite volume operators are simply described in the computational space. The complexities of nonuniform, refined, and smoothly curved meshes are encapsulated in the handling of the discrete metric terms of the non-analytic or analytic mapping \cite{Guzik2015}. While such grids are not presently considered in this study, the general applicability of the numerical methods to Cartesian and mapped, single-block and multiblock, refined and unrefined, structured and unstructured grids is important for the effective simulation of demanding relativistic and multiscale physics. In all cases, these high-order methods satisfy local conservation, guarantee freestream preservation (i.e., the solution to a uniform flow is unaffected by a smooth mapping or discretization), and are compatible with many underlying finite volume methods for computing fluxes in 1D such as WENO and PPM.

We adopt the notation used by \cite{ColellaDorrHittingerMartin2009, ColellaDorrHittingerMartin2011, McCorquodaleColella2011, Guzik2015} and summarized in Section 2 of \cite{Guzik2015}, but we restrict the notation to \(D=2\) dimensions, without loss of generality. Subscripts with the letters \( i,j  \) and integer offsets are used to index cell-centered quantities along the first and second dimensions, respectively,  in the lab-frame coordinate system. Half-integer offsets are used to index the cell faces and corners along these directions. General 1D operators involving cell interfaces will be centered on the lower  \( (i-\frac{1}{2}, j) \) interface, for example. Operators involving 2D cell corners will be centered on the lower left corner, for example, \( (i-\frac{1}{2}, j-\frac{1}{2}) \).

Beyond second-order accuracy, the midpoint approximation traditionally used in finite volume methods must be abandoned. The algorithm must distinguish all cell-/face-/edge-averaged quantities from pointwise cell-/face-/edge-centered approximations, and products of averages are only equal to averages of products to within a second order approximation. Angled brackets are used to denote the spatial averaging of a quantity defined on the mesh. For example, \( \langle Q \rangle_{i,j} \) indicates the cell-averaged value and \( \langle Q \rangle_{i-\frac{1}{2},j}, \langle Q \rangle_{i,j-\frac{1}{2}} \) are face-averaged quantities on faces in the first and second direction, respectively. In contrast, \( Q_{i,j} \) indicates the cell-centered value and \( Q_{i-\frac{1}{2},j}, Q _{i,j-\frac{1}{2}} \) are face-centered quantities.

Conversions between these averaged and pointwise values can be performed at fourth-order accuracy using stencils that approximate multidimensional data locally on the mesh. When integrating a quantity defined on a 2D cell volume, for example, the integrand can be replaced with a Taylor-series expansion about the cell-center. The odd-powered coordinate terms cancel out when integrating across the cell, so a finite difference Laplacian operator (of at least second-order accuracy) can be combined with the cell-centered value to get a fourth-order approximation to the cell-averaged quantity \cite{Guzik2015}. A similar argument holds for the inverse transformation. We use the following approximation to the Laplacian when converting between cell-centered and cell-averaged quantities in 2D:
\begin{equation}
\Delta Q_{i,j} = \frac{1}{h^2} (Q_{i-1,j} - 2Q_{i,j} + Q_{i+1,j}) + \frac{1}{h^2} (Q_{i,j-1} - 2Q_{i,j} + Q_{i,j+1}) \, .\label{eq:laplacian}
\end{equation}

For face-centered and face-averaged conversions, we require Laplacian operators \(\Delta^{\perp,d} \) for each direction \(d\) that only include derivative terms that are transverse to the surface normal vector \(\mathbf{e}^d\):
\begin{subequations}
\begin{align}
\Delta^{\perp,1} Q_{i-\frac{1}{2},j} =& \frac{1}{h^2} (Q_{i-\frac{1}{2},j-1} - 2Q_{i-\frac{1}{2},j} + Q_{i-\frac{1}{2},j+1}) \, ,  \label{eq:laplacian-transverse-x1} \\
\Delta^{\perp,2} Q_{i,j-\frac{1}{2}} =& \frac{1}{h^2} (Q_{i-1,j-\frac{1}{2}} - 2Q_{i,j-\frac{1}{2}} + Q_{i+1,j-\frac{1}{2}}) \, . \label{eq:laplacian-transverse-x2 }
\end{align}
\end{subequations}

\subsection{Temporal integration} \label{subsec:fv-temporal}
While alternative techniques like ADER may be used to resolve flux gradients in time using compact stencils with local timesteps \cite{DumbserZanotti2013}, we follow \cite{ColellaDorrHittingerMartin2009, ColellaDorrHittingerMartin2011, McCorquodaleColella2011, Guzik2015} and use a simple semi-discrete/method of lines formulation of Equation~\eqref{eq:conservation-law}
\begin{equation}
\frac{\mathrm{d}}{\mathrm{d}t} \langle \mathbf{U} \rangle_{i,j} + \frac{1}{h} (\langle \mathbf{F}_1 \rangle_{i+\frac{1}{2},j} - \langle \mathbf{F}_1 \rangle_{i-\frac{1}{2},j}) + \frac{1}{h} (\langle \mathbf{F}_2 \rangle_{i,j+\frac{1}{2}} - \langle \mathbf{F}_2 \rangle_{i,j-\frac{1}{2}}) = 0 \, . \label{eq:fv-div}
\end{equation}
One advantage of the method of lines is the decoupling of the temporal and spatial orders of accuracy of the overall scheme. This separation encapsulates the (often complicated) specifics of the given equations in the treatment of the spatial terms at a single stage. The initial discretization of all terms exclusively in space results in a time-dependent system of autonomous ODEs. The ODE system can be integrated with many general ODE integrators of different orders of accuracy and computational demands. For the purposes of this study, any explicit, multistage one-step integrator that is absolutely stable for the fourth-order central difference operator may be used. See \textsection{4.2} of \cite{ColellaDorrHittingerMartin2011} for the corresponding linear stability analysis that computes the CFL restriction for the purely imaginary eigenvalues of the constant coefficient advection problem.

For the majority of the results presented here, we use a strong-stability preserving (SSP), low-storage variant of the fourth-order accurate Runge-Kutta method. See Gottlieb, Ketcheson, and Shu \cite{GottliebKetchesonShu2009} for
the precise implementation details, and refer to \cite{Ketcheson2008, Ketcheson2010, SpiteriRuuth2002, GottliebShu1998, GottliebShuTadmor2001, GottliebKetchesonShu2009} for the theory of low-storage, SSP RK integrators. This \(\mathcal{O}(\Delta t^4)\) accurate RK variant uses five substages of flux updates and requires the storage of three intermediate solutions of the entire domain (registers) per time step. We will refer to the timestepper as RK4 throughout the paper.
For smooth error convergence studies, we also employ the RK3 variant defined by \cite{ShuOsher1988} Equation 2.19. This method requires three substages and two registers per timestep. Future work will focus on comparing integrators with optimal effective SSP coefficients \cite{Ketcheson2008} in the context of challenging MHD applications.
\subsection{Spatial discretization and numerical fluxes} \label{fv-spatial}

In a finite volume method, the spatial operator must ultimately provide a consistent approximation to the average fluxes on all cell faces for the discrete divergence law in Equation~\eqref{eq:fv-div}. When solving nonlinear systems of hyperbolic PDEs at high-resolution, solution properties such as smoothness, monotonicity, positivity, and stability may become important considerations. These concerns may motivate the augmentation of the numerical flux function with intermediate steps, such as nonlinear variable transformations or projections, the enforcement of variable floors, and application of slope limiters. All of the steps in the algorithm must compute any intermediate approximations at \(\mathcal{O}(\Delta x^4)\) in order to preserve the spatial accuracy of the overall scheme.

For completeness, we reproduce the outline detailed by Guzik, et al. in \textsection 3.2 of \cite{Guzik2015}  for the procedure of computing the hydrodynamic fluxes while describing the differences of our particular implementation.

\subsubsection{Equation of state and variable inversion} \label{subsubsec:eos}
In finite volume methods for hydrodynamics, piecewise polynomial reconstruction of fluid profiles is frequently performed on the set primitive variables \( \mathbf{W} = [\rho, \mathbf{v}, P ]\) instead of the conserved variables \( \mathbf{U} \). Application of slope limiters to primitive reconstructions typically results in less oscillatory solutions, and positivity of density and fluid pressure can be explicitly enforced in the primitive variable space. The sets of variables are related by a nonlinear, invertible transformation \( \mathbf{W}(\mathbf{U}) \).  Because the pointwise transformation is exact, the variable inversion at fourth-order accuracy is performed using approximations to the pointwise cell-centered values. The steps are as follows:
\begin{enumerate}
\item Convert from cell-averaged to cell-centered conserved variables
\begin{equation}
\mathbf{U}_{i,j} = \langle \mathbf{U}\rangle_{i,j}  - \frac{h^2}{24} \Delta \langle \mathbf{U} \rangle_{i,j} \, , \label{eq:U-cell-center}
\end{equation}
via the finite difference Laplacian operator in Equation~\eqref{eq:laplacian}
\item Apply the variable inversion to the \(\mathcal{O}(\Delta x^4)\) approximation to cell-centered conserved variables, resulting in a pointwise \(\mathcal{O}(\Delta x^4)\) primitive variable approximation
\begin{equation}
\mathbf{W}_{i,j} = \mathbf{W}(\mathbf{U}_{i,j}) \, . \label{eq:W-cell-center}
\end{equation}
\item Apply the variable inversion to cell-averaged conserved variables, resulting in a \(\mathcal{O}(\Delta x^2)\) approximation to cell-averaged primitive variables
\begin{equation}
\overline{\mathbf{W}}_{i,j} = \mathbf{W}(\langle \mathbf{U} \rangle_{i,j}) \, .\label{eq:W-bar-cell-ave}
\end{equation}
\item Combine both primitive variable approximations to compute a fourth-order approximation to the cell-averaged primitive variables
\begin{equation}
\langle \mathbf{W} \rangle_{i,j} = \mathbf{W}_{i,j} + \frac{h^2}{24} \Delta \overline{\mathbf{W}}_{i,j} \, . \label{eq:W-cell-ave}
\end{equation}
\end{enumerate}
In addition to applying primitive variable floors to the density and pressure during the variable inversion steps in Equations~\eqref{eq:W-cell-center} and~\eqref{eq:W-bar-cell-ave}, the floors are reapplied after Equation~\eqref{eq:W-cell-ave}.

\subsubsection{Primitive variable reconstruction and limiters} \label{subsubsec:fv-reconstruct}
From the \(\mathcal{O}(\Delta x^4)\) cell-averaged primitive variables, the face-averaged primitive states are reconstructed using the piecewise parabolic method. The averages on each interface are initialized using a four-point stencil along the longitudinal direction. In the \(x_1\) direction, for example, this is computed as
\begin{equation}
\langle \mathbf{W}\rangle_{i-\frac{1}{2},j} =  \frac{7}{12} (\langle \mathbf{W}\rangle_{i-1,j} + \langle \mathbf{W} \rangle_{i,j}) - \frac{1}{12} (\langle \mathbf{W} \rangle_{i+1,j} + \langle \mathbf{W} \rangle_{i-2,j} ) \, . \label{eq:PPM4}
\end{equation}
While the interface approximation is single valued and fourth-order accurate when applied to smooth data, the presence of discontinuities and nonlinear dynamics demands the use of limiters to suppress spurious oscillations. Limiting may introduce multivalued face-averaged interface L/R Riemann states. We have tested more than five limiters when designing the overall scheme, including:
\begin{enumerate}
\item The original PPM limiter of Colella \& Woodward \cite{ColellaWoodward1984}
     \item The variant of Mignone \cite{Mignone2014} which has formulations for spherical and cylindrical coordinate systems
\item The smooth extremum preserving limiter of Colella \& Sekora \cite{ColellaSekora2008}
\item A modification of \cite{ColellaSekora2008} presented by Colella, et al. in \textsection 4.3 of \cite{ColellaDorrHittingerMartin2011}
\item The improved version of \cite{ColellaDorrHittingerMartin2011} by McCorquodale \& Colella in \textsection 2.4 of \cite{McCorquodaleColella2011}
\end{enumerate}
We refer the reader to the original references for the complete implementation details. We have found that the final three limiters, which avoid the clipping of smooth extrema, all produce similar results in the test problems below with only minor differences in numerical dissipation. However, the modifications to the original smooth extremum preserving PPM limiter of \cite{ColellaSekora2008} made in \cite{ColellaDorrHittingerMartin2011} and \cite{McCorquodaleColella2011} cause the algorithm to lose the property of strict monotonicity-preservation. For example, Figure 4 of \cite{McCorquodaleColella2011} shows a non-monotonic solution for the 1D square wave advection problem. In addition, the derivative approximations in the McCorquodale variant lead to a 7-cell stencil, which is prohibitively expensive for the MHD applications under consideration.

Therefore, we implement PPM with fourth-order interface approximation (summarized as PPM4 in \cite{PetersonHammett2013}) and a smooth extremum preserving limiter variant based on the version in \textsection 4.3 \cite{ColellaDorrHittingerMartin2011}. Our implementation eschews the check for monotonicity of the derivative estimates when limiting the initial interface states (Equation 86 of \cite{ColellaDorrHittingerMartin2011}). We find that the additional dissipation relative to the more advanced limiter in \cite{McCorquodaleColella2011} is acceptable, especially in problems with strong discontinuities.

There are several typos in the above PPM limiter literature that we wish to identify here for clarity:
\begin{itemize}
\item In Colella \& Sekora \cite{ColellaSekora2008}, the final term in Equation 19 should have a factor of \(\frac{1}{6}\), not \(\frac{1}{3}\). This was identified in \textsection 4.3.1 of \cite{ColellaDorrHittingerMartin2011}.
\item In Colella \& Sekora \cite{ColellaSekora2008}, Equation 20 is missing an ``or'' conditional when checking two conditions for detecting local extrema. \item In Colella, et al \cite{ColellaDorrHittingerMartin2011}, Equations 85a, 85c,  95b, 95c, 95d should not have the factors of \(\frac{1}{2}\). These second-derivative cell-averaged stencils are not consistent with McCorquodale \textsection 2.4.1 nor Equation 21 of \cite{ColellaDorrHittingerMartin2011}.
\item In the original PPM reference \cite{ColellaWoodward1984}, Equation 1.8 for van Leer limiting of the initial slopes, one of the terms in the \(\min()\) function should be indexed with \(j+1\), not \(j-1\).
\end{itemize}

In certain MHD shock tube tests, we have found that reconstructing characteristic variable profiles instead of primitive variable profiles was necessary to suppress spurious oscillations.  The eigenvectors of Appendix A of Stone, et al. \cite{Stone2008} are used for the characteristic projections, but the projection algorithm is fundamentally different from steps 1-5 in \cite{Stone2008} Section 4.2.2. The reconstruction procedure in characteristic variable space for the two \(\langle \mathbf{W}^{R_1} \rangle _{i-\frac{1}{2},j}, \langle \mathbf{W}^{L_1} \rangle_{i+\frac{1}{2},j}\) primitive Riemann states local to each cell is:
\begin{enumerate}
\item The left and right eigenvectors of the linearized system are computed using the local cell-averaged primitive variables. The eigenmatrices are
\begin{subequations}
\begin{align}
\vec{\bm{\mathrm{L}}}_{i,j} =& \vec{\bm{\mathrm{L}}}(\langle \mathbf{W} \rangle_{i,j} ) \, , \label{eq:L-eigenmatrix} \\
\vec{\bm{\mathrm{R}}}_{i,j} =& \vec{\bm{\mathrm{R}}}(\langle \mathbf{W} \rangle_{i,j} ) \, . \label{eq:R-eigenmatrix}
\end{align}
\end{subequations}
\item The left eigenvectors are applied to all cell-averaged primitive variables within the stencil along the direction of reconstruction
\begin{equation}
\langle \mathbf{Q} \rangle _{i+l,j}  = \vec{\bm{\mathrm{L}}}_{i,j} \langle \mathbf{W} \rangle _{i+l,j} \, , \label{eq:char-projection}
\end{equation}
where \( -2\leq l \leq 2 \) for PPM4.
\item Then, the same interpolation and limiting procedure as in the primitive variable case is followed using these projected quantities, resulting in characteristic Riemann states on the \(x_1\) interfaces
\begin{subequations}
\begin{align}
\langle \mathbf{Q}^{R_1} \rangle _{i-\frac{1}{2},j} \, , \label{eq:char-states-R1} \\
\langle \mathbf{Q}^{L_1} \rangle_{i+\frac{1}{2},j} \, . \label{eq:char-states-L1}
\end{align}
\end{subequations}
\item Finally, the limited reconstructed characteristic variable interface states are converted to primitive interface states
\begin{subequations}
\begin{align}
\langle \mathbf{W}^{R_1} \rangle _{i-\frac{1}{2},j} =& \vec{\bm{\mathrm{R}}}_{i,j} \langle \mathbf{Q}^{R_1} \rangle _{i-\frac{1}{2},j} \, , \label{eq:primitive-projection-R1} \\
\langle \mathbf{W}^{L_1} \rangle_{i+\frac{1}{2},j}  =& \vec{\bm{\mathrm{R}}}_{i,j} \langle \mathbf{Q}^{L_1} \rangle_{i+\frac{1}{2},j} \, , \label{eq:primitive-projection-L1}
\end{align}
\end{subequations}
using the right eigenvectors.
\end{enumerate}

This procedure is summarized in the context of MP5 reconstruction by Equations 30-36 in Section 3.3 of Matsumoto, et al. \cite{Matsumoto2016}. Examples in the context of WENO reconstruction procedures are given in \cite{BalsaraShu2000} for monotonicity preserving MPWENO and \cite{JiangShu1996} for a reduced pressure and entropy projection scheme, WENO-LF-5-PS. In our particular formulation, the characteristic reconstruction process is computationally expensive because the projection step produces local stencils for each cell. The expense grows for high-order and multidimensional schemes \cite{Londrillo2000}. Furthermore, decomposing the wave along characteristics may not be feasible in relativistic simulations or in the presence of complex physics \cite{Matsumoto2016}. When presented with a problem for which primitive reconstruction produces oscillatory results and characteristic decomposition is impractical, alternative techniques such as artificial viscosity, slope flattening, or more aggressive limiting may be pursued \cite{McCorquodaleColella2011}.

\subsubsection{Approximate flux calculation using Riemann solvers} \label{subsubsec:fv-flux}
Finally, fourth-order accurate interface-averaged fluxes \(\langle \mathbf{F}_1\rangle _{i\pm\frac{1}{2}, j}, \langle \mathbf{F}_2\rangle _{i,j\pm\frac{1}{2}}\) are computed through the approximation of the face-centered fluxes. The process is compatible with any approximate or exact Riemann solver \( \mathscr{F}(\mathbf{W}^{L}, \mathbf{W}^{R}) \) that only depends on the L/R primitive states at an interface. In this publication, we use approximate HLL-type Riemann solvers, including HLLE \cite{Einfeldt1988}, HLLC \cite{Toro1994}, and HLLD \cite{MiyoshiKusano2005} (for MHD). The linearized Roe solver \cite{Roe1981} and the Local Lax-Friedrichs one-speed flux approximation are also implemented.
The steps for \(x_1\) faces are:
\begin{enumerate}
\item Convert face-averaged primitive L/R Riemann states to pointwise face-centered states
\begin{subequations}
\begin{align}
\mathbf{W}^{L_1}_{i-\frac{1}{2}, j} =& \langle \mathbf{W}^{L_1} \rangle _{i-\frac{1}{2}, j} - \frac{h^2}{24} \Delta^{\perp,1} \langle \mathbf{W}^{L_1}\rangle_{i-\frac{1}{2}, j} \, ,  \label{eq:WL1-face-center} \\
\mathbf{W}^{R_1}_{i-\frac{1}{2}, j} =& \langle \mathbf{W}^{R_1} \rangle _{i-\frac{1}{2}, j} - \frac{h^2}{24} \Delta^{\perp,1} \langle \mathbf{W}^{R_1}\rangle_{i-\frac{1}{2}, j} \, . \label{eq:WR1-face-center}
\end{align}
\end{subequations}
\item Using a Riemann solver \( \mathscr{F}() \), compute pointwise interface-centered fluxes
\begin{equation}
\mathbf{F}_{1,i-\frac{1}{2},j} = \mathscr{F}(\mathbf{W}^{L_1}_{i-\frac{1}{2},j},\mathbf{W}^{ R_1}_{i-\frac{1}{2},j}) \, ,\label{eq:flux-face-center}
\end{equation}
from interface-centered primitive states.
\item Using a Riemann solver \( \mathscr{F}() \), compute fluxes from the fourth-order accurate interface-averaged primitive states. This results in a \(\mathcal{O}(\Delta x^2) \) accurate approximation to the interface-averaged fluxes
\begin{equation}
\overline{\mathbf{F}}_{1,i-\frac{1}{2},j} = \mathscr{F}(\langle \mathbf{W}^{L_1}\rangle _{i-\frac{1}{2},j},\langle \mathbf{W}^{R_1}\rangle_{i-\frac{1}{2},j} ) \, . \label{eq:flux-bar-face-ave}
\end{equation}
\item Compute the Laplacian of the \(\mathcal{O}(\Delta x^2) \) estimate of the face-averaged fluxes in directions orthogonal to the interface normal. Then, transform the face-centered fluxes to fourth-order accurate face-averaged fluxes
\begin{equation}
\langle \mathbf{F}_1\rangle _{i-\frac{1}{2}, j} = \mathbf{F}_{1,i-\frac{1}{2},j}  -\frac{h^2}{24} \Delta^{\perp,1} \overline{\mathbf{F}}_{1,i-\frac{1}{2},j} \, . \label{eq:flux-face-ave}
\end{equation}
Note, the decisions to apply the Laplacian operator to \( \overline{\mathbf{F}} \) in Equation~\eqref{eq:flux-face-ave} and to \( \overline{\mathbf{W}} \) in Equation~\eqref{eq:W-cell-ave} are made to reduce the stencil size at the cost of an additional Riemann solve and variable inversion, respectively \cite{Guzik2015}.
\end{enumerate}
After the flux averages are known for all faces, the flux divergence is computed and the conserved variables are updated using Equation~\eqref{eq:fv-div}.

\section{Validation of fourth-order finite volume implementation for hydrodynamics subsystem} \label{sec:hydro-tests}
For the purpose of the results in this section, the high-order algorithm described in Section~\ref{sec:fv-hydro} will be designated as RK4+PPM. We begin by validating that RK4+PPM exhibits the expected behavior in the hydrodynamics limit when \(\mathbf{B}=0\). We emphasize comparisons to the second-order counterpart to the high-order algorithm. This scheme, below referred to using the shorthand VL2+PLM, uses:
\begin{itemize}
  \item \( \mathcal{O}(\Delta t^2) \) predictor-corrector time integrator based on the method in \cite{Falle1991} which uses a first-order scheme for the predictor step.
  \item \( \mathcal{O}(\Delta x^2) \) piecewise linear method (PLM) reconstruction of primitive variables using the limiter defined in \textsection 3.1 of  \cite{Mignone2014}.
  \item The midpoint approximation is assumed everywhere in the algorithm; hence the above Laplacian conversions from face-/cell-centered to face-/cell-averaged variables are skipped.
\end{itemize}
VL2+PLM is currently used in Athena++, a new version of the Athena astrophysics code \cite{Stone2008}. We refer the reader to the corresponding Athena method paper \cite{StoneGardiner2009} for a summary of the VL2+PLM scheme for MHD. The HLLC Riemann solver was used in both VL2+PLM and RK4+PPM to produce the hydrodynamics results.

We use \((x_1, x_2)\) to refer to the lab-frame coordinates in the below results. All results are from uniform, Cartesian grids with resolution \((N_{x_1}, N_{x_2})\).  For some problem descriptions and analysis, a rotated vector frame is employed, and \((x,y)\) will refer to the rotated frame coordinates. In such cases, the coordinate transformation will be explicitly given from the lab-frame coordinates.

\subsection{2D slotted cylinder circular advection} \label{subsec:slotted-cylinder}
Before considering the fully nonlinear hydrodynamics regime, the first problem we consider is the slotted cylinder scalar advection test presented in \cite{ColellaDorrHittingerMartin2011} Section 4.4.5. We model scalar advection in our solver by considering a 2D domain governed by an isothermal equation of state, uniform density \(\rho=1\), and rotational flow about \( \mathbf{x}^c = (0.5,0.5) \) defined by
\begin{equation}
(v_1(\mathbf{x}),v_2(\mathbf{x})) = 2\pi \omega (-(x_2 - x_2^c), (x_1 - x_1^c)) \, , \label{eq:cylinder-velocity}
\end{equation}
with \(\omega=1\). Under these assumptions, the out-of-plane velocity component, \(v_3\), is passively advected counterclockwise by the fluid.

\begin{figure}[!ht]
\includegraphics[width=\textwidth]{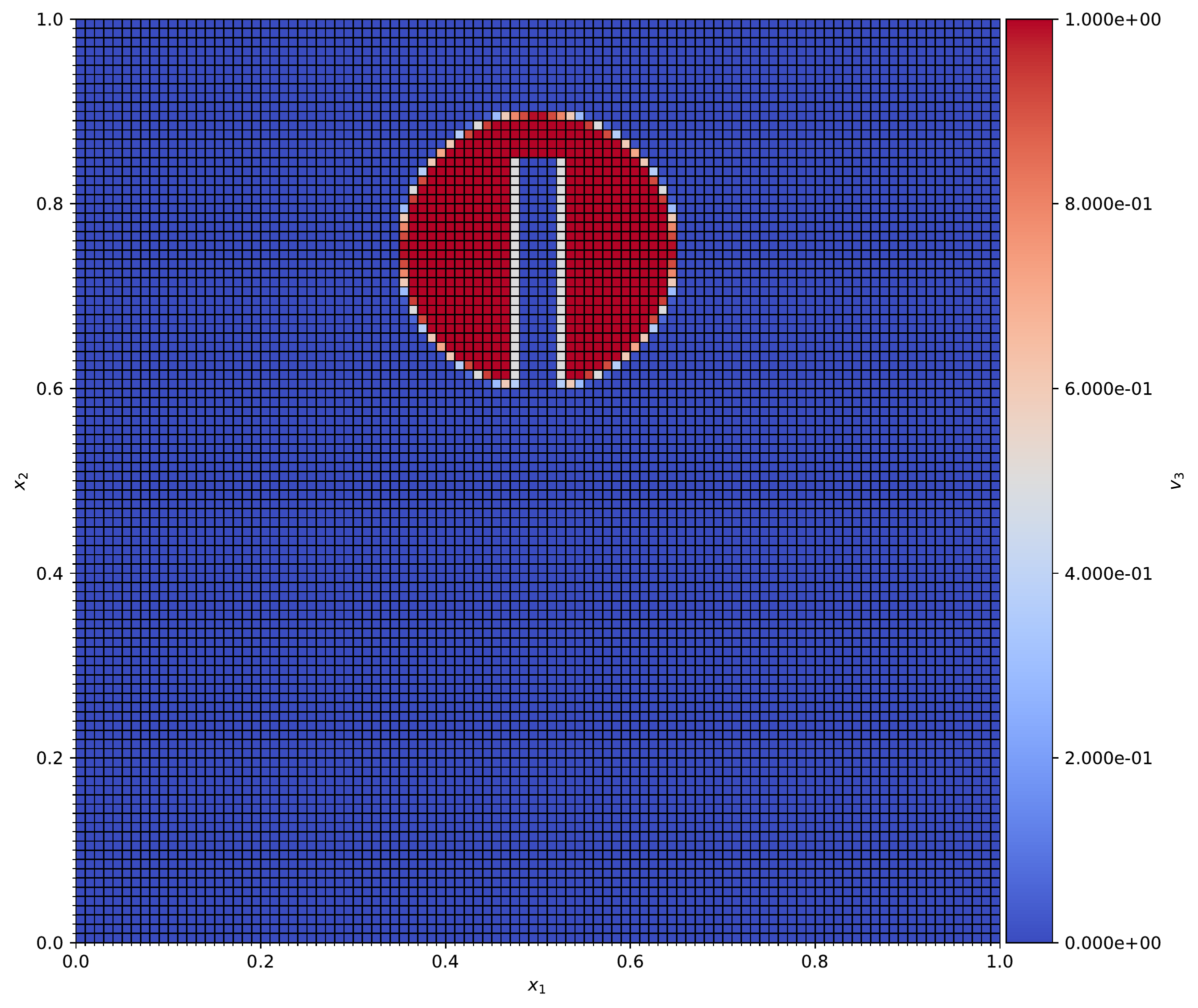}
\caption{The initial condition and reference solution for the slotted cylinder advection test, where \(v_3\) is treated as a passive scalar. The Cartesian grid is composed of \(100 \times 100\) cells, and the boundary conditions are periodic.}
\label{fig:slotted-cylinder-ic-quad}
\end{figure}

Figure~\ref{fig:slotted-cylinder-ic-quad} plots the initial condition of a slotted cylinder with radius \(R=0.15\), slot width \(W=0.05\), and slot height \(H=0.25\) centered on \(\mathbf{x}^* = (0.5, 0.75)\) and
\begin{equation}
v_3(\mathbf{x}) = \left\{
\begin{array}{ll}
      0 & 0 \leq R < r \\
      0 & |2z_1| \leq W \text{ and } 0 < z_2 + R < H \\
      1 & \text{ otherwise } \\
\end{array}
\right. \, , \label{eq:cylinder-profile}
\end{equation}
where \(\mathbf{z} = \mathbf{x} - \mathbf{x}^*, r =|\mathbf{z}| \). At a \(100 \times 100\) resolution, the slot width is exactly five cells wide. We note that the initial condition in \cite{ColellaDorrHittingerMartin2011} is asymmetric; the slot has two cells to the right of and three cells to the left of the center of the domain and cylinder.

\begin{figure}[!ht]
\includegraphics[width=\textwidth]{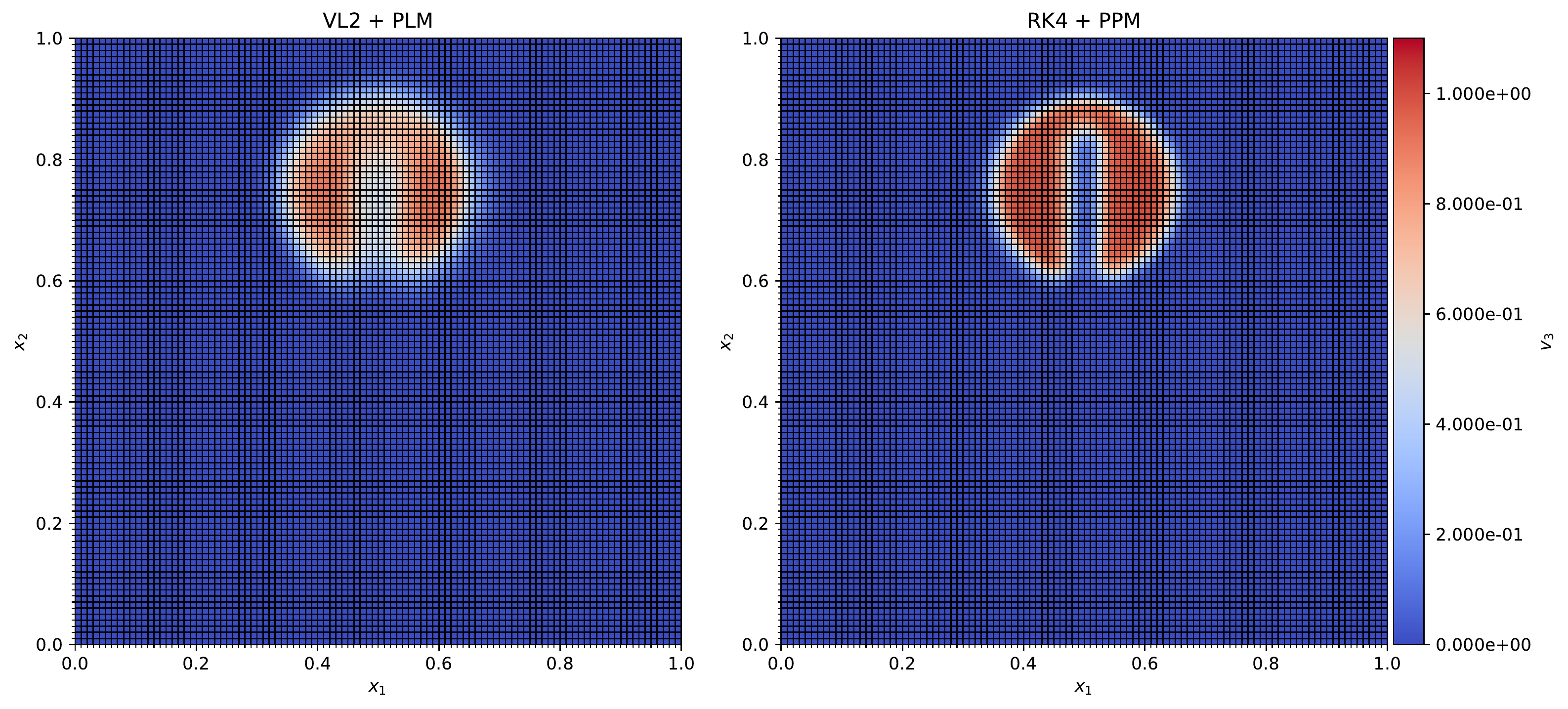}
\caption{Second-order accurate VL2+PLM and the fourth-order accurate RK4+PPM computed solutions at \(t=1.0 \) for the slotted cylinder advection test. While both limiters prevent unphysical oscillations, the high-order solution preserves most of the 5-cell slot while VL2+PLM completely fills it in. }
\label{fig:slotted-cylinder-comparisons-2x}
\end{figure}

Figure~\ref{fig:slotted-cylinder-comparisons-2x} compares the VL2+PLM and RK4+PPM advected solutions after one rotation. Both the PLM and PPM limiters exhibit excellent preservation of the monotonicity of \(v_3\). When an unlimited second- or fourth-order reconstruction is used, unphysical oscillations of the same order of magnitude as the cylinder height appear in the solution. The fourth-order limited solution preserves the discontinuous slot of the cylinder while the second-order limited solution does not. The numerical diffusivity of the two methods can also be compared at the cylinder edge. The smooth transition (shown in white in the diverging color map of Figure~\ref{fig:slotted-cylinder-comparisons-2x}) to the \(v_3=0\) background state is much narrower in the RK4+PPM solution.

\begin{figure}[!ht]
\includegraphics[width=\textwidth]{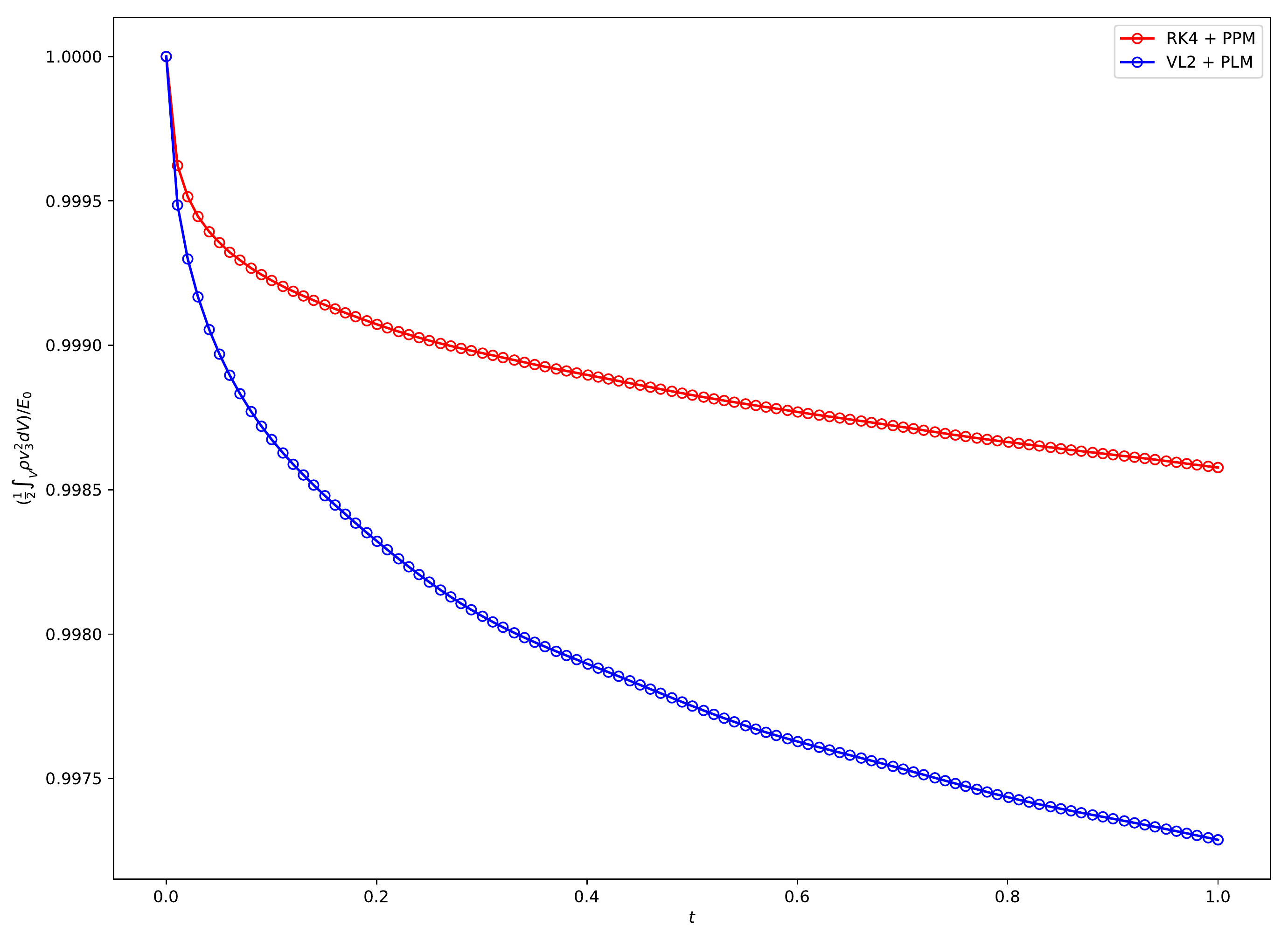}
\caption{Plot of the decay of the volume-averaged concentration of the passive scalar \(v_3^2\) over time. The values are normalized by the initial value, \( E_0\)}
\label{fig:slotted-cylinder-hst}
\end{figure}

Figure~\ref{fig:slotted-cylinder-hst} compares the time-series data of the domain-averaged \(x_3\) component of the kinetic energy. While the total domain \(v_3\) is constant within machine precision throughout the simulation (as is guaranteed by the global conservation property of the finite volume method), the decay of the volume-averaged \(v_3^2\) from the initial reference value provides a measure of the algorithm's numerical dissipation per time step. RK4+PPM causes a much slower initial dissipation of the solution than VL2+PLM, which indicates that high-order reconstruction can be effective even for highly discontinuous data.

\subsection{2D oblique hydrodynamic linear wave convergence} \label{subsec:hydro-linwave}
The next test problem increases the difficulty of the underlying dynamics by introducing the adiabatic hydrodynamics equation of state and smooth waves. However, the full nonlinearity of the Euler system is avoided by restricting the problem to the evolution of planar waves of small, linear amplitude \(\varepsilon\).
The conserved variable profiles are initialized using the exact eigenvectors \(\mathbf{R}_k \) (for each mode \(k\)) of the Euler system linearized about the background state \( \overline{\mathbf{U}}_{k} \). They are best described in the coordinate system that is rotated to be aligned with the wave propagation direction with
\begin{subequations}
\begin{align}
x =& x_1\cos\theta + x_2\sin\theta \, , \label{eq:coord-rotation1} \\
y =& -x_1\sin\theta + x_2\cos\theta \, . \label{eq:coord-rotation2}
\end{align}
\end{subequations}
The conserved quantities vary sinusoidally with \(x\), the coordinate along the parallel rays of the wavefront that are obliquely oriented relative to the \(N_{x_1}\times \frac{N_{x_1}}{2}\) grid. The periodic domain extends from \( 0 \leq x_1 \leq \sqrt{5} \) and \( 0 \leq x_2 \leq \frac{\sqrt{5}}{2} \) to ensure square cells. The wavevector direction is set to \(\theta = \tan^{-1}(2) \approx 63.43^\circ\) inclined with respect to the \(x_1\)-axis, and the wavelength is \( \lambda=1\). With these parameters, exactly one wavelength propagates along each domain boundary in one period. Furthermore, the problem is truly multidimensional, as there is no symmetry between the \( x_1 \) and \(x_2\) fluxes.

The eigenfunctions for the sound and entropy wave modes must be initialized at fourth-order or greater accuracy on the mesh. While the exact cell-averaged initial condition could be calculated analytically in this case, the existing
second-order accurate initialization of the linear wave problem in Athena++ is extended to fourth-order accuracy using a two step process. A similar procedure can be applied to correct other smooth initial conditions approximated by
cell-centered approximations. First, the eigenfunctions are evaluated at the cell center position in the rotated coordinate frame, resulting in a midpoint approximation to the cell-averaged conserved variables,
\begin{equation}
\mathbf{U}^0 = \overline{\mathbf{U}}_{k} + \varepsilon \mathbf{R}_k \cos(2\pi x) \, . \label{eq:linwave-ic}
\end{equation}
Then, the Laplacian operator in Equation~\eqref{eq:laplacian} is applied to get a fourth-order accurate approximation to cell-averaged initial conserved data
\begin{equation}
\langle \mathbf{U}^0 \rangle_{i,j} = \mathbf{U}^0_{i,j}+ \frac{h^2}{24} \Delta \mathbf{U}^0_{i,j} \, . \label{eq:laplacian-ic4}
\end{equation}
We let \(\varepsilon=10^{-6}\) for all the results shown here. The uniform background is \( \rho = 1, P = 3/5, \gamma = 5/3 \), thus the sound speed is \(c_s=1\). For the sound wave, the background flow velocity is \(v_1 = 0 \); for the entropy wave \(v_1 = 1 \).

\begin{figure}[!ht]
\includegraphics[width=\textwidth]{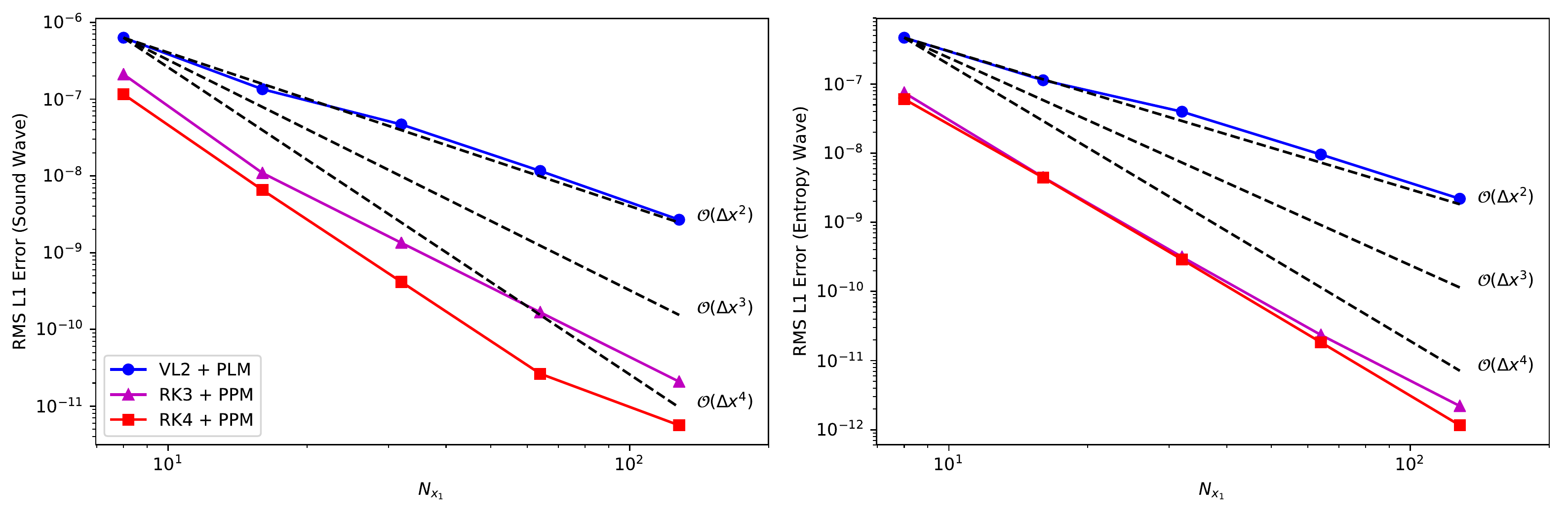}
\caption{Hydrodynamic linear wave convergence plots of the adiabatic sound and entropy modes. The high-order schemes reduce the error nearly by three orders of magnitude relative to the second-order scheme at the largest resolution.}
\label{fig:hydro-linwave}
\end{figure}

Due to the global smoothness of the wave, this problem does not test the behavior of the limiter at discontinuities. Nevertheless, the linear wave test is a discriminating challenge of the algorithm's formal order of accuracy. The wave propagates for one wavelength, and the evolved solution is compared to the initial condition. The vector of \(L_1\) errors of each \(s\) of the \( N_{hydro} \) conserved variables at timestep \(n\) is
\begin{equation}
\delta \mathbf{U}^n = \frac{1}{N_{x_1}N_{x_2}} \sum_{i,j} | \langle \mathbf{U}^n \rangle_{i,j}  - \langle \mathbf{U}^0 \rangle_{i,j} | \, . \label{eq:l1-vector}
\end{equation}
Figure~\ref{fig:hydro-linwave} displays in logarithmic scale the convergence of the root mean square norm of the \(L_1\) error vector
\begin{equation}
\norm{\delta \mathbf{U}^n} = \sqrt{\sum_s^{N_{hydro}} (\delta U^n_s)^2} \, , \label{eq:rms-l1}
\end{equation}
for the sound and entropy wave modes for resolutions spanning \(8\times 4\) to \(128\times 64 \) cells. Example second, third, and fourth-order convergence rates are juxtaposed as dashed lines. In Figure~\ref{fig:hydro-linwave}, the plots show that the errors for both the RK3+PPM and RK4+PPM methods converge much faster than for VL2+PLM. The RK4+PPM solution on a \(16\times8\) grid has similar accuracy as the VL2+PLM solution at the \(128\times 64 \) resolution.

Despite initially converging at fourth-order, the RK3+PPM sound wave error converges at only third-order with the fixed CFL number of 0.4 for most resolutions, which indicates that the temporal error dominates for this test and solver configuration. The third-order convergence turn-over point decreases to before the smallest \(N_{x_1}=8\) when the CFL number is increased to 0.8. The RK3+PPM error lines nearly exactly match the RK4+PPM errors when the CFL is decreased to 0.2. At the largest resolution considered, the convergence stops as nonlinear steepening effects invalidate the linear approximation. This occurs at errors near \( L_1 \approx \varepsilon^2 = 10^{-12} \). Floating-point round-off error may affect the error convergence at smaller wave amplitudes and larger resolutions.

\subsection{1D Shu-Osher shock tube} \label{subsec:shu-osher}
The final hydrodynamics validation test is a fully nonlinear shock tube problem. The Shu-Osher problem involves the interaction of a discontinuous shock front propagating from \(x_1=-0.8\) with a sinusoidal smooth flow \cite{ShuOsher1989}. The adiabatic index is \( \gamma=\frac{7}{5} \) and the domain spans \( x_1 \in [-1, 1] \), with initial condition to the left and right of \(x_1=-0.8\)   given by
\begin{equation}
\begin{pmatrix}
\rho^L \\
v_1^L \\
v_2^L \\
v_3^L \\
P^L
\end{pmatrix}
=
\begin{pmatrix}
3.857143 \\
2.629369 \\
0 \\
0 \\
10.3333
\end{pmatrix}  \, ,
\begin{pmatrix}
\rho^R \\
v_1^R \\
v_2^R \\
v_3^R \\
P^R
\end{pmatrix}
=
\begin{pmatrix}
1+ 0.2\sin(5\pi x_1)\\
0 \\
0 \\
0 \\
1.0
\end{pmatrix}
\, . \label{eq:shu-osher-states}
\end{equation}
The interaction produces a density profile containing both discontinuities and smooth structure composed of a large range of wavelengths.

\begin{figure}[!ht]
\includegraphics[width=\textwidth]{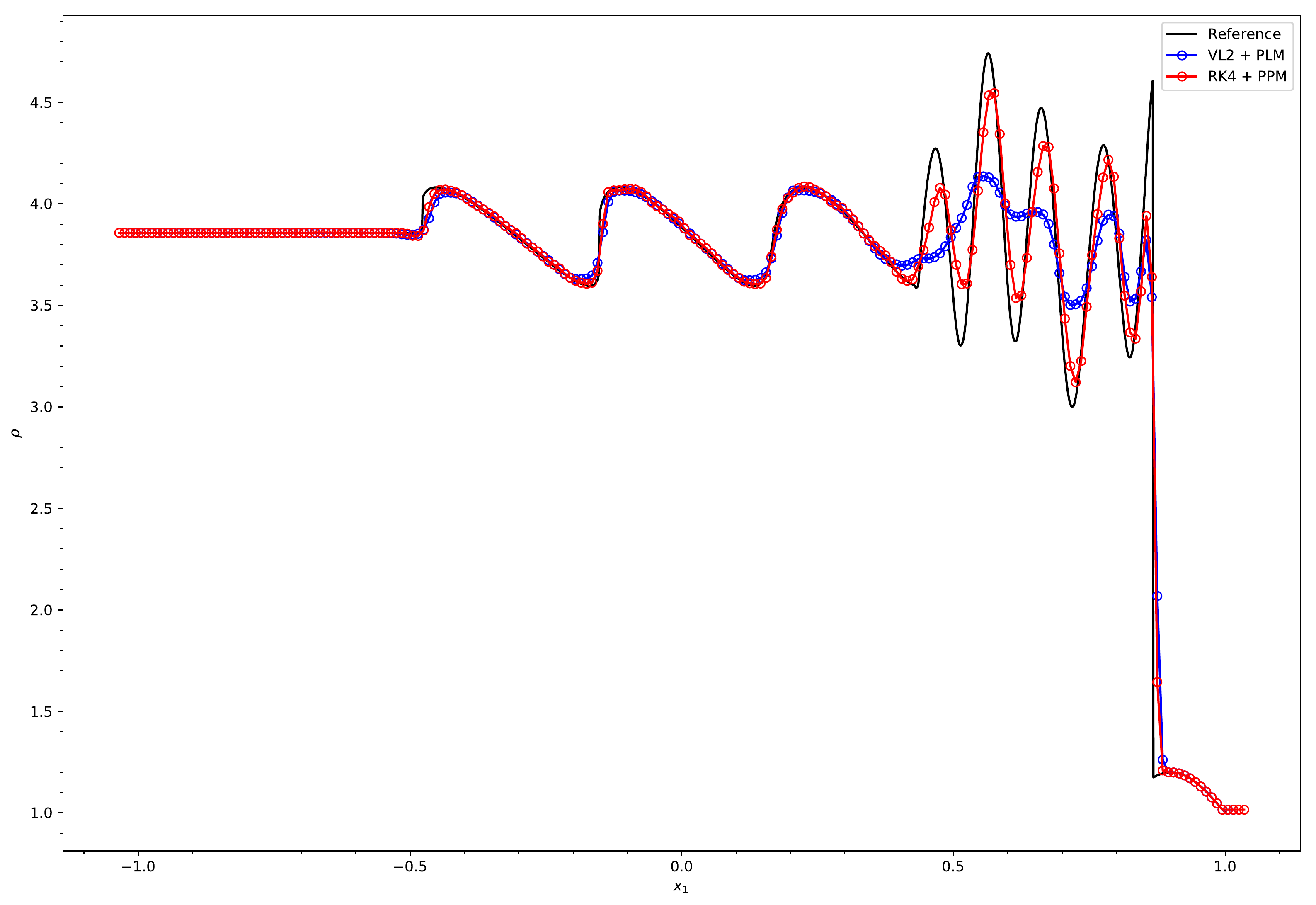}
\caption{Second-order and fourth-order accurate solutions of the density in the Shu-Osher shock tube at \(t_f=0.47\) are shown for \(N_{x_1}=200 \) cells are shown in comparison to the \(N_{x_1}=8000\) reference solution. Both algorithms capture the long wavelength smooth features, but the fourth-order method yields the largest improvements in solution accuracy in regions where the profile changes rapidly over a few grid cells. }
\label{fig:shu-osher}
\end{figure}

Figure~\ref{fig:shu-osher} compares the \(N_{x_1}=200\) low-resolution results from the second-order algorithm VL2+PLM with the results from the fourth-order algorithm RK4+PPM. The solid line is a high-resolution \(N_{x_1}=8000\) reference solution computed with the RK4+PPM method. The solutions are shown at \(t_f=0.47\). The smooth extremum preserving PPM limiters are essential to prevent the clipping of the many short wavelength peaks at the low-resolution. The second-order VL2+PLM solution experiences large dissipation near the short wavelength structures due to the frequent and severe extremum clipping of the sinusoidal profile at this resolution.

\section{Upwind constrained transport implementation} \label{sec:uct}
Having validated the fourth-order finite volume method for the hydrodynamics subsystem, Equation~\eqref{eq:conservation-law}, we reintroduce the full MHD system of equations and focus on the treatment of the magnetic field. The induction equation enters the conservative system of equations of MHD, but it is not evolved using the techniques from Section~\ref{sec:fv-hydro}. In differential form, the equation is
\begin{equation}
\frac{\partial \mathbf{B}}{\partial t} + \nabla \times \mathscr{E} = 0 \, , \label{eq:induction}
\end{equation}
where \(\mathscr{E} = -\mathbf{v} \times \mathbf{B} \) is the electric field under the assumption of ideal MHD. Constrained transport ensures a strictly solenoidal magnetic field by evolving field quantities averaged on cell faces. By applying Stoke's theorem to Equation~\eqref{eq:induction} in 2D, the following difference formulas in semi-discrete form are derived
\begin{subequations}
\begin{align}
\frac{\mathrm{d}}{\mathrm{d}t} \langle B_{1} \rangle_{i-\frac{1}{2},j}& =- \frac{1}{h} (\langle \mathscr{E}_{3} \rangle_{i-\frac{1}{2},j+\frac{1}{2}} - \langle \mathscr{E}_{3} \rangle_{i-\frac{1}{2},j-\frac{1}{2}} )\label{eq:induction-difference-1}\\
\frac{\mathrm{d}}{\mathrm{d}t} \langle B_{2} \rangle_{i,j-\frac{1}{2}}& = \frac{1}{h} (\langle \mathscr{E}_{3} \rangle_{i+\frac{1}{2},j-\frac{1}{2}} - \langle \mathscr{E}_{3} \rangle_{i-\frac{1}{2},j-\frac{1}{2}} ) \, . \label{eq:induction-difference-2}
\end{align}
\end{subequations}
In this formulation, the magnetic flux is a conserved quantity while the fluxes are line-averaged corner (edges in 3D) electric fields \(\langle \mathscr{E}_{3} \rangle \), or emf; therefore, exact maintenance of the divergence-free condition in Equation~\eqref{eq:div-b} is a result of the local conservation property of the numerical method. While this advantage of the CT technique is well known, the staggered discretization of the field on cell faces introduces a dual, independent representation of the magnetic field quantities. Approaches to coupling CT with upwinded, cell-averaged quantities of the Godunov scheme are highly varied and may not prevent the onset of numerical monopoles in Equation~\eqref{eq:div-b} \cite{Londrillo2004}.

Gardiner and Stone \cite{GardinerStone2005} (hereafter GS05) present a 2D CT scheme that couples to the underlying corner transport upwind (CTU) method by transversely upwinding the electric fields of the flux vectors returned by the Godunov-type method. The algorithm was extended to 3D for CTU in \cite{GardinerStone2008} (hereafter GS08) and to a simplified Godunov-type method, VL2 in \cite{StoneGardiner2009}. The multidimensional construction of CT schemes in \textsection 3.2 of GS05 proceeds by considering schemes that reduce to the analytic solution for plane-parallel, grid-aligned flow. They begin by pointing out that the viscous flux contribution of the CT algorithm based on arithmetic averaging of the emf \cite{BalsaraSpicer1999} must be doubled for stability. The modified scheme is referred to as \(\mathscr{E}_z^\circ \). The authors then construct two other novel CT schemes \(\mathscr{E}_z^\alpha \) and \(\mathscr{E}_z^c \), motivated by the Local Lax-Friedrichs and upwinding methods, respectively, applied to the emf derivatives in the differentiated induction equations. We refer the reader to Section 3.2.2 of GS05 for the derivations \cite{GardinerStone2005}. After a rigorous comparison of the three proposed CT schemes, \(\mathscr{E}_z^\circ, \mathscr{E}_z^\alpha, \mathscr{E}_z^c, \) the authors concluded that \(\mathscr{E}_z^c \), which upwinds the emf by the contact mode direction, produces a stable, non-oscillatory CT scheme with optimal numerical viscosity.

In this method, the upwinding of the face-averaged emf contained in the Godunov fluxes is completed from the nearest four cell faces and averaged for each corner. Equation 41 of GS05 expresses the spatial average of the four \(\mathcal{O}(\Delta x^2)\) estimates to the corner emf as
\begin{multline}
\mathscr{E}_{3,i-\frac{1}{2},j-\frac{1}{2}} = \frac{1}{4} (  \mathscr{E}_{3,i,j-\frac{1}{2}} + \mathscr{E}_{3,i-1,j-\frac{1}{2}} +\mathscr{E}_{3,i-\frac{1}{2},j} +\mathscr{E}_{3,i-\frac{1}{2},j-1} )  \\
+\frac{h}{8} \left( \left(\frac{\partial \mathscr{E}_3}{\partial x_2} \right)_{i-\frac{1}{2},j-\frac{3}{4}} - \left(\frac{\partial \mathscr{E}_3}{\partial x_2} \right)_{i-\frac{1}{2},j-\frac{1}{4}} \right)
+\frac{h}{8} \left( \left(\frac{\partial \mathscr{E}_3}{\partial x_1} \right)_{i-\frac{3}{4},j-\frac{1}{2}} - \left(\frac{\partial \mathscr{E}_3}{\partial x_1} \right)_{i-\frac{1}{4},j-\frac{1}{2}} \right) \, .  \label{eq:gs05-emf-average}
\end{multline}
The approximations to the derivatives are upwinded in the transverse direction, and the upwind directions are based on the fluid contact mode for both \(x_1\) and \(x_2\). For example, the expression for upwinding the \(\partial_2\) spatial derivatives to \(x_1\) faces is
\begin{equation}
\left(\frac{\partial \mathscr{E}_3}{\partial x_2} \right)_{i-\frac{1}{2},j-\frac{1}{4}} =
\begin{cases}
\left(\frac{\partial \mathscr{E}_3}{\partial x_2} \right)_{i-1,j-\frac{1}{4}} & v_{1,i-\frac{1}{2}} > 0 \\
\left(\frac{\partial \mathscr{E}_3}{\partial x_2} \right)_{i,j-\frac{1}{4}} & v_{1,i-\frac{1}{2}} < 0  \\
\frac{1}{2} \left[ \left(\frac{\partial \mathscr{E}_3}{\partial x_2} \right)_{i-1,j-\frac{1}{4}}  + \left(\frac{\partial \mathscr{E}_3}{\partial x_2} \right)_{i-1,j-\frac{1}{4}} \right] & \text{ otherwise}
\end{cases} \, . \label{eq:gs05-ct-upwind}
\end{equation}

The \(\mathscr{E}_z^c \) GS05 CT scheme is the exclusive CT discretization used in subsequent publications by the authors \cite{GardinerStone2008, Stone2008, StoneGardiner2009} and in the Athena astrophysics code \cite{Stone2008}. We remark that \(\mathscr{E}_z^\circ \) has been occasionally misidentified as the final GS05 CT scheme in subsequent literature such as \cite{Balsara2014a}.\footnote{See Balsara (2014) \textsection 9.5 for results from a field loop advection test generated by the GS05 \(\mathscr{E}_z^\circ \) method.} We also note that there are typos in subsequent formulations of GS05 Equation 41 due to a change in indexing from the upper corner \(\mathscr{E}_{z,i+\frac{1}{2},j+\frac{1}{2}}\) to the lower corner \(\mathscr{E}_{z,i-\frac{1}{2},j-\frac{1}{2}}\). The signs of the derivative terms in \cite{Stone2008} Equation 79 and \cite{StoneGardiner2009} Equation 22 are all incorrect.

While this CT scheme is consistent with the underlying Godunov-type algorithm for 1D plane parallel solutions, the consistency and accuracy are at most second-order in spatial resolution. The algorithm contains several steps and assumptions that limit the overall spatial accuracy of the scheme to \(\mathcal{O}(\Delta x^2)\), even when combined with a higher-order base finite volume scheme:
\begin{enumerate}
\item By truncating higher-order derivative terms, Equation 40 of GS05 provides a single spatial estimate for the emf at a grid cell corner using a second-order Taylor-series expansion across a face
\begin{equation}
\mathscr{E}_{3,i-\frac{1}{2},j-\frac{1}{2}} = \mathscr{E}_{3,i-\frac{1}{2},j} - \frac{h}{2}\left(\frac{\partial \mathscr{E}_3}{\partial x_2} \right)_{i-\frac{1}{2},j-\frac{1}{4}} + \mathcal{O}(\Delta x^2) \, . \label{eq:gs05-taylor-series}
\end{equation}
\item The face-centered emf quantities in Equations~\eqref{eq:gs05-emf-average} and~\eqref{eq:gs05-taylor-series} are midpoint approximations.
\item The cell-averaged magnetic field components, necessary for the transverse reconstruction steps in the Godunov-type finite volume scheme, are derived at second-order accuracy using the average of the longitudinal face-averaged values. Equations 19 and 20 of GS05 define this final step in the CT scheme after the update of the face-averaged magnetic field components in induction Equations~\eqref{eq:induction-difference-1} and~\eqref{eq:induction-difference-2}:
\begin{subequations}
\begin{align}
\langle B_{1} \rangle_{i,j} =& \frac{1}{2}(\langle B_{1} \rangle_{i+\frac{1}{2},j} + \langle B_1 \rangle_{i-\frac{1}{2},j}) \, , \label{eq:average-b1-face} \\
\langle B_{2} \rangle_{i,j} =& \frac{1}{2}(\langle B_{2} \rangle_{i,j+\frac{1}{2}} + \langle B_2 \rangle_{i,j-\frac{1}{2}}) \, . \label{eq:average-b2-face}
\end{align}
\end{subequations}
\item Most subtly, the \( \mathscr{E}_z^c \) scheme only captures dimensionally split approximations to the multidimensional Riemann fan at the cell corner. The upwinded emf slopes in Equation~\eqref{eq:gs05-ct-upwind} are approximations given by GS05 Equation 45 as
\begin{equation}
\left(\frac{\partial \mathscr{E}_3}{\partial x_1} \right)_{i-\frac{3}{4},j} = \frac{2}{h} (\mathscr{E}_{3,i-\frac{1}{2},j} - \mathscr{E}^r_{3,i-1,j}) \, , \label{eq:emf-slope}
\end{equation}
where \( \mathscr{E}_{3,i-\frac{1}{2},j} \) is the magnetic flux returned by an approximate 1D Riemann solver and \(\mathscr{E}^r_{3,i,j} = v_{2,i,j}B_{1,i,j} - v_{1,i,j}B_{2,i,j} \) is the cell-centered reference electric field. Regardless of the order of accuracy of the reconstruction method, this difference between the cell-centered reference electric field and the face-centered magnetic flux limits the approximation to second-order accuracy.
\end{enumerate}
While the first three second-order accurate assumptions of the \( \mathscr{E}_z^c \) scheme can all be addressed by replacing them with higher-order approximations, the final limitation cannot be generalized to fourth-order accuracy in a straightforward fashion. See~\ref{sec:appendix-gs05-uct} for detailed analysis, including a proof demonstrating that the upwinding of fluxes from approximate Riemann solutions on \(x_1,x_2\) faces (instead of smooth approximations to \(\mathscr{E}_3\)) in each dimension causes the scheme to fail to reduce to the 2D planar wave modes along the \(x-y, x+y\) cell diagonals.

The upwind constrained transport framework of Londrillo and Del Zanna, developed in \cite{Londrillo2000} (hereafter LD2000) and generalized in \cite{Londrillo2004} (hereafter LD2004), is an approach to implementing a CT discretization that is not subject to the \(\mathcal{O}(\Delta x^2)\) limitations above. In \ref{sec:appendix-gs05-uct}, we show that the \( \mathscr{E}_z^c \) CT scheme is consistent to within second-order approximations to UCT, but the derivation of the scheme in GS05 is not extensible to higher-order for the above reasons.

UCT defines two phases for the treatment of the MHD subsystem of Equation~\eqref{eq:induction}: the reconstruction phase and the upwind phase \cite{Londrillo2000}. This dichotomy is analogous to the steps for computing fluxes in a high-order generalization of Godunov's scheme for hydrodynamics \cite{Londrillo2000}. Whereas the reconstruction and upwind phases are typically formulated to approximate the fluxes at cell faces in hydrodynamics, the UCT analogy is made to compute the fluxes (electric fields in induction Equation~\eqref{eq:induction}) at cell corners in 2D. Furthermore, the UCT steps require special handling to ensure that the divergence-free and field line continuity constraints of MHD are maintained.

Therefore, the implementation of a fourth-order accurate UCT method consistent with the finite volume hydrodynamics subsystem of Section~\ref{sec:fv-hydro} requires specification of two main algorithmic components:
\begin{enumerate}
\item Reconstruction of quantities along cell faces to cell corners
\item Corner upwinding procedure that approximates the solution to a multidimensional Riemann problem for the magnetic fluxes
\end{enumerate} As in the steps of the hydrodynamics subsystem of Section~\ref{sec:fv-hydro}, each component of the UCT implementation must treat all approximations at fourth-order accuracy. In the following subsections, we specify these techniques used in our overall scheme at the end of each integrator substage, after the calculation of the Godunov fluxes in Section~\ref{subsubsec:fv-flux} but before the flux-divergence is applied to update the cell-averaged conserved variables in Equation~\eqref{eq:fv-div}. For the 2D algorithm, \(\langle B_3\rangle\) is evolved using the finite volume techniques of Section~\ref{sec:fv-hydro}.

\subsection{Reconstruction phase} \label{subsec:uct-reconstruction}
\begin{figure}[!ht]
\begin{center}
\includegraphics[width=0.5\textwidth]{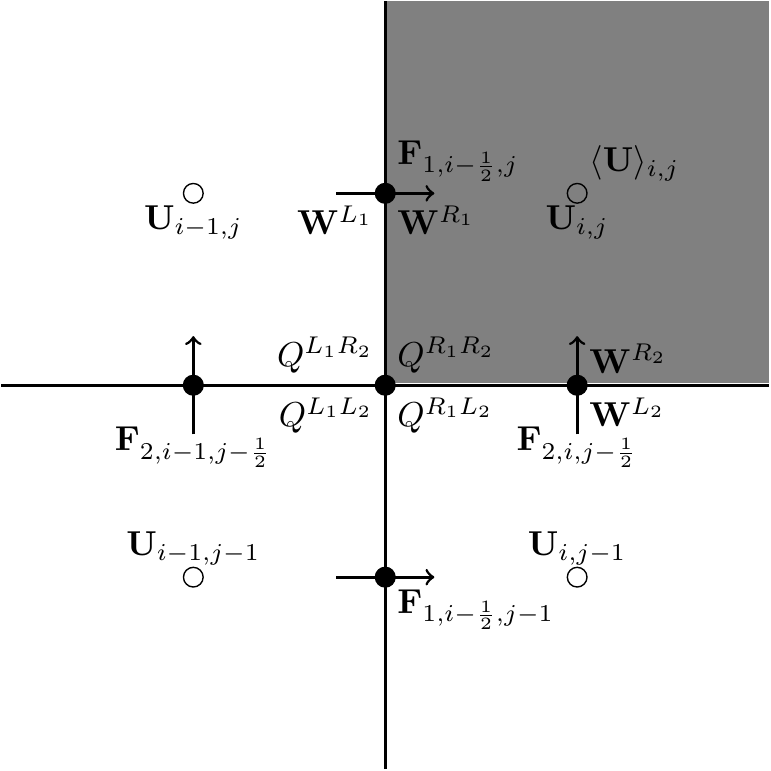}
\end{center}
\caption{2D slice of \(x_1-x_2\) plane that shows the locations of cell-centered conserved variables, face-centered 1D Riemann states and upwind flux components, and the four-state function of a quantity \(Q\) reconstructed in 2D for each of the nearest cells at the \(i-\frac{1}{2},j-\frac{1}{2}\) corner. The cell-averaged conserved variables are shown in the \(i,j\) cell as a shaded gray square to emphasize that all of the other displayed cell-/face-centered quantities are distinct from the corresponding cell-/face-averaged quantities (not shown).}
\label{fig:corner-states-diagram}
\end{figure}

Just as limiting may introduce dual \(L/R \) states collocated at cell faces, limited piecewise polynomial reconstruction may introduce four discontinuous states at cell corners. LD2004 establishes notation for the four-state functions by referring to the orientation of each state relative to the center of its reconstructed cell: \(Q^{NW}, Q^{NE}, Q^{SE}, Q^{SW} \) \cite{Londrillo2000, Londrillo2004}. Note, these are counter-intuitive if you consider the cardinal directions relative to the corner, and the \(N/S\) states corresponding to discontinuities in the \(x_2\) direction come before the \(E/W\) states used for jumps across \(x_1\) in the superscript. Therefore, we adopt the notation in Equation 39 of \cite{Amano2015}, which uses superscripted states relative to \(x_1, x_2 \) interfaces. Figure~\ref{fig:corner-states-diagram} summarizes the locations of these states. The states equivalent to the above LD2004 states are \(Q^{R_1L_2}, Q^{L_1L_2}, Q^{L_1R_2}, Q^{R_1R_2} \), respectively.

Unlike normal reconstructed quantities in hydrodynamics, the magnetic field quantities cannot be freely represented by the basis of piecewise polynomials. The components of the field must have single-valued states at longitudinal faces in order to ensure that the field lines are continuous \cite{Londrillo2000}. For example,
\begin{equation}
\langle B_1^{L_1} \rangle = \langle B_1^{R_1} \rangle = \langle B_1 \rangle \,  \label{eq:B1-face-continuity}
\end{equation}
must be satisfied at an \(x_1\) face. At an \( x_1 \)-\(x_2\) corner, discontinuous states of \(B_1(x_1,x_2)\) can only occur across the \(x_2\) jump as
\begin{subequations}
\begin{align}
\langle B_1^{L_1R_2} \rangle =& \langle B_1^{R_1R_2} \rangle = \langle B_1^{R_2} \rangle \, , \label{eq:B1R2-corner-continuity} \\
\langle B_1^{L_1L_2} \rangle =& \langle B_1^{R_1L_2} \rangle = \langle B_1^{L_2} \rangle \, .  \label{eq:B1L2-corner-continuity}
\end{align}
\end{subequations}
In UCT the longitudinal face-averages for each magnetic field component, such as \(\langle B\rangle_{i\pm\frac{1}{2},j} \) for the \( x_1\)-component, become the primary representation of the field; these quantities are never reconstructed. From stencils of these quantities, both the cell-averaged \(\langle {B_1}\rangle_{i,j}\) required by the finite volume hydrodynamics subsystem in Section~\ref{sec:fv-hydro} and the corner states \( \langle B_1^{R_2} \rangle_{i-\frac{1}{2},j-\frac{1}{2}}, \langle B_1^{L_2} \rangle_{i-\frac{1}{2},j-\frac{1}{2}} \) required by the CT scheme below are reconstructed.

\subsubsection{Reconstruction of corner electric fields} \label{subsubsec:uct-reconstruction-emf}
After completing the reconstruction step of the hydrodynamics subsystem in Section~\ref{subsubsec:fv-reconstruct} for all face-averages, we compute fourth-order accurate reconstructions of the emf at each cell corner in 2D from the necessary velocity and magnetic field components using
\begin{equation}
\langle \mathscr{E}_{3} \rangle_{i-\frac{1}{2},j-\frac{1}{2}} =  \langle v_{2} \rangle_{i-\frac{1}{2},j-\frac{1}{2}} \langle B_{1} \rangle_{i-\frac{1}{2},j-\frac{1}{2}} - \langle v_{1} \rangle_{i-\frac{1}{2},j-\frac{1}{2}} \langle B_{2} \rangle_{i-\frac{1}{2},j-\frac{1}{2}} \, . \label{eq:corner-emf-unlimited}
\end{equation}
We continue to use the angled bracket notation in Equation~\eqref{eq:corner-emf-unlimited} to emphasize that the emfs are line-averaged along cell edges in 3D and to distinguish the quantities from their second-order counterparts from the GS05 CT scheme.

\begin{figure}[!ht]
\begin{center}
\includegraphics[width=0.5\textwidth]{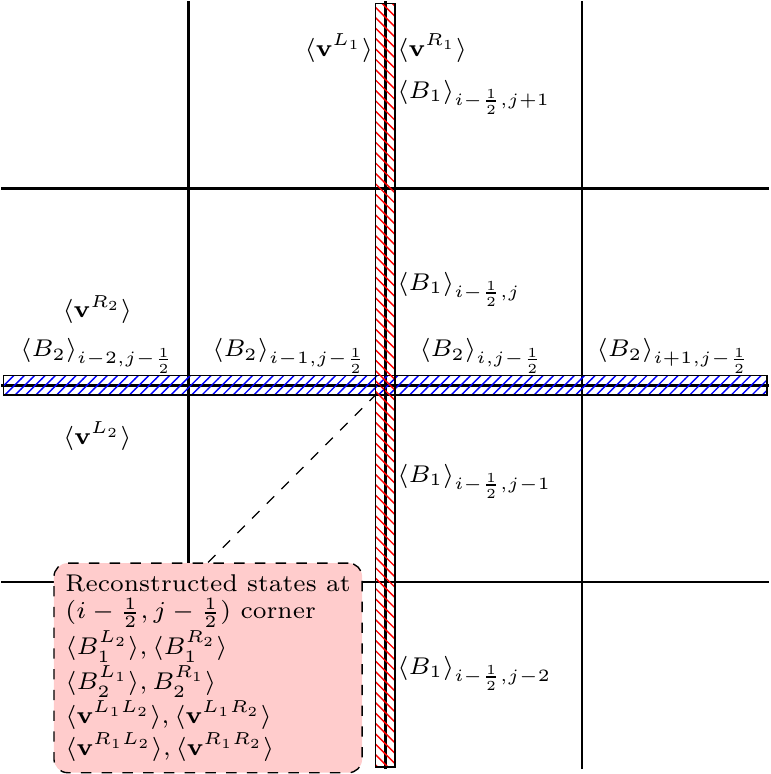}
\end{center}
\caption{The stencils of face-averaged quantities used to compute the four-state emf at a corner to fourth-order accuracy using PPM4. The magnetic field components are single-valued on longitudinal faces because these face-averages are never reconstructed, but the limiting in the transverse reconstruction may introduce two discontinuous Riemann states at the corner. Unlike \(\mathbf{B}\), the 1D reconstructions of face-averaged velocity components are dual-valued at all faces, so the subsequent application of PPM4 results in a four-state function of \(\mathbf{v} \) at the corner for both \(x_1\) and \(x_2\) face-averages. The independent corner reconstructions of velocity are averaged in our method.}
\label{fig:corner-ppm-diagram}
\end{figure}

The above calculation at each corner of each cell is accomplished by performing transverse reconstructions of previously reconstructed \(L/R\) Riemann states of \(\mathbf{v},\mathbf{B}\) along cell faces. We apply PPM4 as in Section~\ref{subsubsec:fv-reconstruct} to suppress spurious oscillations that may arise from physically admissible discontinuities. However, there is nothing inherent to this UCT formulation that requires the use of PPM for these reconstructions. Future work will compare the computational efficiency and accuracy of results from using alternative methods for non-oscillatory reconstructions such as WENO in this step.
For example, the  \(\langle \mathscr{E}^{R_1R_2}_3\rangle_{i-\frac{1}{2}, j-\frac{1}{2}}  = \langle \mathscr{E}^{SW}_3\rangle_{i-\frac{1}{2}, j-\frac{1}{2}}  \) state at the \(i-\frac{1}{2},j-\frac{1}{2} \) corner in the \(i,j\) cell is approximated by:
\begin{enumerate}
\item \( \langle B_1^{L_2}\rangle_{i-\frac{1}{2}, j-\frac{1}{2}}, \langle B_1^{R_2}\rangle_{i-\frac{1}{2}, j-\frac{1}{2}} \) are reconstructed along the \(x_2 \) coordinate from \(x_1\)-face-averaged field data, \( \langle B_1 \rangle_{i-\frac{1}{2}, j} \).
\item \( \langle B_2^{L_1}\rangle_{i-\frac{1}{2}, j-\frac{1}{2}}, \langle B_2^{R_1}\rangle_{i-\frac{1}{2}, j-\frac{1}{2}}  \) are reconstructed along the \(x_1 \) coordinate from \(x_2\)-face-averaged field data, \( \langle B_2 \rangle_{i, j-\frac{1}{2}} \).
\item Both components of \(\mathbf{v} \) are independently reconstructed along both \(x_1\) and \( x_2 \) face-averaged data. Because limited 1D reconstruction operations may not commute, the independent velocity estimates are averaged to approximate \(\langle v_1^{R_1R_2} \rangle_{i-\frac{1}{2}, j-\frac{1}{2}}, \langle v_2^{R_1R_2}\rangle_{i-\frac{1}{2}, j-\frac{1}{2}}\).
\item Finally, the approximation to the emf is computed using
\begin{equation}
\langle \mathscr{E}^{R_1R_2}_3\rangle_{i-\frac{1}{2}, j-\frac{1}{2}} = \langle v_2^{R_1R_2} \rangle_{i-\frac{1}{2}, j-\frac{1}{2}} \langle B_1^{R_2}\rangle_{i-\frac{1}{2}, j-\frac{1}{2}} - \langle v_1^{R_1R_2}\rangle_{i-\frac{1}{2}, j-\frac{1}{2}}\langle B_2^{R_1} \rangle_{i-\frac{1}{2}, j-\frac{1}{2}} \, ,\label{eq:corner-emf-limited}
\end{equation}
the limited version of Equation~\eqref{eq:corner-emf-unlimited}.
\end{enumerate}
The quantities used to compute these corner reconstructions are illustrated in Figure~\ref{fig:corner-ppm-diagram}. After the four-state emf function is approximated at each cell corner, the UCT upwinding step discussed in Section~\ref{subsec:uct-upwind} is used to select a single valued \( \langle \mathscr{E}^{U}_3\rangle_{i-\frac{1}{2}, j-\frac{1}{2}} \).

\subsubsection{Reconstruction of cell-averaged \(\langle\mathbf{B} \rangle_{i,j} \)} \label{uct-reconstruction-B}
Before discussing the UCT corner upwinding procedure, we discuss the final reconstruction step necessary to relate the primary field representations of longitudinal face averages \( \langle B_1 \rangle_{i\pm\frac{1}{2}, j}, \langle B_2 \rangle_{i, j\pm\frac{1}{2}} \) to the derived cell-averaged field at \(\mathcal{O}(\Delta x^4)\) accuracy. While this is the final UCT step in a single integration substage and occurs after the content of Section~\ref{subsec:uct-upwind}, it is a reconstruction procedure, so we provide the details in this section.

After evolving the face-averaged magnetic field quantities in the induction difference Equations~\eqref{eq:induction-difference-1} and~\eqref{eq:induction-difference-2}, the cell-averaged \(\langle\mathbf{B} \rangle_{i,j} \) must be updated to be consistent with the evolved field. At second-order accuracy, this consistency relationship is typically maintained by taking the average of the longitudinal face-averaged components as seen in Equations~\eqref{eq:average-b1-face} and~\eqref{eq:average-b2-face}. Using the techniques from Section~\ref{subsubsec:eos} and the finite difference Laplacian operator in Equation~\eqref{eq:laplacian}, we can perform an ``inverse-reconstruction'' of the magnetic field at fourth-order accuracy. For example, the procedure for the \(x_1\) field component follows:
\begin{enumerate}
\item Using the Laplacian operator consisting of transverse derivatives, convert the face-averaged field to an approximation of the face-centered field on longitudinal faces
\begin{equation}
B_{1,i-\frac{1}{2},j} = \langle B_1 \rangle_{i-\frac{1}{2},j}  - \frac{h^2}{24} \Delta^{\perp,1} \langle B_1 \rangle_{i-\frac{1}{2},j} \, . \label{eq:B1-face-center}
\end{equation}
\item Interpolate along longitudinal face-centers to cell-centered field components
\begin{equation}
B_{1,i,j} = -\frac{1}{16} (B_{1,i-\frac{3}{2},j} + B_{1,i+\frac{3}{2},j} ) + \frac{9}{16}(B_{1,i-\frac{1}{2},j} + B_{1,i+\frac{1}{2},j} ) \, . \label{eq:B1-cell-center}
\end{equation}
\item Apply Laplacian operator to convert face-centered fields to cell-averaged fields
\begin{equation}
\langle B_1 \rangle_{i,j}  = B_{1,i,j}  +  \frac{h^2}{24} \Delta B_{1,i,j} \, . \label{eq:B1-cell-average}
\end{equation}
\end{enumerate}
As in the unlimited equation of state conversions in Section~\ref{subsubsec:eos}, no limiting is used in the above conversions. Future work may consider using non-oscillatory second-derivative approximations as in LD2000.

\begin{figure}[!ht]
\begin{center}
\includegraphics[width=1.0\textwidth]{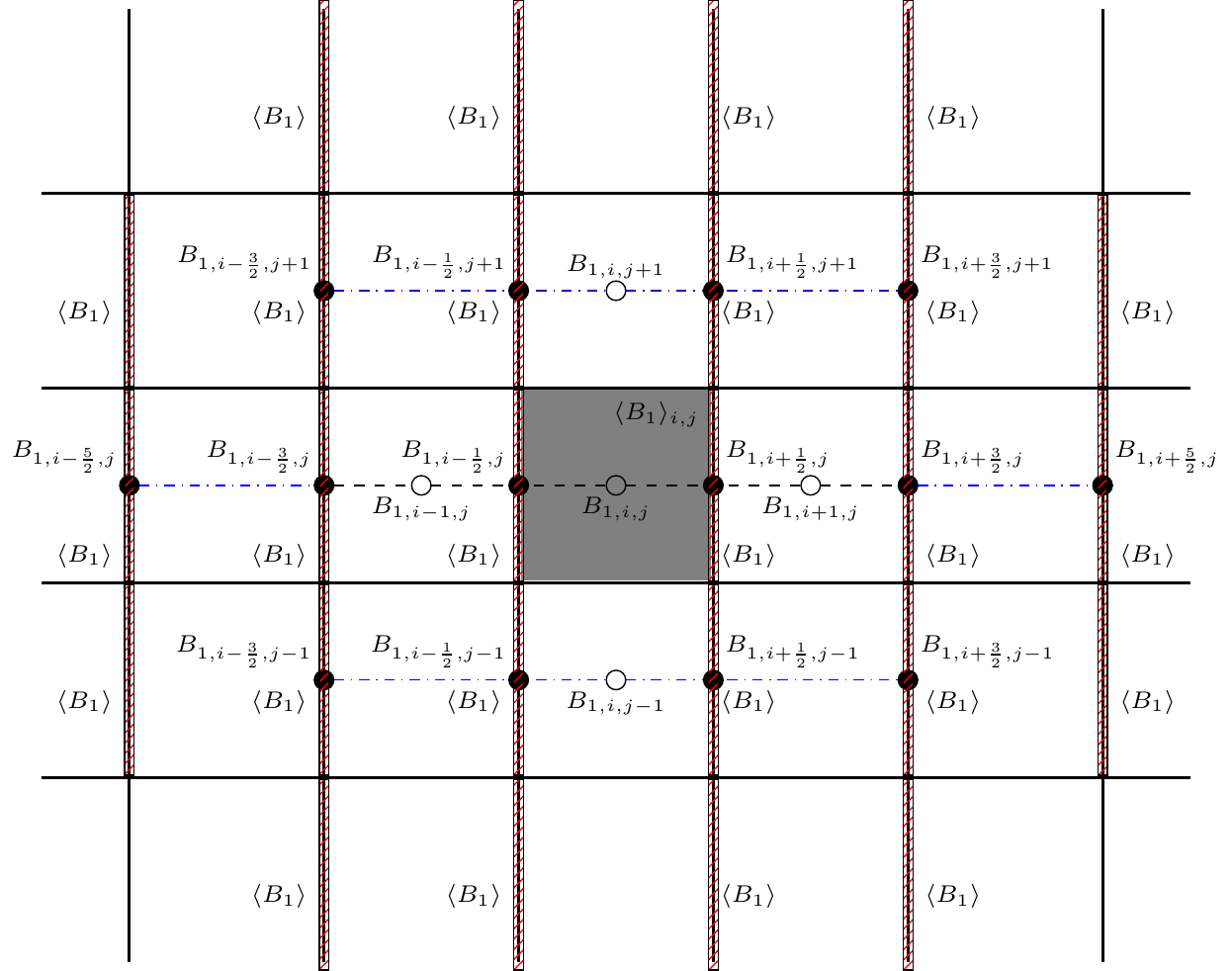}
\end{center}
\caption{The red shaded \(x_1\) interfaces indicate the requisite longitudinal face-averaged \(\langle B_1\rangle \) input data (indices suppressed in diagram) for the \(\mathcal{O}(\Delta x^4) \) inverse reconstruction of cell-averaged \(\langle B_1\rangle_{i,j} \) using the procedure described in Section~\ref{uct-reconstruction-B}. The empty circles are the cell-centered \(B_{1,i,j} \) used in Equation~\eqref{eq:B1-cell-average} when approximating the 2D Laplacian. The filled circles are the face-centered \( B_{1,i-\frac{1}{2},j} \) used in the longitudinal interpolation in Equation~\eqref{eq:B1-face-center}. The black dashed line connects the face-centered points necessary to interpolate the cell-centered \(B_{1,i,j}\) at fourth-order accuracy. The blue dash-dotted lines denote the additional stencil points necessary to interpolate to the four nearest cell centers.}
\label{fig:cell-averaged-B1-reconstruction}
\end{figure}

Figure~\ref{fig:cell-averaged-B1-reconstruction} summarizes the \(x_1\) face-averaged magnetic field input data and intermediate approximations computed in Equations~\eqref{eq:B1-face-center},  \eqref{eq:B1-cell-center} and~\eqref{eq:B1-cell-average} necessary for the inverse reconstruction of a single cell-averaged \(\langle B_1 \rangle_{i,j} \) at fourth-order accuracy. While this wide stencil contains many quadrature points for the approximation of a single cell-averaged value, the intermediate quantities are reused in calculations of the surrounding cell-averages, as is expected for such approaches to high-order accuracy. However, optimizing data reuse and on-node performance of these stencils with finite cache sizes generally requires large box sizes and careful loop scheduling techniques \cite{Olschanowsky2014}. The stencil is wider in \(x_1\) than in \(x_2\) because the upwind constrained transport framework treats the longitudinal faces as primary representations of the magnetic field. The transverse stencil quantities are used for the Laplacian corrections in Equations~\eqref{eq:B1-face-center} and~\eqref{eq:B1-cell-center}, which is only approximated at second-order accuracy. In contrast, the pointwise interpolation in the longitudinal direction \(x_1\) must be performed at \(\mathcal{O}(\Delta x^4)\).

\subsection{Upwind phase: collocated corner flux functions for Roe-type and HLL-type Riemann solvers} \label{subsec:uct-upwind}
The upwinding phase for Godunov schemes for nonlinear systems of conservation laws is a generalization of the trivial upwinding procedure of the scalar advection equation. In Godunov-type schemes for hydrodynamics, the solution of the Riemann problem typically encapsulates the selection of the upwind state and the calculation of the single valued flux from the reconstructed L/R Riemann states collocated at the interface between two cells. Since solving Riemann problems typically encapsulates the majority of the computational demand of a Godunov-type scheme, approximate Riemman solvers are used to simplify the problem while maintaining accuracy. Approximate Riemann solvers provide a \emph{direct} approximation to the numerical flux (as opposed to approximating a state and then evaluating the flux function). Two popular classes of approximate Riemann solvers include:
\begin{enumerate}
\item Roe-type solvers are based on the linearization of the flux Jacobian of the system of equations. The Roe solver requires the characteristic decomposition of the variables at the interface, which may be expensive \cite{Roe1981}. Also, Roe-type solvers may return unphysical states in certain pathological cases \cite{Einfeldt1991}. \item Introduced by Harten, van Leer, and Lax \cite{HartenLaxLeer1983}, HLL-type Riemann solvers compute the fluxes by averaging over an approximate Riemann fan. These Riemann solvers are computationally efficient and guarantee positivity but may produce less accurate solutions than the Roe Riemann solver. However, the greater numerical dissipation of simple approximate Riemann solvers can be ameliorated by high-order reconstructions; for such numerical methods, the one-state HLL or even the Local Lax-Friedrichs (LLF) Riemann solvers may be sufficient to produce the desired level of accuracy. An illustration of this phenomenon is shown in Section~\ref{subsec:mhd-rj2a}.
\end{enumerate}
While they were originally developed for hydrodynamics, both types of Riemann solvers have been extended to the MHD system. The Roe Riemann solver was extended to MHD by Cargo and Gallice \cite{CargoGalice1997}. Miyoshi and Kusano \cite{MiyoshiKusano2005} developed HLLD, which restores the Alfv\'{e}n wave and contact modes in the MHD Riemann fan approximation.

As discussed above, the Godunov fluxes for the magnetic field variables should not be evaluated at the face-centers in the constrained transport context. Rather, the cell corners are the proper locations for evaluating the Riemann problem of the MHD subsystem. UCT generalizes this upwinding procedure from cell faces to cell corners in 2D (edges in 3D). Upwinding at cell-corners requires extending the underlying Riemann solver from a two-state to a four-state selection rule \cite{Londrillo2000}.

The UCT formulation for these corner magnetic fluxes depends on the type of approximate Riemann solver used in the underlying finite volume subsystem. LD2000 first presents a UCT formulation based on Roe-linearized fluxes \cite{Londrillo2000}. The derivation of the formula requires averaging only the dissipative fluxes in each direction using a flux-vector splitting (FVS) formalism. LD2004 \textsection 3.2 provides two central-upwind implementations of the UCT framework based on the one-state HLL Riemann solver: a second-order accurate scheme, MC-HLL-UCT, and a third-order accurate scheme, CENO-HLL-UCT \cite{Londrillo2004}.

For our applications, we are primarily interested in the one-state HLL-type UCT formulation. Averaging the two overlapping \(x_1\) and \(x_2\) approximate Riemann fans at the corner results in the flux formula in Equation 56 of LD2004
 \begin{multline}
\langle \mathscr{E}^U_{3} \rangle_{i-\frac{1}{2},j-\frac{1}{2}} = \frac{\alpha^+_1\alpha^+_2 \langle \mathscr{E}^{L_1L_2}_{3}\rangle_{i-\frac{1}{2},j-\frac{1}{2}} + \alpha^-_1 \alpha^+_2 \langle \mathscr{E}_{3}^{R_1L_2}  \rangle_{i-\frac{1}{2},j-\frac{1}{2}} + \alpha^+_1\alpha^-_2 \langle \mathscr{E}_{3}^{L_1R_2} \rangle_{i-\frac{1}{2},j-\frac{1}{2}} + \alpha^-_1 \alpha^-_2 \langle \mathscr{E}_{3}^{R_1R_2} \rangle_{i-\frac{1}{2},j-\frac{1}{2}}}{(\alpha^+_1 + \alpha^-_1)(\alpha^+_2 + \alpha^-_2)}  \\
 - \frac{\alpha^+_2 \alpha^-_2}{\alpha^+_2 + \alpha^-_2}( \langle {B}_1^{R_2} \rangle_{i-\frac{1}{2},j-\frac{1}{2}}  - \langle {B}_1^{L_2} \rangle_{i-\frac{1}{2},j-\frac{1}{2}}  )+ \frac{\alpha^+_1 \alpha^-_1}{\alpha^+_1 + \alpha^-_1}(\langle {B}_2^{R_1} \rangle_{i-\frac{1}{2},j-\frac{1}{2}}  - \langle {B}_2^{L_1} \rangle_{i-\frac{1}{2},j-\frac{1}{2}} ) \, , \label{eq:uct-hll}
\end{multline}
where \(\alpha^\pm_1\) are the nonnegative dissipative terms computed from \(S^{L_1}, S^{R_1} \), the minimum and maximum wavespeed estimates in the \(x_1\) direction, for example. See \ref{sec:appendix-gs05-uct} for more details. The wavespeed estimates are properly evaluated at the \(i-\frac{1}{2},j-\frac{1}{2}\) corner along with the reconstructed emf states. However, in practice they are computed by taking the extrema of the existing wavespeed estimates from the four 1D Riemann solutions computed at the nearest faces for the Godunov fluxes in Section~\ref{sec:fv-hydro}. The implementations of Equation~\eqref{eq:uct-hll} in LD2004 use simple wavespeed estimates from Davis, Equation 4.5 of \cite{Davis1988}. We use the Einfeldt wavespeed estimates which reference the Roe-averaged wavespeeds \cite{Einfeldt1988}.
This is a generalization of the GS05 four-state upwinding and averaging procedure encapsulated in Equation~\eqref{eq:gs05-emf-average}.

The HLL-UCT solver is a simple multidimensional Riemann solver applied to the induction equation. Recent work has involved the development of multidimensional Riemann solvers for MHD \cite{Balsara2010, Balsara2012}. It automatically reduces to the 1D HLL fluxes for the magnetic fluxes for plane parallel solutions.

\section{MHD numerical results} \label{sec:mhd-tests}
In this section, we present the results from a series of numerical experiments designed to test the accuracy, stability, shock-capturing ability, and numerical monopole suppression of the overall upwind constrained transport finite volume scheme, RK4+PPM. As in Section~\ref{sec:hydro-tests}, we emphasize comparisons to the second-order counterpart to the high-order algorithm. This scheme, again referred to using the shorthand VL2+PLM, uses the second-order hydrodynamics scheme augmented with the second-order UCT scheme from GS05 \cite{GardinerStone2005, StoneGardiner2009}. All MHD results are generated using the HLLD Riemann solver for the hydrodynamic subsystem fluxes unless otherwise noted. The CFL number for the MHD tests is defined by
\begin{equation}
C_0 = \frac{\Delta t}{ \min \left( \frac{h}{|\lambda_1^{max}|}, \frac{h}{|\lambda_2^{max}|} \right)} \, , \label{eq:cfl-mhd}
\end{equation}
where \(\lambda_{1,2}^{max} \) is the fastest wave mode speed in the \(x_1\)- or \(x_2\)- direction over all of the cells. For a single cell, the wavespeed estimate is defined using the cell-averaged states as \( \lambda_1 = |\langle v_1\rangle| + c_1^f \), where \( c_1^f \left(\langle \mathbf{W} \rangle, \langle \mathbf{B} \rangle \right) \) is the fast magnetosonic wavespeed in the \(x_1\)-direction, for example. The same  linear stability analysis as in \cite{ColellaDorrHittingerMartin2011} was used as reference to estimate the CFL restriction for the MHD problems, but a conservative value of \( C_0 = 0.4 \) was used for all of the following tests.

For the convergence plots in this section, the root mean square error metric of Equation~\eqref{eq:rms-l1} of the \(L_1\) error vector is extended to \(N_{hydro} + N_{field} \) cell-averaged variables:
\begin{equation}
\norm{\delta \mathbf{U}^n} = \sqrt{\sum_s^{N_{hydro}} (\delta U^n_s)^2 + \sum_s^{N_{field}} (\delta B^n_s)^2} \, . \label{eq:rms-l1-mhd}
\end{equation}

\subsection{2D oblique MHD linear wave convergence} \label{subsec:mhd-linwave}

As in the hydrodynamics linear wave test of Section~\ref{subsec:hydro-linwave}, the conserved variable profiles are initialized along an oblique direction on a 2D Cartesian grid using the exact eigenvectors of the linearized MHD system. The domain size, wavelength, and wave propagation direction remain unmodified from the earlier section. The same uniform background fluid variables as the hydrodynamics test are used with \( \rho = 1, P = 3/5, \gamma = 5/3 \), and a background magnetic field is introduced with \(\mathbf{B} = (1, \sqrt{2}, 1/2)\) specified in the wavevector rotated frame. The background velocity \(v_1=1\) only for the entropy wave mode test, and it is \(\mathbf{v}=0\) for all others.

With these parameters, the wavespeeds are: \( c_f=2,  c_{a,x}=1, c_s=1/2, c_{v_1}=1 \) for the fast magnetosonic, Alfv\'{e}n, slow magnetosonic, and entropy wave modes, respectively. See Appendix A of GS05 for the exact eigenvectors used for each wave family \cite{GardinerStone2005}. We note that the 3D MHD linear wave convergence results presented in \cite{Stone2008} Section 8.6 and Section 6.1 of \cite{StoneGardiner2009} both reference GS08 Appendix A \cite{GardinerStone2008} for the eigenvectors used in the test, but the correct eigenvector values are those in GS05 Appendix A \cite{GardinerStone2005}. The earlier references are incorrect since GS08 eigenvectors are derived from the linearization around a different wavevector-frame background field, \(\mathbf{B} = (1, \frac{3}{2}, 0)\) \cite{GardinerStone2008}.

The \(\mathcal{O}(\Delta x^4) \) accurate initialization of the fluid variables follows Equations~\eqref{eq:linwave-ic} and~\eqref{eq:laplacian-ic4}. The average magnetic field is initialized for each component on longitudinal cell faces using the differences of the analytic magnetic vector potential computed at cell corners, which ensures that the initial condition satisfies \( \nabla \cdot \mathbf{B}=0\) to machine precision. However, the initialization of the face-averaged fields \( \langle B_1 \rangle_{i\pm \frac{1}{2}, j, k} , \langle B_2 \rangle_{i, j\pm \frac{1}{2}, k} , \langle B_3 \rangle_{i, j,k\pm \frac{1}{2}} \) required some care to ensure fourth-order accurate solutions at errors near machine precision. We originally observed that the errors prematurely ceased converging when naively applying Stoke's theorem to \(\mathbf{B} = \nabla \times \mathbf{A} \) on cell faces after computing analytic \(\mathbf{A}\) at cell corners. The convergence issues were caused by a numerical loss of significance at double precision in the differencing operations. For example, when initializing the \(x_2\) face-averaged longitudinal field component with
\begin{equation}
\langle B_2 \rangle_{i, j\pm \frac{1}{2}, k} = \frac{1}{\Delta x_3} (\langle A_1 \rangle_{i,j\pm \frac{1}{2},k+\frac{1}{2}} - \langle A_1 \rangle_{i,j\pm \frac{1}{2},k-\frac{1}{2}} ) - \frac{1}{\Delta x_1} (\langle A_3 \rangle_{i+\frac{1}{2},j\pm \frac{1}{2},k} - \langle A_3 \rangle_{i-\frac{1}{2},j\pm \frac{1}{2},k} ) \, , \label{eq:stokes-linwave}
\end{equation}
two differences are evaluated. The particularly large relative error of this operation in finite-precision arithmetic for the linear wave can be attributed to the relative sizes
\begin{equation} \label{eq:linwave-terms}
\begin{split}
A_1(\mathbf{x}) &= \bar{A}_1(\mathbf{x}) + \tilde{A}_1(\mathbf{x})  \\
                          &\approx O(1) + O(\varepsilon)
\end{split}
\end{equation}
of the linear and perturbative terms in the vector potential profiles. The loss of significance due to the rounding of intermediate floating-point values was ameliorated by changing the order of operations. For example, the first term in Equation~\eqref{eq:stokes-linwave} was reordered as
\begin{equation}\label{eq:reordered-linwave-diff}
\frac{1}{\Delta x_3} \left((\langle \bar{A}_1 \rangle_{i,j\pm \frac{1}{2},k+\frac{1}{2}} - \langle \bar{A}_1 \rangle_{i,j\pm \frac{1}{2},k-\frac{1}{2}} ) + (\langle \tilde{A}_1 \rangle_{i,j\pm \frac{1}{2},k+\frac{1}{2}} - \langle \tilde{A}_1 \rangle_{i,j\pm \frac{1}{2},k-\frac{1}{2}} ) \right)  \,
\end{equation}
in order to first difference similarly sized quantities.

\begin{figure}[!ht]
\includegraphics[width=\textwidth]{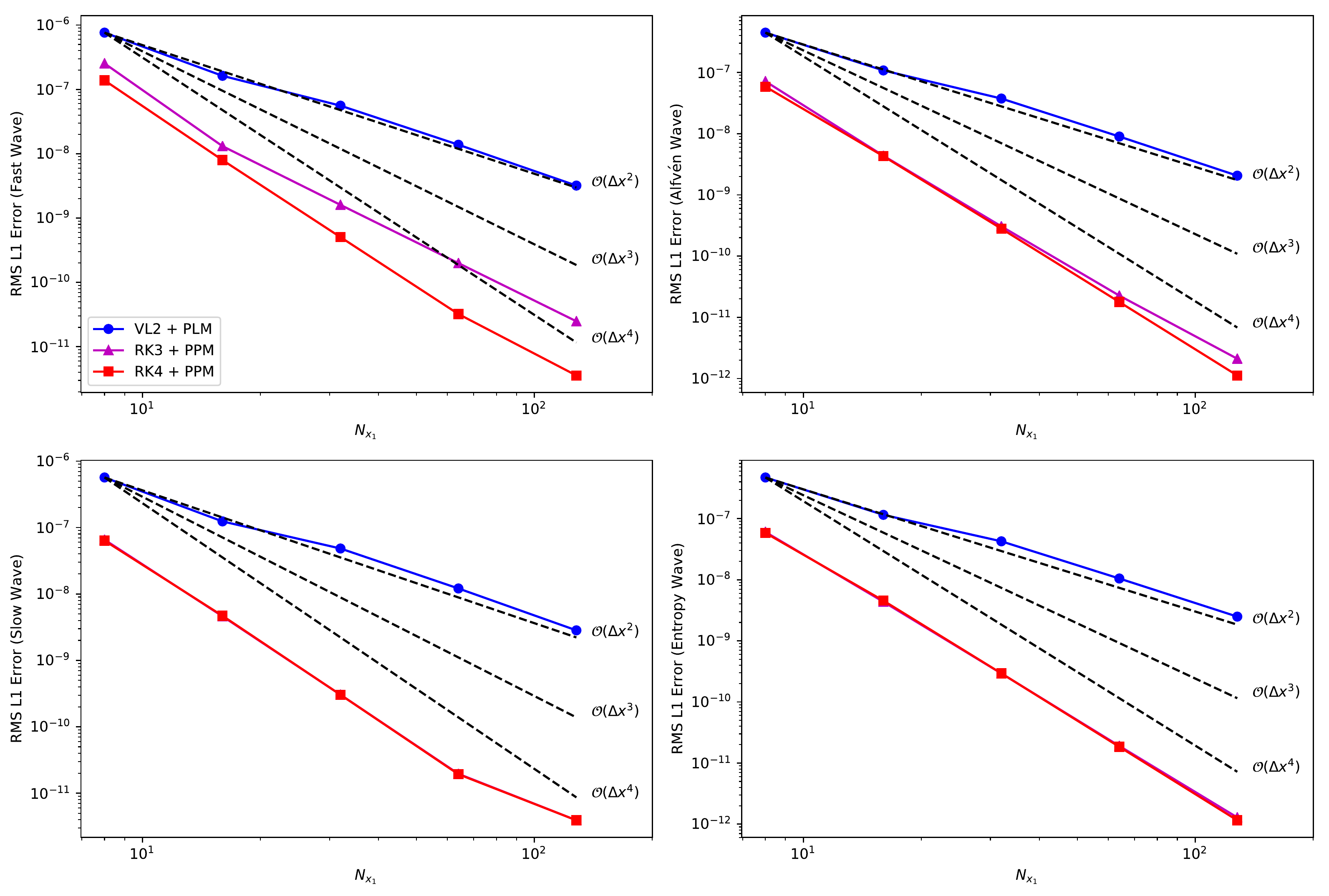}
\caption{MHD linear wave convergence plots for the fast magnetosonic, Alfv\'{e}n, slow magnetosonic, and entropy wave modes. With nearly four orders of magnitude smaller error than the second-order VL2+PLM results at \(N_{x_1}=128\), the high-order results demonstrate even greater improvement than in the hydrodynamics test of Figure~\ref{fig:hydro-linwave}. The RK3+PPM error lines are obscured by the RK4+PPM lines in the slow and entropy wave plots.}
\label{fig:mhd-linwave}
\end{figure}

Figure~\ref{fig:mhd-linwave} demonstrates formal fourth-order convergence of the UCT implementation in conjunction with the hydrodynamics finite volume subsystem. The root mean square of the \(L_1\) error vector is shown for resolutions spanning \(8 \times 4 \) to \(128 \times 64\) for all MHD wave modes. When combined with the fourth-order spatially accurate single stage algorithm, the third-order temporally accurate integrator RK3 produces errors that are nearly identical to the fourth-order RK4 results. The only significant difference occurs for the fast magnetosonic mode, for which the RK3+PPM solution converges at only third-order for the majority of resolutions. Approximately four times fewer timesteps are required to evolve the wave for one period in the fast magnetosonic wave test when compared to the slow magnetosonic wave test. Therefore, the spatial truncation error associated with application of the operators in single stage contributes (in total) approximately four times more error for the evolution of the slow magnetosonic wave. For this fixed CFL of \(0.4\), the spatial error then dominates the global solution error at all resolutions in the slow magnetosonic wave test.

The difference in behavior of the integrators illustrates the flexibility of the semi-discrete formulation. The same complicated single stage algorithm can easily be combined with many potential temporal integrators depending on the demands of the particular application. RK3 can be used at (potentially significantly) reduced computational expense relative to RK4 if the dynamics indicate that the error will likely be dominated by sources other than the finite temporal resolution.

\subsection{3D and 2D oblique circularly polarized Alfv\'{e}n waves} \label{subsec:mhd-cpaw}
The circularly polarized Alfv\'{e}n wave test was first described by T\'{o}th in Section 6.3.1 of \cite{Toth2000}. The same domain setup, wavelength, and propagation direction as in the 2D MHD and hydrodynamics linear wave tests of Sections~\ref{subsec:mhd-linwave}, \ref{subsec:hydro-linwave} are used. However, no restriction to small amplitude perturbations is made for the initial condition. Unlike the linear wave test, these wave profiles are exact nonlinear solutions to the ideal MHD equations. We use the parameters from the subsequent formulation in GS05 Section 3.3.2 \cite{GardinerStone2005}, using background \(\rho=1, P=0.1\) with velocity and magnetic field components
\begin{subequations}
\begin{align}
\mathbf{B} =& (1, 0.1\sin(2\pi x), 0.1\cos(2\pi x)) \, , \label{eq:cpaw-B} \\
\mathbf{v} =& (v_x, 0.1\sin(2\pi x), 0.1\cos(2\pi x)) \, , \label{eq:cpaw-v}
\end{align}
\end{subequations}
specified in the wavevector rotated coordinate system. These parameters produce a circularly polarized wave that is not subject to a parametric instability that may cause other numerically evolved Alfv\'{e}n waves to decay into magnetosonic waves \cite{GardinerStone2005}.

As in the MHD linear wave test, the average magnetic field components are initialized on longitudinal cell faces using the differences of the analytic magnetic vector potential at cell corners. The background flow velocity is set to \(v_x=0\) for the traveling wave test. For a standing wave profile, the background flow is \(v_x=1\) to exactly oppose the wave propagation to the left. The evolution of the standing wave is a challenging variant of the test because the multidimensional operators for updating the face-averaged magnetic field must exactly cancel to preserve the wave field profile.

\begin{figure}[!ht]
\includegraphics[width=\textwidth]{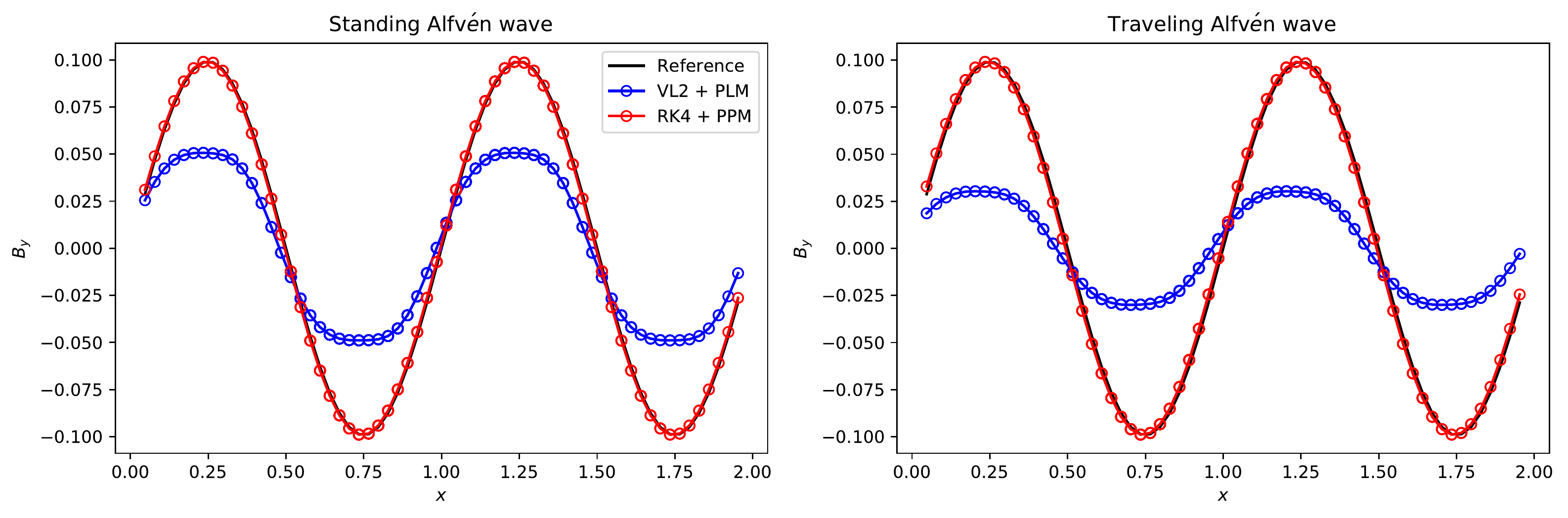}
\caption{Scatter plots of one component of the transverse magnetic field of all points in the \(32 \times 16\) resolution 2D circularly polarized Alfv\'{e}n wave test. All cells are shown according to their cell center positions in the wavevector-aligned coordinate frame.}
\label{fig:cpaw-By}
\end{figure}

Figure~\ref{fig:cpaw-By} plots the transverse, in-plane magnetic field component \(B_y\) in the wavevector-frame of all cells for a \(32 \times 16\) grid at \(t=5\). RK4+PPM nearly exactly reproduces the initial condition, whereas VL2+PLM results in diffusion of more than half of the peak height. The smooth extrema preserving PPM limiter was instrumental in preventing such dissipation in the fourth-order solver. At lower resolutions such as \(N_{x_1}= 16\), dispersion error dominates the second-order solutions. In contrast, the high-order solutions have negligible dispersion error for the all tested resolutions.

Unlike line plots showing a subset of cells produced by a 1D slice of the domain, a scatter plot of all cells based on the cell-centered positions along the wave, \(x\), can reveal the presence of multidimensional grid noise in the solution. VL2+PLM and RK4+PPM both produce solutions with negligible spread in the distribution of \(B_y\) samples from nearby phases. Therefore, both the second- and fourth-order schemes preserve uniformity along the planar wavefronts. The fourth-order results in the right plot of Figure~\ref{fig:cpaw-By} can be compared to Figure A.2 of \cite{Mignone2010a}, which also showed nearly perfect recovery of \(B_y\) with the fifth-order WENO-Z and MP5 schemes  for the traveling circularly polarized Alfv\'{e}n wave test at \(32\times16 \times 16\) resolution.

\begin{figure}[!ht]
\includegraphics[width=\textwidth]{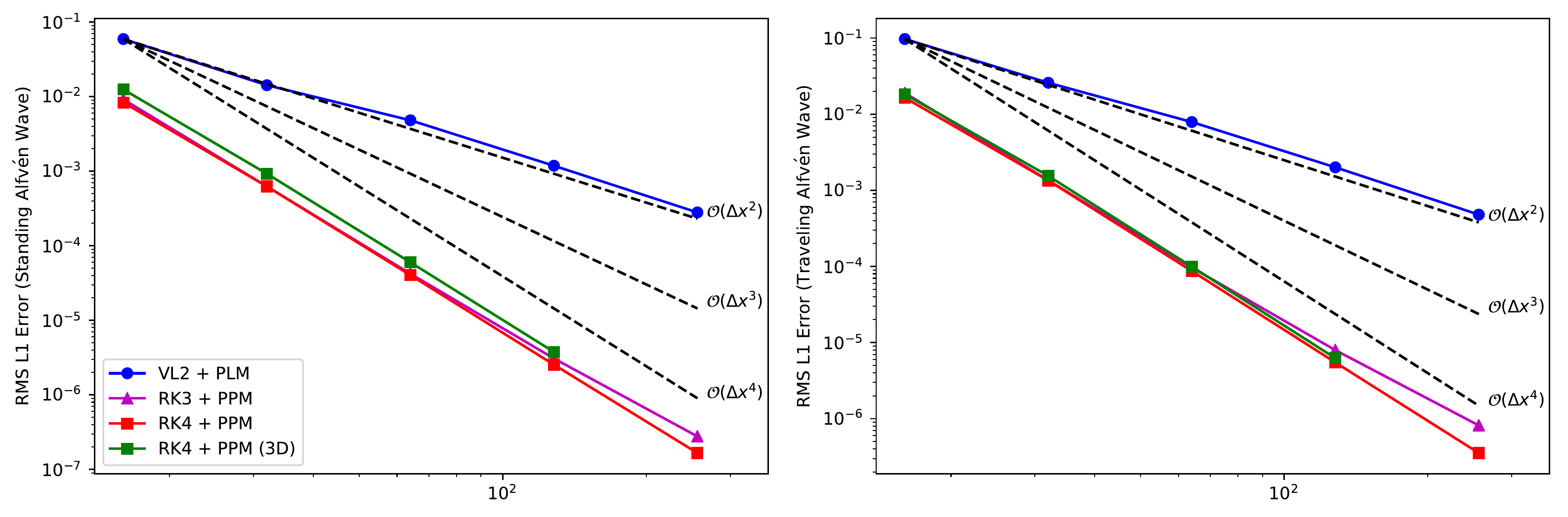}
\caption{Convergence of the root mean square \(L_1\) error of the circularly polarized Alfv\'{e}n plane wave profiles after the traveling wave has propagated for one wavelength. The fourth-order convergence of the RK4+PPM errors of a 3D formulation of the problem are juxtaposed in green; the errors are uniformly larger than the 2D problem due to the increased spatial error in also resolving the wave's oblique orientation relative to \(x_3\). As observed in Figure 4 of GS08, the traveling wave mode errors are larger than the standing wave mode errors, and the increase is fairly uniform over the components of the error vector (not shown). \cite{GardinerStone2008}. }
\label{fig:cpaw-convergence-t1}
\end{figure}

Because the wave is globally smooth, the errors should converge at fourth-order as the mesh is resolved with fixed CFL number, as in the linear wave tests. Figure~\ref{fig:cpaw-convergence-t1} plots the convergence of the RMS-L1 error for the standing and traveling circularly polarized waves at \(t=1\) from \(16\times 8 \) to \(256 \times 128\) cells.  When compared to the linear Alf\'{e}n wave convergence results in Figure~\ref{fig:mhd-linwave}, the curves in Figure~\ref{fig:cpaw-convergence-t1} are nearly identical when scaled by \(10^{-5}\), the ratio of wave amplitudes. As Stone and Gardiner identified in Section 6.2 of \cite{StoneGardiner2009} for the VL2+PLM scheme, the relative dissipation of the waves does not depend on the wave amplitude or the presence of nonlinear effects. Figure~\ref{fig:cpaw-convergence-t1} confirms that the same invariance holds for the fourth-order RK4+PPM scheme; the resolution of the grid is the only factor that controls the numerical diffusivity of the overall scheme in these tests.

Figure~\ref{fig:cpaw-convergence-t1} also juxtaposes the errors of the RK4+PPM method applied to a 3D variant of the problem. We refer the reader to Section 5.3 of GS08 \cite{GardinerStone2008} for the details on the problem setup. The same fixed CFL number of 0.4 was used, and a range of resolutions from \( 16 \times 8 \times 8 \) to \( 128\times 64 \times 64 \) cells was tested. The results again demonstrate fourth-order convergence of the method in both the standing and traveling wave tests. For the standing wave case, the errors are all between 48-50\% greater than their 2D counterparts. For the traveling wave, the errors grew by 13-15\% for this problem and solver configuration. There are no new algorithmic components in the 3D method, but the computational expense relative to VL2+PLM is significantly greater than for 2D problems. The growth in the performance cost from 2D to 3D for the fourth-order MHD algorithm is largely dominated by the additional transverse PPM reconstructions and upwinding of \(\mathscr{E}_1, \mathscr{E}_2 \) necessary to compute the induction equation.

Low resolution, large \(t_f=5\) comparisons are displayed in the Figure~\ref{fig:cpaw-By} scatter plots to highlight the differences in numerical diffusivity between the second-order and fourth-order algorithms. However, shorter \(t_f=1\) test results are used in the error convergence plots of Figure~\ref{fig:cpaw-convergence-t1}, since we follow the conventions of earlier publications \cite{GardinerStone2005, GardinerStone2008, Stone2008, StoneGardiner2009, Londrillo2004}. We have observed (plot not shown) that the second-order VL2+PLM solution fails to converge at second-order for longer \(t_f=5\) propagation tests at the initial resolutions due to the large dispersion error. RK4+PPM does not suffer from such a slow transition to the asymptotic fourth-order convergence regime. The same difference in convergence behavior was observed when comparing the second-order and fourth-order schemes in Figure 1 of Susanto (2013) \cite{Susanto2013}.

\subsection{3D diagonal advection of a field loop} \label{mhd-field-loop}
The dynamic advection of a field loop, described for a 2D domain in GS05 and for a 3D domain in GS08, is a rigorous test of the CT discretization \cite{GardinerStone2005, GardinerStone2008}. In this test, a cylinder of constant, weak magnetic pressure \( P_B \) is passively transported by the fluid for two periods along the domain diagonal. We let \(B_z=0\) everywhere, and initialize a uniform poloidal magnetic field in a cylinder with radius \(R=0.3\) centered on the origin via the out-of-plane component of the vector potential
\begin{equation}
A_3(x_1, x_2) \equiv \left\{
\begin{array}{ll}
      A_0(R-r) & r\leq R \\
      0 & r > R \\
\end{array}
\right. \, , \label{eq:field-loop-a3}
\end{equation}
where \(A_0= 10^{-3}\). This potential corresponds to a plasma \(\beta = \frac{2P}{B^2} = 2 \times 10^{-6} \) inside the cylinder. Again, the average magnetic field is initialized on longitudinal cell faces by differencing this analytic magnetic vector potential across cell corners.

Following GS05, we initialize a \(2N \times N \) uniform 2D grid with periodic boundary conditions, where \(N=64\) \cite{GardinerStone2005}. The uniform fluid has \(\rho=1, P=1\), with an adiabatic MHD equation of state and \(\gamma=5/3\). Unlike GS05 or GS08, the grid spans \( -1 \leq x_1 \leq 1, -0.5 \leq x_2 \leq 0.5, -1 \leq x_3 \leq 1 \), and the velocity vector points along the space diagonal of the 3D rectangular domain with \( \mathbf{v} = (2, 1, 2) \). Hence, \( |v| = 3 \) and the cylinder returns to its original position every \(t=1\) periods.

\begin{figure}[!ht]
\includegraphics[width=\textwidth]{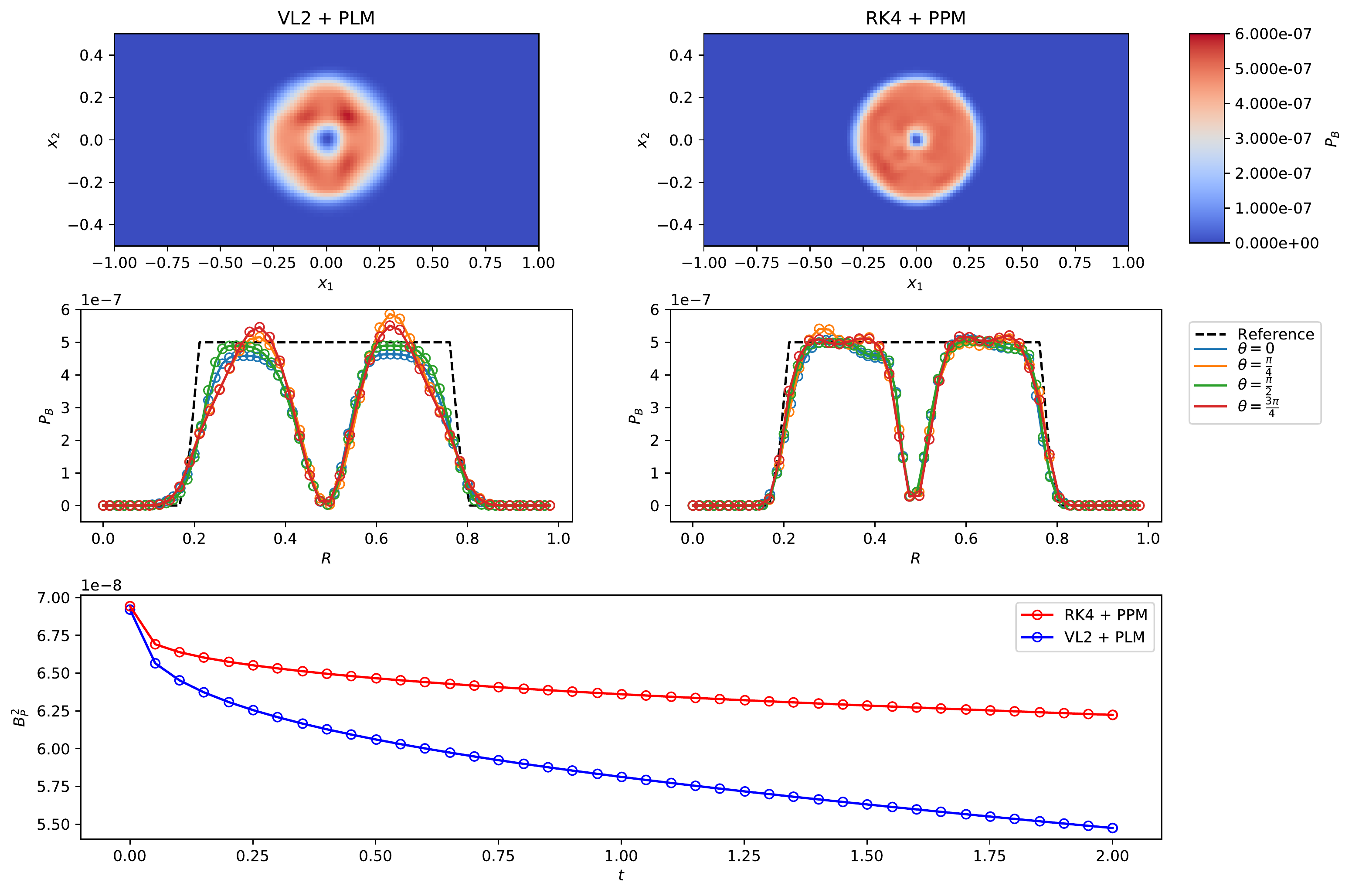}
\caption{Comparisons of magnetic field quantities advected by the second-order and fourth-order schemes at \(t_f=2\). The high-order algorithm demonstrates improved preservation of the original cylinder symmetry and overall domain magnetic energy density.}
\label{fig:field-loop-comparison}
\end{figure}

Several comparisons of the advection test results at \(t_f=2\) produced by the second-order and fourth-order schemes are shown in Figure~\ref{fig:field-loop-comparison}. The first row displays 2D plots of the magnetic pressure represented with the same color scale. Overall, the fourth-order scheme produces a more uniform advected \(P_B\) cylinder than the second-order counterpart. This comparison also highlights the sharper resolution of the cylinder edge in the RK4+PPM result. Additionally, the hole that has formed due to magnetic reconnection in the center of the field loop is much larger in the VL2+PLM result.

The second row considers a subset of the \(P_B\) data that is produced by taking 1D slices through the center of the domain along four directions of particular interest: the two coordinate axes and the two diagonals of the uniform Cartesian mesh. The corresponding slicing of the analytic reference solution (equivalent to the initial condition) is shown as the dashed black line. The results show reflective symmetry about the domain diagonal that is parallel to the fluid velocity vector. The leading edge of the cylinder (upper right) overshoots the initial maximum by nearly 20\% in the VL2+PLM result. In contrast, the RK4+PPM solution has a much smaller maximum overshoot on the trailing edge (bottom left), which is approximately 8\% greater than the initial condition. Both solutions exhibit small leading-edge versus trailing-edge asymmetries that may occur in solutions to advection problems; this can be observed by comparing the \( R > 0.5 \) and \( R <0.5\) profiles along the \( \theta = \frac{\pi}{2} \) slice in Figure~\ref{fig:field-loop-comparison}.

The improved resolution of field discontinuities, namely the magnetically-reconnected hole and the outer cylinder boundary, can be quantified using these profiles. Along a diagonal ray, for example, the edges are resolved by 6-8 cells in the VL2+PLM solution, whereas RK4+PPM resolves the edges with only 4 cells. Although minor oscillations are present in the fourth-order solution, they are relatively minor, and the fourth-order UCT scheme better approximates the flat-top \(P_B\) profile of the cylinder despite the oscillations.

Finally, the third row of Figure~\ref{fig:field-loop-comparison} provides an MHD counterpart to Figure~\ref{fig:slotted-cylinder-hst} of the slotted cylinder advection test. The time-series of the domain-averaged magnetic energy \(B_p^2\) provides a measure of the numerical diffusivity of the UCT implementation. The rate of decay is significantly slower with the fourth-order scheme than in VL2+PLM with the second-order UCT implementation. The PPM reconstruction of field quantities at cell-corners can be credited for significantly decreasing the diffusion of the \(P_B=5\times 10^{-7}\) profile in the UCT upwinding step.

\begin{figure}[!ht]
\includegraphics[width=\textwidth]{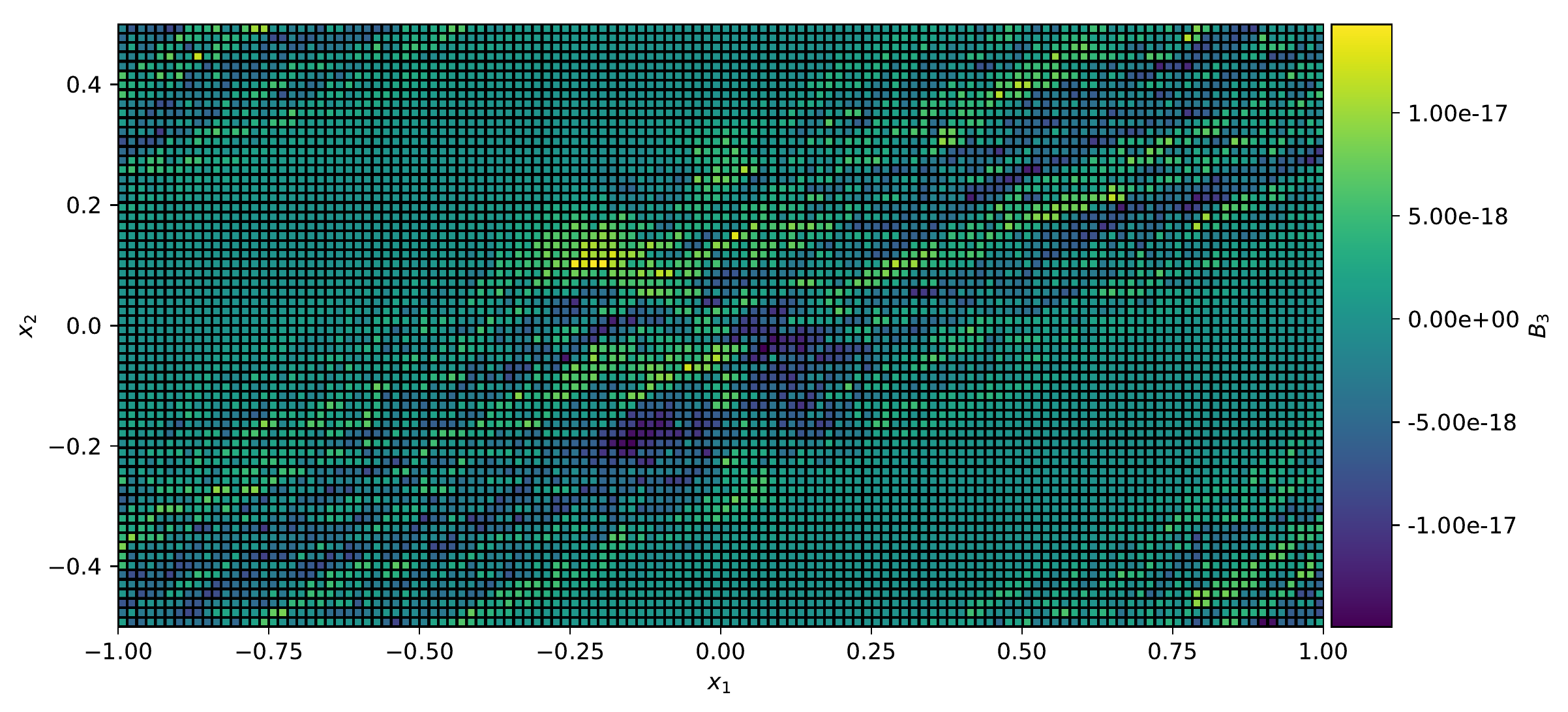}
\caption{Plot of the out-of plane \(B_3\) at \(t_f=2\). The differences in the induction equation updates of \(B_3\) are guaranteed to be 0.0 within floating-point round-off error by the \(\nabla \cdot \mathbf{B}=0\) preservation of the constrained transport method.}
\label{fig:field-loop-Bcc3}
\end{figure}

\begin{figure}[!ht]
\includegraphics[width=\textwidth]{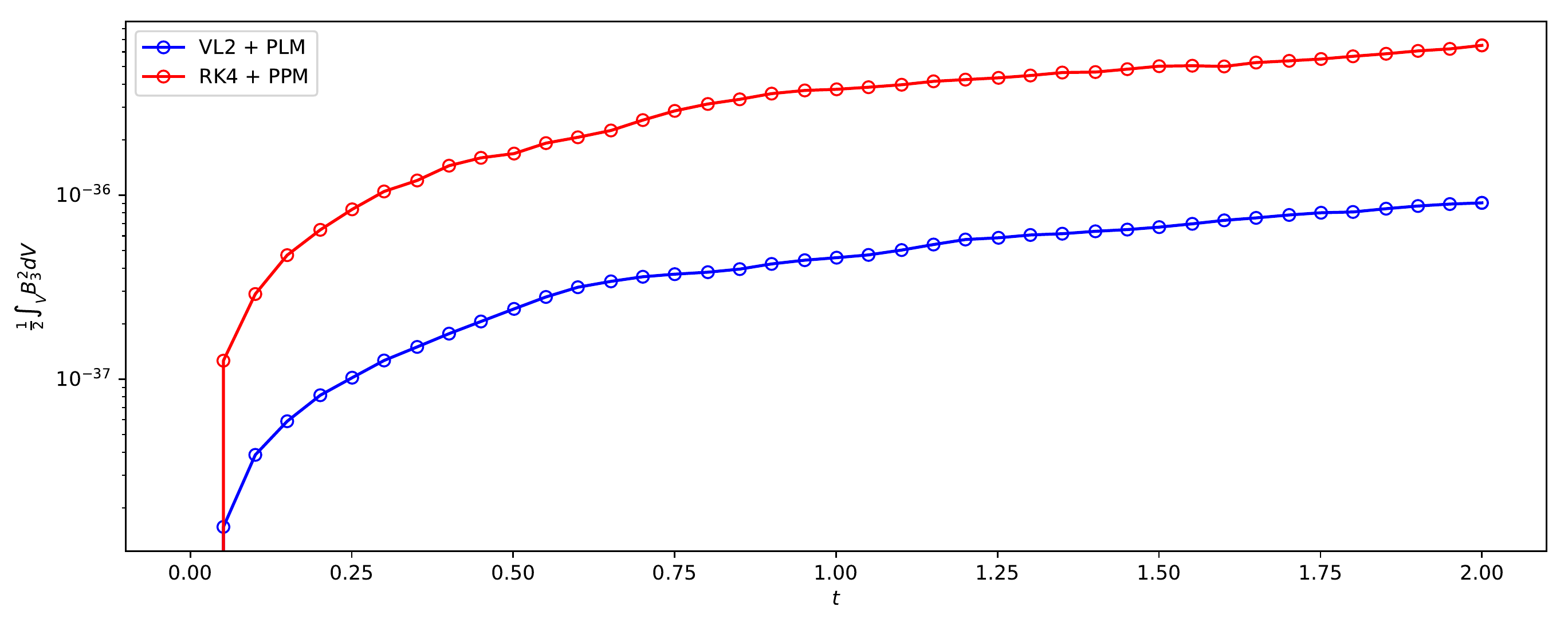}
\caption{Time-series growth of the out-of-plane component of the magnetic energy averaged over the domain.}
\label{fig:field-loop-me3-t}
\end{figure}

Most importantly, \(B_z\) remains zero to within round-off error for the lifetime of the simulation even as \(v_z \neq 0\) and nonzero terms enter the induction equation updates. Figure~\ref{fig:field-loop-Bcc3} illustrates the final spatial distribution of the deviations from \(B_3=0.0\). The nonzero values remain largely concentrated near the path of the field loop where \(B_1,B_2 \neq 0 \), but they remain far smaller in magnitude than double-precision machine epsilon \(\approx 2.22 \times 10^{-16} \). Figure~\ref{fig:field-loop-me3-t} compares the time-series of the domain-integrated \(B^2_3\) of the RK4+PPM and VL2+PLM solutions. While initially the RK4+PPM value grows much faster owing to the greater number of stages and induction equation evaluations per timestep, the number of cells with nonzero \(B_3\) grows at a similar rate to those in the VL2+PLM solution.

Advection in the \(x_3\) direction of a poloidal field is nontrivial for general-purpose constrained transport algorithms \cite{Stone2008}. This test confirms the built-in divergence-free property of the fourth-order UCT implementation. While not shown here, the fourth-order UCT implementation also preserves the geometry of the concentric field lines.

\subsection{1D Brio-Wu shock tube} \label{subsec:mhd-bw}

\begin{figure}[!ht]
\includegraphics[width=\textwidth]{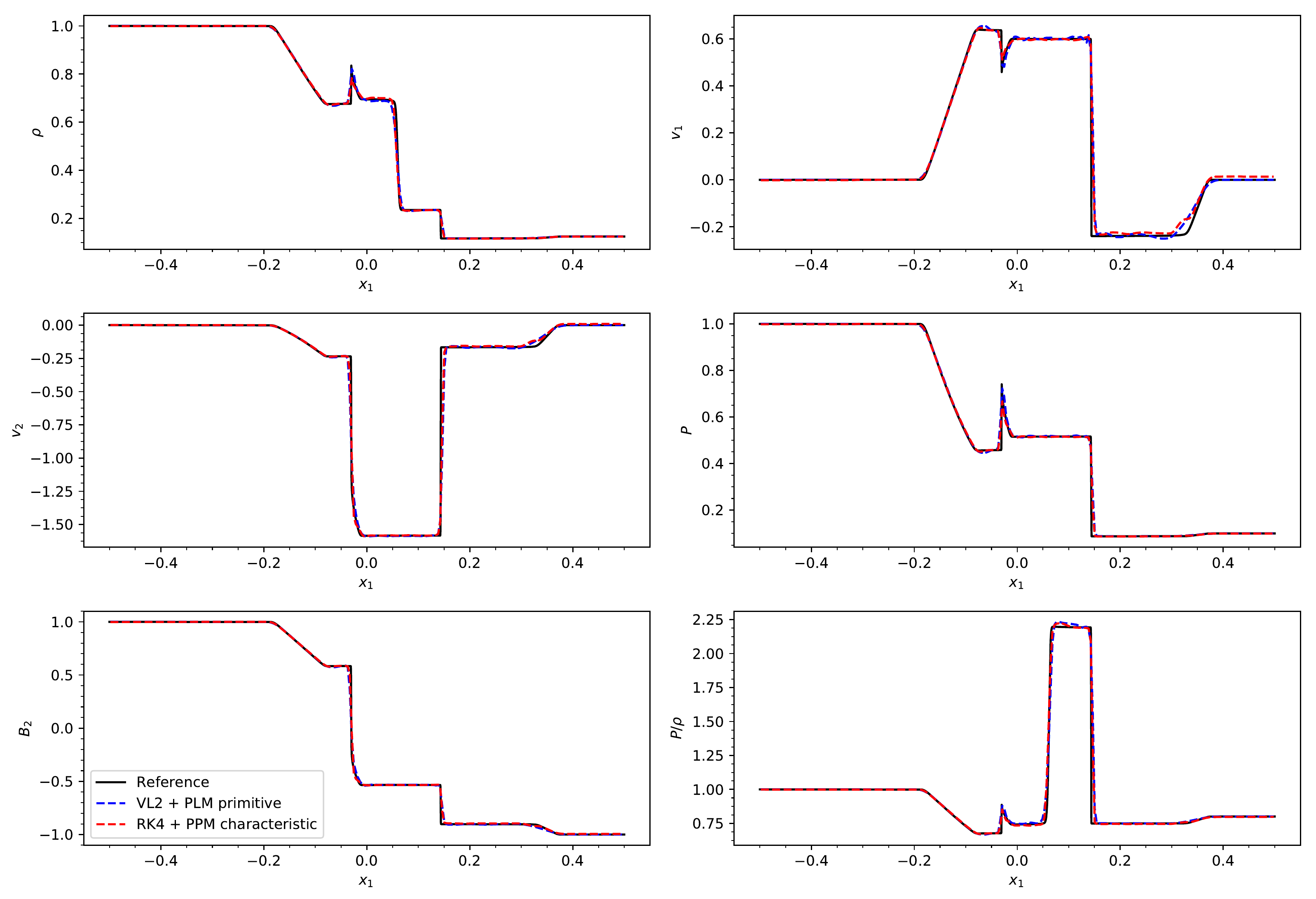}
\caption{Density, pressure, non-constant velocity and magnetic field components, and specific internal energy density scaled by \((\gamma -1)\) profiles of the Brio-Wu shock tube problem at \(t=0.1\). The dashed lines show for \(N_{x_1}=256 \) solutions of the VL2+PLM second-order scheme (blue) and the RK4+PPM fourth-order scheme (red). The reference solution is shown with a solid black line and was computed with RK4+PPM at a resolution of 8192 cells.}
\label{fig:bw-shock-1a}
\end{figure}
\begin{figure}[!ht]
\includegraphics[width=\textwidth]{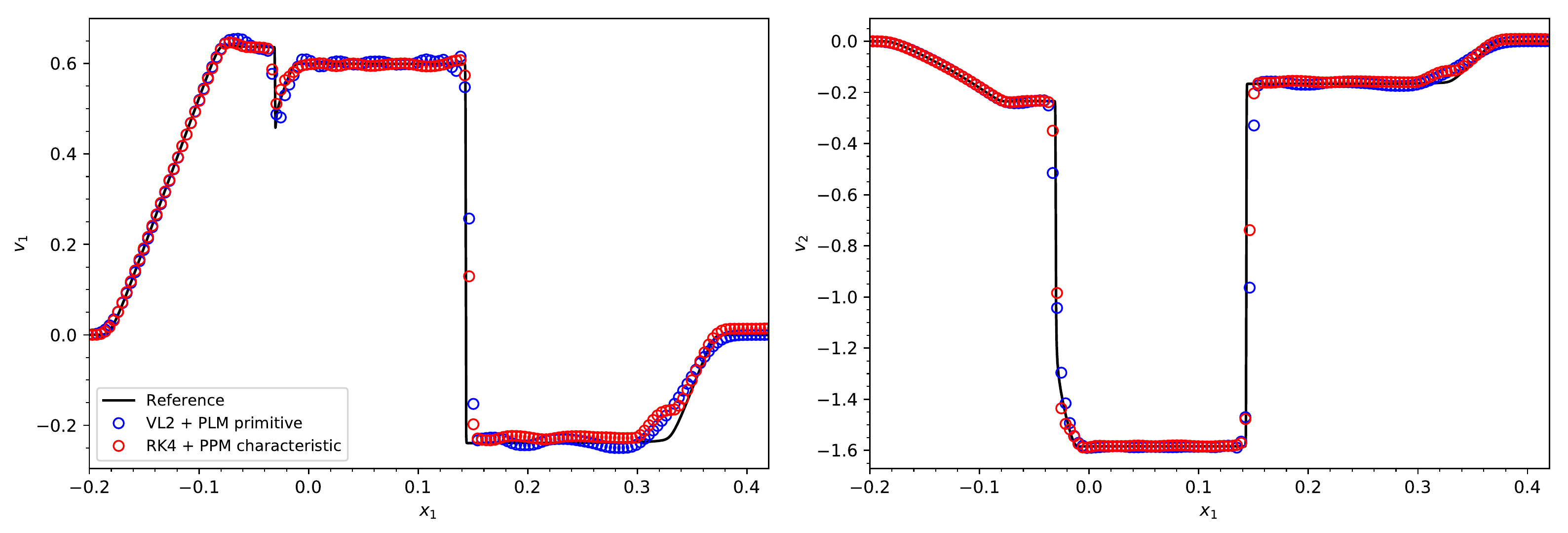}
\caption{Comparison of the Brio-Wu shock tube results from the second-order algorithm using piecewise linear reconstruction of primitive MHD variables and results from the fourth-order algorithm using piecewise parabolic reconstruction of characteristic MHD variables. The largest spurious oscillations occur in the velocity component profiles when using primitive reconstruction.}
\label{fig:bw-shock-1b}
\end{figure}

In following the analogous increasing complexity of the hydrodynamics tests of Section~\ref{sec:hydro-tests}, we now introduce an MHD Riemann problem to test the solver's ability to capture shocks and complex nonlinear waves. The Brio-Wu shock tube is an MHD analog to the classical Sod shock tube of hydrodynamics \cite{BrioWu1988, Sod1978}. For this shock tube problem, the background longitudinal magnetic field is \(B_1 = 0.75 \) and \(\gamma=2\). The left and right states are given by
\begin{equation}
\begin{pmatrix}
\rho^L \\
v_1^L \\
v_2^L \\
v_3^L \\
P^L \\
B_1^L \\
B_2^L \\
B_3^L \\
\end{pmatrix}
=
\begin{pmatrix}
1 \\
0 \\
0 \\
0 \\
1 \\
\frac{3}{4} \\
1 \\
0 \\
\end{pmatrix}  \, ,
\begin{pmatrix}
\rho^R \\
v_1^R \\
v_2^R \\
v_3^R \\
P^R \\
B_1^R \\
B_2^R \\
B_3^R \\
\end{pmatrix}
=
\begin{pmatrix}
0.125 \\
0 \\
0 \\
0 \\
0.1 \\
\frac{3}{4} \\
-1 \\
0 \\
\end{pmatrix}
\, . \label{eq:bw-states}
\end{equation}

Figure~\ref{fig:bw-shock-1a} compares the global solutions of the second-order and fourth-order schemes to a high-resolution reference solution at \(t=0.1\). Characteristic reconstruction as described in Section~\ref{subsubsec:fv-reconstruct} was used to produce the RK4+PPM results in Figures~\ref{fig:bw-shock-1a} and~\ref{fig:bw-shock-1b}. In this test, the characteristic projection procedure was necessary for the fourth-order scheme to avoid spurious oscillations that exceeded 10\% of the solution range. Even at second order, primitive PLM reconstruction causes nonphysical oscillations to appear.

Figure~\ref{fig:bw-shock-1b} provides a closer view of all solutions. The velocity profile in between the slow shock front at \(x_1\approx 0.14\) and the fast rarefaction at \(x_1\approx 0.36\) exhibited the worst oscillations. This phenomenon is well-known for high-order reconstruction; Figure 5 of  \cite{Matsumoto2016} provides an analogous comparison with fifth-order MP5 reconstruction. The RK4+PPM results shown here avoid the anomalous staircasing present in the compound slow wave near \(x_1\approx -0.02\) produced by the MP5 method. Figure~\ref{fig:bw-shock-1b} also shows slight improvement in the resolution of the slow shock front with the RK4+PPM solver.

\subsection{1D RJ2a shock tube} \label{subsec:mhd-rj2a}
The next MHD test we consider is the shock tube problem introduced by Ryu \& Jones in Figure 2a (RJ2a) \cite{RyuJones1995}. In this test, all 7 MHD wave modes propagate from the discontinuous initial data given by
\begin{equation}
\begin{pmatrix}
\rho^L \\
v_1^L \\
v_2^L \\
v_3^L \\
P^L \\
B_1^L \\
B_2^L \\
B_3^L \\
\end{pmatrix}
=
\begin{pmatrix}
1.08 \\
1.2 \\
0.01 \\
0.5 \\
0.95 \\
0.5641895835477562\\
1.0155412503859613 \\
0.5641895835477562 \\
\end{pmatrix} \, ,
\begin{pmatrix}
\rho^R \\
v_1^R \\
v_2^R \\
v_3^R \\
P^R \\
B_1^R \\
B_2^R \\
B_3^R \\
\end{pmatrix}
=
\begin{pmatrix}
1 \\
0 \\
0 \\
0 \\
1 \\
0.5641895835477562\\
1.1283791670955125 \\
0.5641895835477562 \\
\end{pmatrix}
\, . \label{eq:rj2a-states}
\end{equation}

\begin{figure}[!ht]
\includegraphics[width=\textwidth]{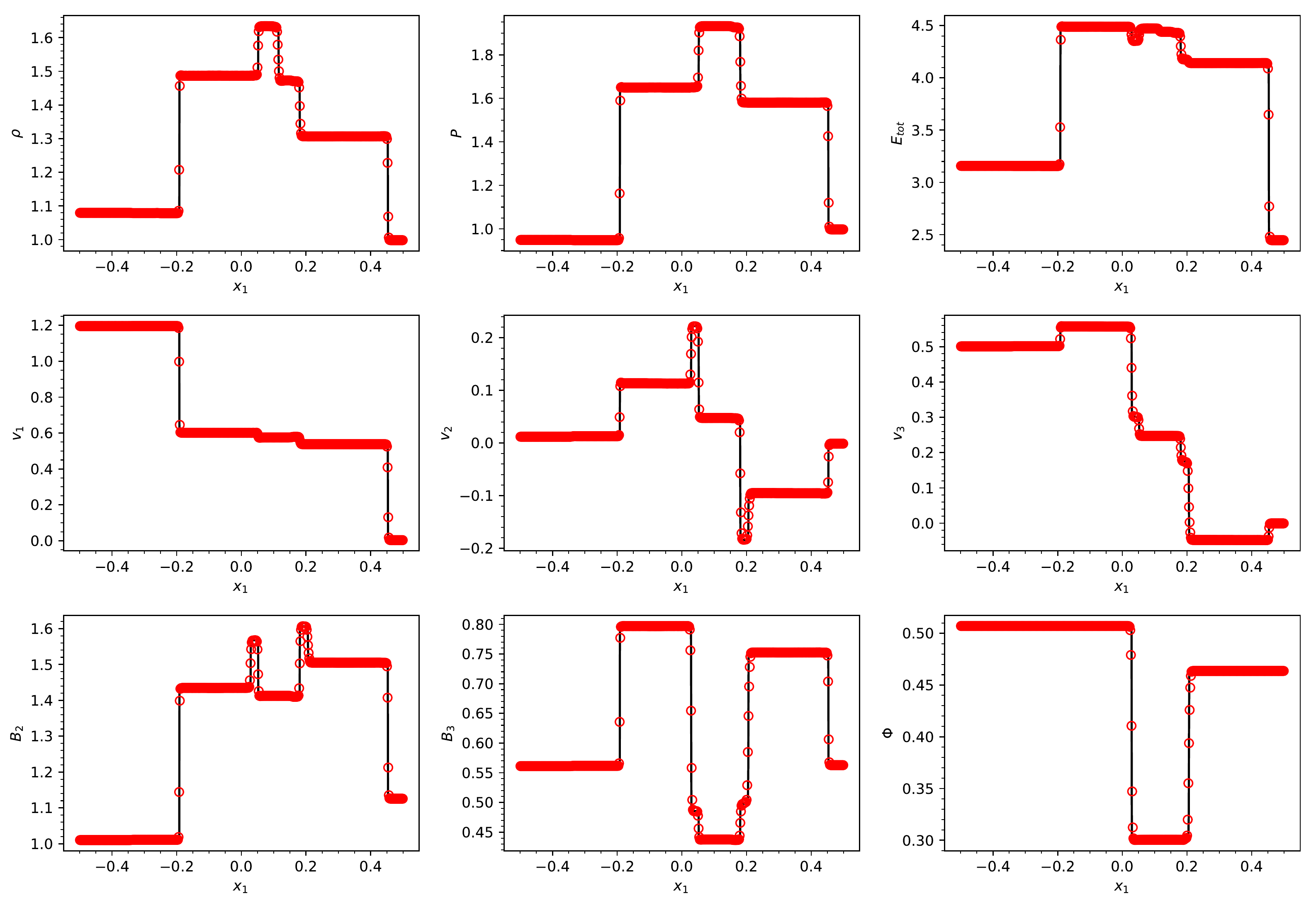}
\caption{RJ2a profiles of density, pressure, total energy, velocity components, transverse magnetic field components, and rotation angle \(\Phi=\tan^{-1}(\frac{\langle B_3 \rangle}{\langle B_2 \rangle}) \) of the magnetic field produced by the fourth-order scheme with \(N_{x_1}=512 \). The low-resolution solution is superimposed on a reference solution produced by the same scheme at high-resolution \(N_{x_1}=8192\) . All MHD discontinuities are resolved with 2-5 cells.}
\label{fig:rj2a-1}
\end{figure}

\begin{figure}[!ht]
\includegraphics[width=\textwidth]{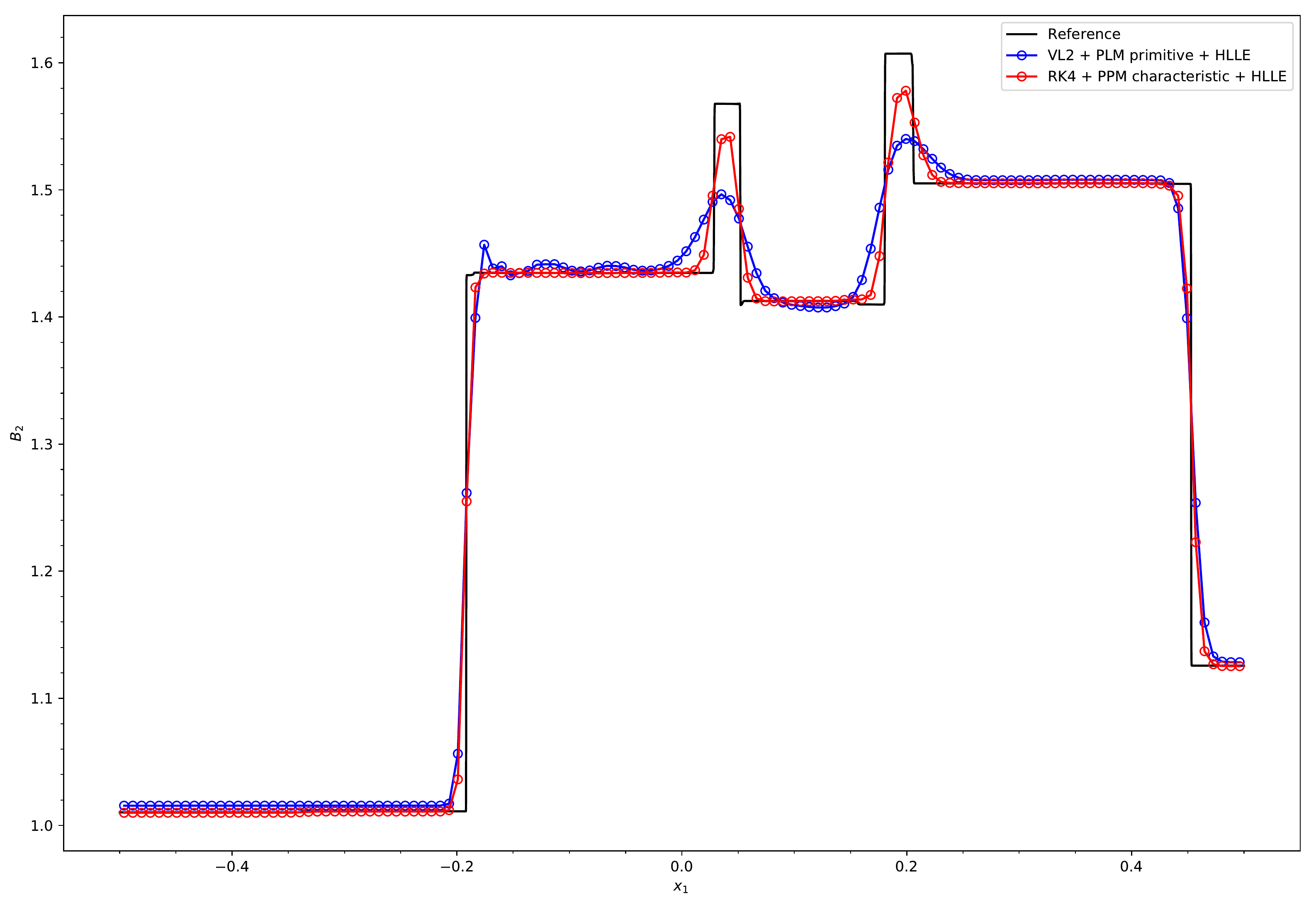}
\caption{The transverse magnetic field component \(B_2\) profile of \( N_{x_1}=128\) cells illustrates the advantages of using the high-order scheme in highly diffusive settings. The HLLE Riemann solver is used to generate both the second-order VL2+PLM and fourth-order RK4+PPM results. The high-order solution shows much better resolution of the rotational discontinuities.}
\label{fig:rj2a-2}
\end{figure}

Figure~\ref{fig:rj2a-1} shows that the fourth-order scheme captures all of the features and resolves the discontinuities with at most 5 cells for the resolution \(N_{x_1}=512\). Figure~\ref{fig:rj2a-2} considers a single profile, \(B_2\), given a lower resolution mesh and a more diffusive Riemann solver, HLLE, instead of the default HLLD solver. The RK4+PPM result compares favorably to the second-order VL2+PLM result, especially when comparing the rotational discontinuities. If stability or computational constraints necessitate the use of HLLE in a low-resolution mesh, the low numerical diffusivity of high-order schemes may significantly improve the results, even in a problem dominated by discontinuities.

Reconstruction of characteristic variables was again used for the fourth-order RJ2a results in both Figure~\ref{fig:rj2a-1} and~\ref{fig:rj2a-2}. Unlike the Brio-Wu shock tube problem in Section~\ref{subsec:mhd-bw}, RK4+PPM with primitive reconstruction produced tolerable oscillations on the same order of magnitude as the VL2+PLM results in Figure~\ref{fig:rj2a-2}.

\subsection{2D Orszag-Tang vortex} \label{subsec:orszag-tang}
The vortex problem of Orszag \& Tang \cite{OrszagTang1979} is a common test of the robustness of MHD schemes. The turbulence that results in this problem tests the ability of the numerical method to resolve the MHD shock-shock interactions while maintaining strict suppression of magnetic monopoles. The initial condition is described in Section 8.4 of \cite{Stone2008} and elsewhere; they are simply rescaled here for a domain \( [-0.5, 0.5]^2\) .

\begin{figure}[!ht]
\includegraphics[width=\textwidth]{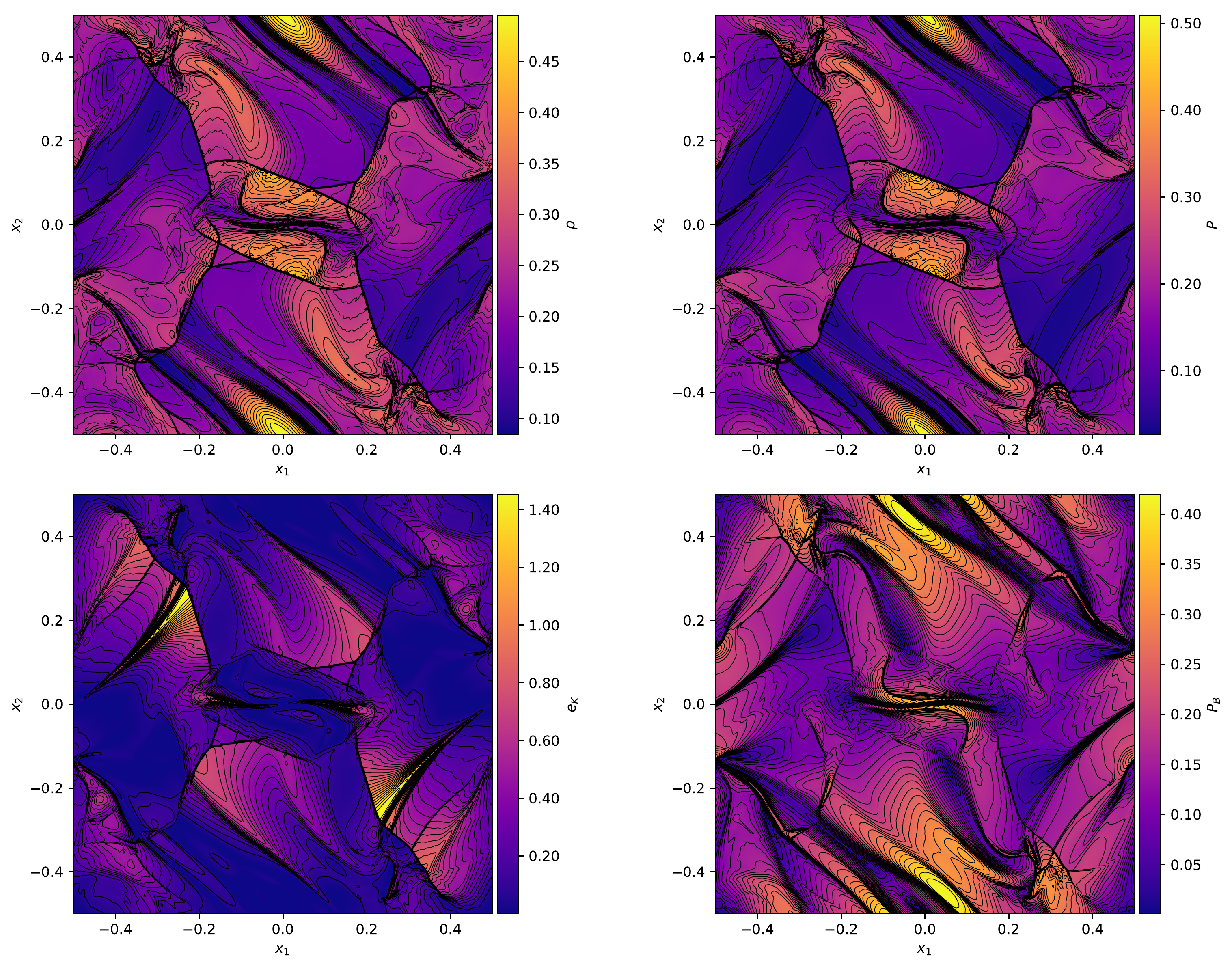}
\caption{Clockwise from top-left subplot: density, pressure, specific kinetic energy, and magnetic pressure of the Orszag-Tang vortex at \(t=\frac{1}{2}\). Thirty contours, linearly spaced between the minimum and maximum values, are overlaid on each plot.}
\label{fig:ot-half}
\end{figure}

\begin{figure}[!ht]
\includegraphics[width=\textwidth]{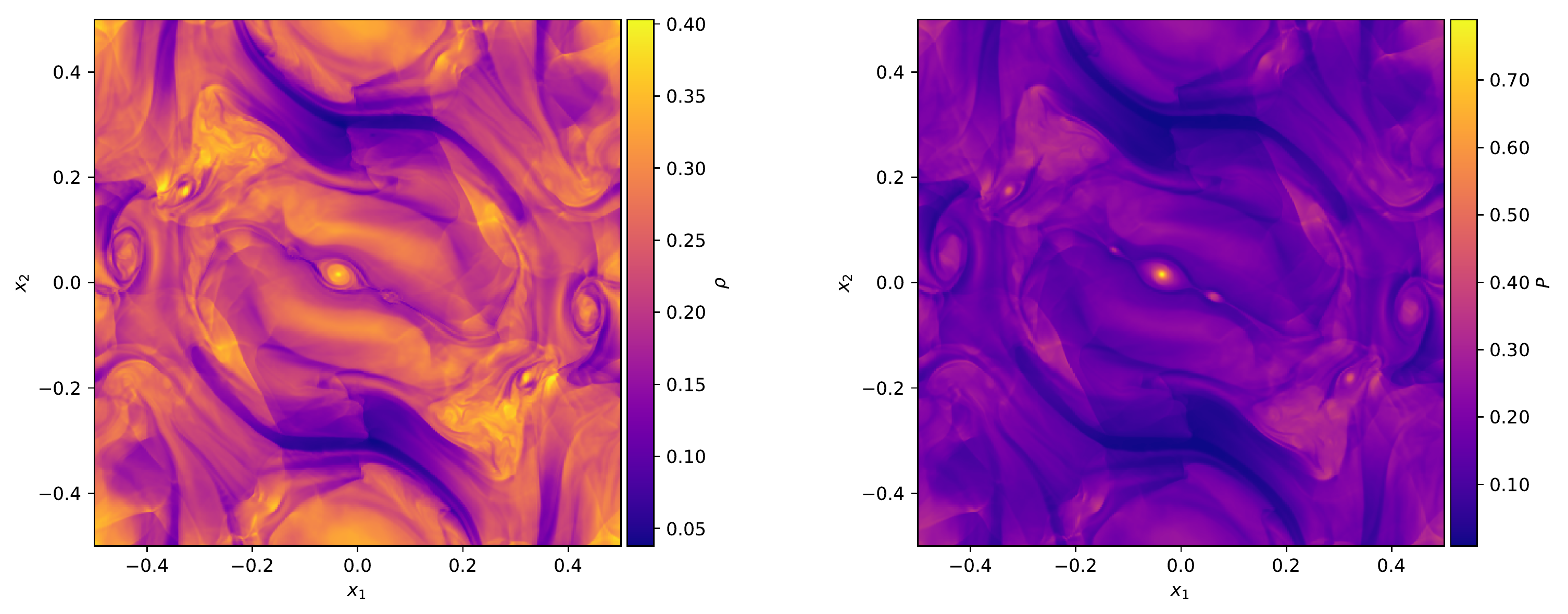}
\caption{Density and pressure of the Orszag-Tang vortex solution at \(t=1 \). Since the late-time evolution of this problem develops into highly turbulent features, no contours are shown as they would obscure much of the plot.}
\label{fig:ot-1}
\end{figure}

Figure~\ref{fig:ot-half} shows the evolution of the vortex at \( t =0.5\) evolved by the fourth-order RK4+PPM scheme with UCT on a uniform grid of \(500^2\) cells. Figure~\ref{fig:ot-1} shows only the pressure and density at a later \(t=1.0\). When compared with the second-order solution (not shown), the fourth-order solution shows improved resolution of the vortex at the origin.

\subsection{2D MHD blast wave} \label{subsec:mhd-blast}
In this test problem, a strongly magnetized medium with uniform \(B_0=1\) aligned with the main diagonal,
\begin{equation}
\mathbf{B} =
\begin{pmatrix}
\frac{B_0}{\sqrt{2}} \\
\frac{B_0}{\sqrt{2}} \\
0
\end{pmatrix}
\, , \label{eq:b-blast}
\end{equation}
of a square periodic domain spanning \( [-0.5, 0.5]^2 \) is initialized with \(\rho=1\) and \( \gamma=5/3 \). The ambient \(P=0.1\) while an overpressure \(P=10\) region is set for cells within a radius \(r=0.1\) of the origin.

\begin{figure}[!ht]
\includegraphics[width=\textwidth]{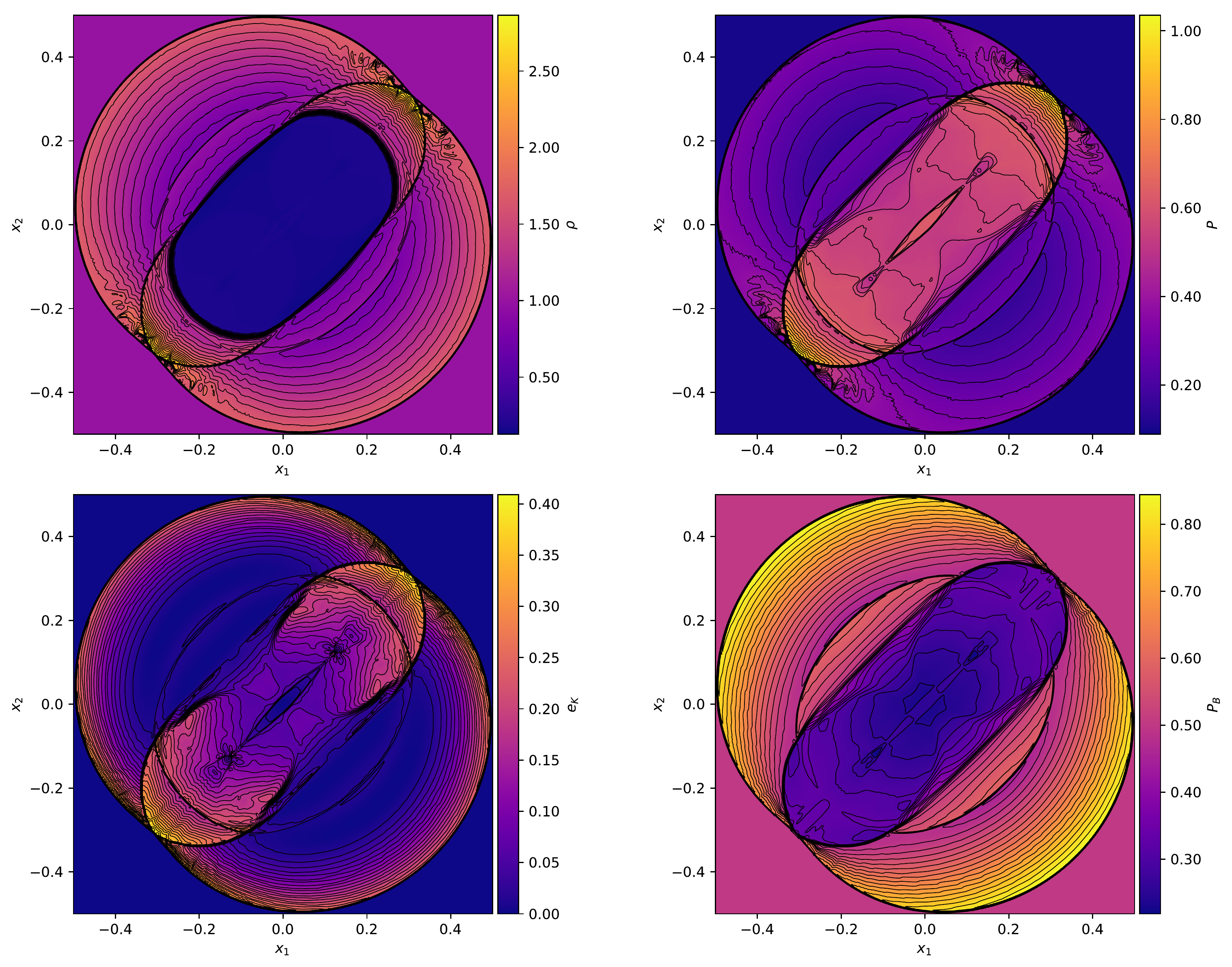}
\caption{Clockwise from top-left subplot: density, pressure, specific kinetic energy, and magnetic pressure of the MHD blast wave at \(t=0.2 \). Thirty linearly spaced contours between the minimum and maximum values of each quantity are shown.}
\label{fig:mhd-blast}
\end{figure}

A resolution of \(500^2\) cells is used for this test. Figure~\ref{fig:mhd-blast} displays the blast wave at \(t=0.2\), right before the shock wave has crossed the periodic boundary. The results can be compared to the second-order accurate results in Figure 28 of the Athena method paper \cite{Stone2008} and Figure 8 of the VL2+PLM method paper \cite{StoneGardiner2009}. The expanding shell is correctly collimated into an ellipse of low density gas oriented with the background magnetic field. The density and pressure contours are well-resolved relative to the second-order results.

Perpendicular to the in-plane \(\mathbf{B}\), the outermost blast wave is a fast-mode that is dominated by the magnetic pressure, and it quickly establishes a large separation from the contact discontinuity of the initial overpressure cylinder. Along the domain diagonal parallel to the magnetic field, the slow-mode shock front is closer to the elongated contact discontinuity, and the fast-mode wave front disappears from the density and gas pressure plots. For the background plasma \(\beta = 0.2 \) in this blast wave setup, the fast-mode wave becomes a non-compressional Alfv\'{e}n mode along the magnetic field lines and appears with only a slight separation from the slow-mode shock front in the \(P_B\) plot of Figure~\ref{fig:mhd-blast}. The correct resolution of these features is a important test for a numerical MHD scheme.

\subsection{2D MHD rotor} \label{subsec:mhd-rotor}
The 2D MHD rotor test was introduced by Balsara and Spicer \cite{BalsaraSpicer1999}; it considers the creation of strong rotational discontinuities in the magnetic field resulting from the shearing of a rapidly rotating disk of dense fluid. We use the same initial conditions used to generate the Athena method paper's Figure 25 results, also described as ``Rotor Test 1'' in 
\cite{Toth2000}. Uniform background density \(\rho=1\), pressure \(P=1\), and \(B_1=\frac{5}{2\sqrt{\pi}}\) are initialized with \( \gamma=7/5\). Within a radius of \(r=0.1\) of the origin, a dense gas of \(\rho=10\) is set to rotate with initial angular velocity \(\omega=20\). No smoothing is used for the density nor velocity of the initial condition.

\begin{figure}[!ht]
\includegraphics[width=\textwidth]{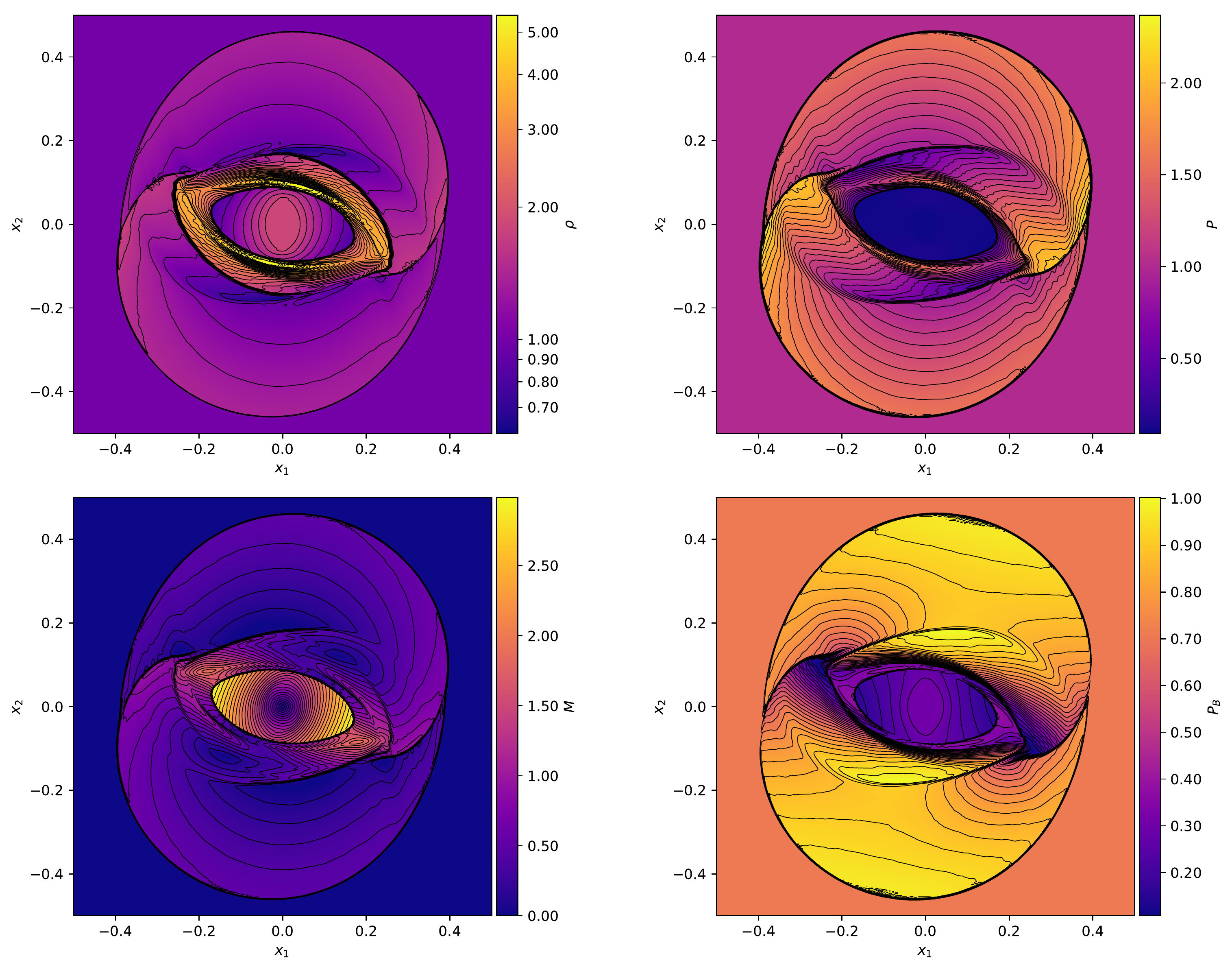}
\caption{Clockwise from top-left subplot: density, pressure, Mach number, and magnetic pressure of the rotor problem at \(t=0.15\). Thirty contours, linearly spaced between the minimum and maximum values, are overlaid on each plot. Note, density is shown using a logarithmic color scale and linear contours to clearly show the density maxima.}
\label{fig:mhd-rotor}
\end{figure}

Figure~\ref{fig:mhd-rotor} shows the result at \(t=0.15\). Again, a resolution of \(500^2\) cells is used; symmetry is well-maintained in the solution, especially for the Mach number plot's concentric circles of the rarefaction from the origin. 

\section{Conclusion} \label{sec:conclusion}
We have presented a fourth-order accurate method of lines method for the numerical solution of the ideal MHD equations.  Using the upwind constrained transport framework, we were able to implement a divergence-free staggered-mesh CT scheme that is consistent with a fourth-order accurate finite volume scheme for the hydrodynamics subsystem. The underlying finite volume scheme is based the quadrature rules and \(\mathcal{O}(\Delta x^4) \) intermediate calculations of McCorquodale and Colella \cite{McCorquodaleColella2011} for nonlinear systems of hyperbolic conservation laws. In comparison to the second-order constrained transport scheme of GS05, the fourth-order method yields orders of magnitude of improvement in error in globally smooth linear and nonlinear MHD problems. The overall scheme also exhibits excellent robustness for discontinuous features in multidimensional tests.

Future work will involve the extension of the four-state flux upwinding rule of HLL-UCT in Section~\ref{subsec:uct-upwind} to be consistent with the multidimensionally-averaged approximate Riemann fan of the HLLD solver. Comprehensive analysis will be made in comparing these methods to the broader multidimensional Riemann solver literature.  Additionally, the extension of our fourth-order accurate UCT implementation to adaptive mesh refinement is a high priority for the practical use in demanding astrophysics simulations. To that end, we will formulate the necessary \(\mathcal{O}(\Delta x^4) \) prolongation and restriction operators that are compatible with AMR and the mapped grid formalism discussed in Section~\ref{sec:fv-hydro}.

High-order methods offer the greatest advantages relative to conventional second-order accurate methods when applied to demanding problems with long simulation times and complex smooth features. Therefore, the fourth-order method will soon be applied to the simulation of shearing box approximations of accretion disk dynamics. The high arithmetic intensity of the fourth-order method can ameliorate the increasing performance and power costs of memory references relative to floating-point operations on modern computing architectures. These performance and accuracy tradeoffs and the scaling trends will be quantitatively evaluated on emerging manycore architectures.

\section*{Acknowledgments} \label{sec:acknowledgements}
The authors thank Thomas Gardiner for discussion and comments on an early draft of this manuscript. K.G.F. was supported by the Department of Energy Computational Science Graduate Fellowship (CSGF), grant number DE-FG02-97ER25308. J.M.S. was supported by National Science Foundation, grant number AST-1715277.

\appendix
\section{\(\mathscr{E}_z^c \) scheme of GS05 scheme is equivalent to a second-order accurate UCT method} \label{sec:appendix-gs05-uct}
This appendix directly compares the \(\mathscr{E}_z^c \) constrained transport algorithm of GS05 \cite{GardinerStone2005} and the UCT framework of LD2004 \cite{Londrillo2004}. In particular, we consider several limiting cases to show that the UCT methodology, when implemented with an \(\mathcal{O}(\Delta x^2) \) reconstruction method, returns upwind corner emf values \(\mathscr{E}_z\) that are consistent with the GS05 algorithm when coupled to an underlying Godunov-type method based on HLL fluxes. The analysis below highlights the differences between the GS05 and LD2004 approaches to constructing a constrained transport discretization, and it provides further context for the \(\mathcal{O}(\Delta x^2) \) limitations that are inherent to the GS05 derivation, as discussed in Section~\ref{sec:uct}.

We begin by assuming that the two-speed flux approximation is made by using the one-state HLL Riemann solver to compute the Godunov fluxes at all faces in the GS05 implementation. Then, the numerical fluxes are approximated by a single intermediate (subsonic) and two supersonic states
\begin{equation}
\mathbf{F}^{HLL}_1 = \begin{cases}
\mathbf{F}^{L_1}_1 & S^{L_1} \geq 0 \\
\mathbf{F}^*_1 & S^L \leq 0 \leq S^R  \\ \mathbf{F}^{R_1}_1 & S^{R_1} \leq 0
\end{cases} \, , \label{eq:hll-states}
\end{equation}
for fluxes in the \(x_1\) direction. For further detail, we refer the reader to Equation 11 of Miyoshi and Kusano \cite{MiyoshiKusano2005}, which uses the same notation but with subscripted \(L/R\). In particular, the nontrivial intermediate state of the \( B_2 \) flux component at an \(x_1\)-face (an {\bf upwinded} approximation to the emf \(\mathscr{E}_{3} \)) can be written as
\begin{equation}
\mathbf{e}_{B_2} \cdot \bm{\mathrm{F}}_1^* = \mathscr{E}_{3,i-\frac{1}{2},j} = \frac{S^{R_1} \mathscr{E}_{3,i-\frac{1}{2},j}^{L_1} - S^{L_1} \mathscr{E}_{3,i-\frac{1}{2},j}^{R_1}}{S^{R_1} - S^{L_1}} - \frac{S^{R_1} S^{L_1}}{S^{R_1} - S^{L_1}} (\langle B_2^{R_1} \rangle_{i-\frac{1}{2},j}  - \langle B_2^{L_1} \rangle_{i-\frac{1}{2},j}) \,, \label{eq:hll-star-emf-x1}
\end{equation}
where we have separated the expression into two terms: the first term consists of smooth flux approximations and the second term encapsulates the explicit numerical dissipation of the HLL solver. A well-known trick to unify Equations~\eqref{eq:hll-states}, \eqref{eq:hll-star-emf-x1} is to replace the wavespeed estimates with nonnegative quantities that are both nonzero only for the subsonic case \cite{MiyoshiKusano2005}.
We use the notation for the dissipative terms in LD2004 Equation 55 \cite{Londrillo2004}. By letting \( \alpha^+_1 \equiv \max(0, S^{R_1}), \alpha^-_1 \equiv -\min(0, S^{L_1}) \), we can express the HLL flux \( F^{HLL} \) for all possible cases of wavespeed estimates with a single expression
\begin{equation}
\mathscr{E}_{3,i-\frac{1}{2},j} = \frac{\alpha^+_1 \mathscr{E}_{3,i-\frac{1}{2},j}^{L_1} + \alpha^-_1 \mathscr{E}_{3,i-\frac{1}{2},j}^{R_1}}{\alpha^+_1 + \alpha^-_1} + \frac{\alpha^+_1 \alpha^-_1}{\alpha^+_1 + \alpha^-_1} (\langle B_2^{R_1} \rangle_{i-\frac{1}{2},j}  - \langle B_2^{L_1} \rangle_{i-\frac{1}{2},j}) \, . \label{eq:hll-emf-x1}
\end{equation}
Due to the antisymmetry of the curl operator, the counterpart of the previous equation for the emf component of the HLL flux on \(x_2\)-faces is
\begin{equation}
\mathscr{E}_{3,i,j-\frac{1}{2}} = \frac{\alpha^+_2 \mathscr{E}_{3,i,j-\frac{1}{2}}^{L_2} + \alpha^-_2 \mathscr{E}_{3,i,j-\frac{1}{2}}^{R_2}}{\alpha^+_2 + \alpha^-_2} - \frac{\alpha^+_2 \alpha^-_2}{\alpha^+_2 + \alpha^-_2} (\langle B_1^{R_2} \rangle_{i,j-\frac{1}{2}}  - \langle B_1^{L_2} \rangle_{i,j-\frac{1}{2}}) \, . \label{eq:hll-emf-x2}
\end{equation}
We now separately consider two possibilities for the upwind directions at the four 1D interfaces used in the \(\mathscr{E}_z^c \) scheme.

\subsection{Stationary domain \(\mathbf{v}=0\)} \label{app:gs05-uct-stationary}
The comparison of GS05 and LD2004 in the case of a stationary domain is nontrivial due to the upwinding of the 1D conserved variable fluxes that occurs in GS05 in Equation~\eqref{eq:gs05-ct-upwind} but does not occur in UCT. Nevertheless, the analysis is greatly simplified because all smooth reconstructions of the emf in both the 2D upwinding in UCT Equation~\eqref{eq:uct-hll} and the 1D upwinding in Equations~\eqref{eq:hll-emf-x1} and~\eqref{eq:hll-emf-x2} are zero since \(\mathscr{E}_{3} = v_{2}B_{1} - v_{1}B_{2} =0 \). By symmetry, the estimates of the minimum and maximum wavespeed bounds must be equal in magnitude separately in each direction, with \(\alpha_1 \equiv \alpha_1^+ = \alpha_1^-\) and \( \alpha_2 \equiv \alpha_2^+ = \alpha_2^-\). Hence, only the HLL explicit dissipation terms remain in the GS05 and UCT Riemann solver fluxes:
\begin{align}
\mathscr{E}_{3,i-\frac{1}{2},j} =&  \frac{\alpha_1}{2} (\langle B_2^{R_1} \rangle_{i-\frac{1}{2},j}  - \langle B_2^{L_1} \rangle_{i-\frac{1}{2},j}) \, , \label{eq:hll-emf-x1-stationary} \\
\mathscr{E}_{3,i,j-\frac{1}{2}} =& - \frac{\alpha_2}{2} (\langle B_1^{R_2} \rangle_{i,j-\frac{1}{2}}  - \langle B_1^{L_2} \rangle_{i,j-\frac{1}{2}}) \, , \label{eq:hll-emf-x2-stationary} \\
\langle \mathscr{E}^U_{3} \rangle_{i-\frac{1}{2},j-\frac{1}{2}} =& - \frac{\alpha_2}{2}( \langle {B}_1^{R_2} \rangle_{i-\frac{1}{2},j-\frac{1}{2}}  - \langle {B}_1^{L_2} \rangle_{i-\frac{1}{2},j-\frac{1}{2}}  ) \nonumber \\
&+ \frac{\alpha_1 }{2}(\langle {B}_2^{R_1} \rangle_{i-\frac{1}{2},j-\frac{1}{2}}  - \langle {B}_2^{L_1} \rangle_{i-\frac{1}{2},j-\frac{1}{2}} ) \, . \label{eq:uct-hll-stationary}
\end{align}
The GS05 method's upwinding of the emf derivatives in Equation~\eqref{eq:gs05-ct-upwind} reduces to averaging in both the \(x_1,x_2\) directions
\begin{align}
\left(\frac{\partial \mathscr{E}_3}{\partial x_1} \right)_{i-\frac{1}{4},j-\frac{1}{2}} =& \frac{1}{2}\left[\left(\frac{\partial \mathscr{E}_3}{\partial x_1} \right)_{i-\frac{1}{4},j-1} + \left(\frac{\partial \mathscr{E}_3}{\partial x_1} \right)_{i-\frac{1}{4},j}\right] \, , \label{eq:gs05-ave-x1}\\
\left(\frac{\partial \mathscr{E}_3}{\partial x_2} \right)_{i-\frac{1}{2},j-\frac{1}{4}} =& \frac{1}{2}\left[\left(\frac{\partial \mathscr{E}_3}{\partial x_2}\right)_{i-1, j-\frac{1}{4}} + \left(\frac{\partial \mathscr{E}_3}{\partial x_2}\right)_{i, j-\frac{1}{4}} \right] \, , \label{eq:gs05-ave-x2}
\end{align}
and the four-way average in Equation~\eqref{eq:gs05-emf-average} becomes
\begin{equation} \label{eq:gs05-emf-average-stationary}
\begin{split}
\mathscr{E}_{3,i-\frac{1}{2},j-\frac{1}{2}} =& \frac{1}{4} (  \mathscr{E}_{3,i,j-\frac{1}{2}} + \mathscr{E}_{3,i-1,j-\frac{1}{2}} +\mathscr{E}_{3,i-\frac{1}{2},j} +\mathscr{E}_{3,i-\frac{1}{2},j-1} ) \\
+&\frac{h}{16} \left(
\left(\frac{\partial \mathscr{E}_3}{\partial x_2}\right)_{i-1, j-\frac{3}{4}} + \left(\frac{\partial \mathscr{E}_3}{\partial x_2}\right)_{i, j-\frac{3}{4}}
-\left(\frac{\partial \mathscr{E}_3}{\partial x_2}\right)_{i-1, j-\frac{1}{4}} - \left(\frac{\partial \mathscr{E}_3}{\partial x_2}\right)_{i, j-\frac{1}{4}}
\right)  \\
+&\frac{h}{16} \left(
\left(\frac{\partial \mathscr{E}_3}{\partial x_1} \right)_{i-\frac{3}{4},j-1} + \left(\frac{\partial \mathscr{E}_3}{\partial x_1} \right)_{i-\frac{3}{4},j}
-\left(\frac{\partial \mathscr{E}_3}{\partial x_1} \right)_{i-\frac{1}{4},j-1} - \left(\frac{\partial \mathscr{E}_3}{\partial x_1} \right)_{i-\frac{1}{4},j}
 \right) \, .
\end{split}
\end{equation}
These averaged slopes are approximated at \(\mathcal{O}(\Delta x^2) \) accuracy in GS05 by Equation~\eqref{eq:emf-slope}. Due to the sharing of the same quadrature point of the face-centered flux, the difference of two linear slope stencils at the same index in the transverse direction can be slightly condensed as
\begin{equation} \label{eq:gs05-slope-diff}
\begin{split}
\left(\frac{\partial \mathscr{E}_3}{\partial x_1} \right)_{i-\frac{3}{4},j} - \left(\frac{\partial \mathscr{E}_3}{\partial x_1} \right)_{i-\frac{1}{4},j} =& \frac{2}{h} (\mathscr{E}_{3,i-\frac{1}{2},j} - \mathscr{E}^r_{3,i-1,j} ) -  \frac{2}{h} (\mathscr{E}^r_{3,i,j}  - \mathscr{E}_{3,i-\frac{1}{2},j} )  \\
= & \frac{2}{h} (2\mathscr{E}_{3,i-\frac{1}{2},j}  - \mathscr{E}^r_{3,i-1,j} - \mathscr{E}^r_{3,i,j} ) \, .
\end{split}
\end{equation}
In this specific case of a stationary domain, the cell-centered reference electric fields are 0, but the HLL upwinded quantities may not be zero, so Equation~\eqref{eq:gs05-slope-diff} is reduced to
\begin{equation}
\left(\frac{\partial \mathscr{E}_3}{\partial x_1} \right)_{i-\frac{3}{4},j} - \left(\frac{\partial \mathscr{E}_3}{\partial x_1} \right)_{i-\frac{1}{4},j} = \frac{4}{h}\mathscr{E}_{3,i-\frac{1}{2},j} \, . \label{eq:gs05-slope-diff-stationary}
\end{equation}
Thus, the GS05 expression for the corner emf in Equation~\eqref{eq:gs05-emf-average-stationary} can be written solely in terms of the 1D face-centered fluxes as
\begin{equation}
\mathscr{E}_{3,i-\frac{1}{2},j-\frac{1}{2}} = \frac{1}{2} (  \mathscr{E}_{3,i,j-\frac{1}{2}} + \mathscr{E}_{3,i-1,j-\frac{1}{2}} +\mathscr{E}_{3,i-\frac{1}{2},j} +\mathscr{E}_{3,i-\frac{1}{2},j-1} ) \, , \label{eq:gs05-emf-average-stationary2}
\end{equation}
which is the directionally unbiased formula with correct numerical viscosity, corresponding to GS05 Equation 39 under the assumption of a stationary domain. Substituting the expressions for the face-centered fluxes in Equations~\eqref{eq:hll-emf-x1-stationary} and~\eqref{eq:hll-emf-x2-stationary}, we get
\begin{equation}\label{eq:gs05-emf-average-stationary-dissipation}
\begin{split}
\mathscr{E}_{3,i-\frac{1}{2},j-\frac{1}{2}} =& \frac{\alpha_1}{4} (\langle B_2^{R_1} \rangle_{i-\frac{1}{2},j}  - \langle B_2^{L_1} \rangle_{i-\frac{1}{2},j}
+ \langle B_2^{R_1} \rangle_{i-\frac{1}{2},j-1}  - \langle B_2^{L_1} \rangle_{i-\frac{1}{2},j-1})  \\
-& \frac{\alpha_2}{4} (\langle B_1^{R_2} \rangle_{i,j-\frac{1}{2}}  - \langle B_1^{L_2} \rangle_{i,j-\frac{1}{2}}
+ \langle B_1^{R_2} \rangle_{i-1,j-\frac{1}{2}}  - \langle B_1^{L_2} \rangle_{i-1,j-\frac{1}{2}}) \, .
\end{split}
\end{equation}
Finally, to show equivalence to the UCT expression in Equation~\eqref{eq:uct-hll-stationary}, we use the second-order reconstruction assumption to relate the cell-corner state of \(\langle {B}_1^{L_2} \rangle_{i-\frac{1}{2},j-\frac{1}{2}}\) to the transverse face-centered reconstructed states \(\langle B_1^{L_2} \rangle_{i,j-\frac{1}{2}}, \langle B_1^{L_2} \rangle_{i-1,j-\frac{1}{2}}\). While the magnetic field is never explicitly reconstructed at cell-corners in GS05, the continuity of \(B_1\) along \(x_1\) demands that
\begin{equation}
\langle {B}_1^{L_2} \rangle_{i-\frac{1}{2},j-\frac{1}{2}} = \frac{1}{2} \left(\langle B_1^{L_2} \rangle_{i,j-\frac{1}{2}} + \langle B_1^{L_2} \rangle_{i-1,j-\frac{1}{2}}\right) + \mathcal{O}(\Delta x^2) \, .  \label{eq:gs05-corner-b}
\end{equation}
Hence, Equation~\eqref{eq:gs05-emf-average-stationary-dissipation} is equivalent to
\begin{equation} \label{eq:gs05-equiv-uct2}
\begin{split}
\mathscr{E}_{3,i-\frac{1}{2},j-\frac{1}{2}} =& \frac{\alpha_1}{2} \left(\frac{\langle B_2^{R_1} \rangle_{i-\frac{1}{2},j} + \langle B_2^{R_1} \rangle_{i-\frac{1}{2},j-1} }{2} - \frac{\langle B_2^{L_1} \rangle_{i-\frac{1}{2},j} - \langle B_2^{L_1} \rangle_{i-\frac{1}{2},j-1}}{2}\right)   \\
-& \frac{\alpha_2}{2} \left(\frac{\langle B_1^{R_2} \rangle_{i,j-\frac{1}{2}}  + \langle B_1^{R_2} \rangle_{i-1,j-\frac{1}{2}}}{2} - \frac{\langle B_1^{L_2} \rangle_{i,j-\frac{1}{2}} + \langle B_1^{L_2} \rangle_{i-1,j-\frac{1}{2}}}{2} \right)  \\
=& \frac{\alpha_1 }{2}(\langle {B}_2^{R_1} \rangle_{i-\frac{1}{2},j-\frac{1}{2}}  - \langle {B}_2^{L_1} \rangle_{i-\frac{1}{2},j-\frac{1}{2}} ) -  \frac{\alpha_2}{2}( \langle {B}_1^{R_2} \rangle_{i-\frac{1}{2},j-\frac{1}{2}}  - \langle {B}_1^{L_2} \rangle_{i-\frac{1}{2},j-\frac{1}{2}}  ) + \mathcal{O}(\Delta x^2)   \\
=& \langle\mathscr{E}_{3}^U \rangle_{i-\frac{1}{2},j-\frac{1}{2}}  + \mathcal{O}(\Delta x^2) \, .
\end{split}
\end{equation}

\subsection{Grid-aligned plane-parallel flow: \(v_1\neq 0, v_2=0\)} \label{app:gs05-uct-x1-flow}
Cases involving non-stationary background flows follow similar lines of reasoning as \ref{app:gs05-uct-stationary}, but the GS05 formulas are significantly more complicated owing to the introduction of \(L/R\) reconstructed emf terms in the HLL expressions. Here, we allow for grid-aligned flow in the positive \(x_1\) direction while assuming no motion in the \(x_2\) direction. Hence, smooth approximations to the emf satisfy \(\mathscr{E}_{3} = - v_{1}B_{2} \). Initially, we consider a magnetic field \(\mathbf{B}(x_1,x_2)\) that may have any physically admissible discontinuities.

By symmetry, the estimates of the minimum and maximum wavespeed bounds must be equal in magnitude in the \(x_2\) direction, with \( \alpha_2 \equiv \alpha_2^+ = \alpha_2^-\). The fluxes upwinded by the Riemann solver in the two methods can be written as
\begin{align}
\mathscr{E}_{3,i-\frac{1}{2},j} =&  \frac{\alpha^+_1 \mathscr{E}_{3,i-\frac{1}{2},j}^{L_1} + \alpha^-_1 \mathscr{E}_{3,i-\frac{1}{2},j}^{R_1}}{\alpha^+_1 + \alpha^-_1} + \frac{\alpha^+_1 \alpha^-_1}{\alpha^+_1 + \alpha^-_1} (\langle B_2^{R_1} \rangle_{i-\frac{1}{2},j}  - \langle B_2^{L_1} \rangle_{i-\frac{1}{2},j}) \, , \label{eq:hll-emf-x1-x1-flow} \\
\mathscr{E}_{3,i,j-\frac{1}{2}} =& \frac{1}{2} \left(\mathscr{E}_{3,i,j-\frac{1}{2}}^{L_2} +\mathscr{E}_{3,i,j-\frac{1}{2}}^{R_2} \right)
- \frac{\alpha_2}{2} (\langle B_1^{R_2} \rangle_{i,j-\frac{1}{2}}  - \langle B_1^{L_2} \rangle_{i,j-\frac{1}{2}}) \, , \label{eq:hll-emf-x2-x1-flow} \\
\langle \mathscr{E}^U_{3} \rangle_{i-\frac{1}{2},j-\frac{1}{2}} =& \frac{
\alpha^+_1\left( \langle \mathscr{E}^{L_1L_2}_{3}\rangle_{i-\frac{1}{2},j-\frac{1}{2}} + \langle \mathscr{E}_{3}^{L_1R_2} \rangle_{i-\frac{1}{2},j-\frac{1}{2}} \right) +  \alpha^-_1\left(\langle \mathscr{E}_{3}^{R_1L_2}  \rangle_{i-\frac{1}{2},j-\frac{1}{2}} + \langle \mathscr{E}_{3}^{R_1R_2} \rangle_{i-\frac{1}{2},j-\frac{1}{2}} \right)  }{2(\alpha^+_1 + \alpha^-_1)} \nonumber \\
&  - \frac{\alpha_2}{2}( \langle {B}_1^{R_2} \rangle_{i-\frac{1}{2},j-\frac{1}{2}}  - \langle {B}_1^{L_2} \rangle_{i-\frac{1}{2},j-\frac{1}{2}}  ) + \frac{\alpha^+_1 \alpha^-_1}{\alpha^+_1 + \alpha^-_1}(\langle {B}_2^{R_1} \rangle_{i-\frac{1}{2},j-\frac{1}{2}}  - \langle {B}_2^{L_1} \rangle_{i-\frac{1}{2},j-\frac{1}{2}} ) \, .\label{eq:uct-hll-x1-flow}
\end{align}
The GS05 upwinding of the emf derivatives in Equation~\eqref{eq:gs05-ct-upwind} reduces to the selection of the lower index of the approximations of \(\partial_{2}\) in the \(x_1\) direction
\begin{equation}
\left(\frac{\partial \mathscr{E}_3}{\partial x_2} \right)_{i-\frac{1}{2},j-\frac{1}{4}} = \left(\frac{\partial \mathscr{E}_3}{\partial x_2} \right)_{i-1, j-\frac{1}{4}} \, , \label{eq:gs05-lower-x1select-x1-flow}
\end{equation}
and central averaging of the \(\partial_1\) approximations in the \(x_2\) direction
\begin{equation}
\left(\frac{\partial \mathscr{E}_3}{\partial x_1} \right)_{i-\frac{1}{4},j-\frac{1}{2}} = \frac{1}{2}\left[\left(\frac{\partial \mathscr{E}_3}{\partial x_1} \right)_{i-\frac{1}{4},j-1} + \left(\frac{\partial \mathscr{E}_3}{\partial x_1} \right)_{i-\frac{1}{4},j}\right] \, .\label{eq:gs05-ave-x2-x1-flow}
\end{equation}
The four-way average in Equation~\eqref{eq:gs05-emf-average} becomes
\begin{equation}  \label{eq:gs05-emf-average-x1-flow}
\begin{split}
\mathscr{E}_{3,i-\frac{1}{2},j-\frac{1}{2}} =& \frac{1}{4} (  \mathscr{E}_{3,i,j-\frac{1}{2}} + \mathscr{E}_{3,i-1,j-\frac{1}{2}} +\mathscr{E}_{3,i-\frac{1}{2},j} +\mathscr{E}_{3,i-\frac{1}{2},j-1} )  \\
+&\frac{h}{8} \left( \left(\frac{\partial \mathscr{E}_3}{\partial x_2} \right)_{i-1,j-\frac{3}{4}} - \left(\frac{\partial \mathscr{E}_3}{\partial x_2} \right)_{i-1,j-\frac{1}{4}} \right)  \\
+&\frac{h}{16} \left(
\left(\frac{\partial \mathscr{E}_3}{\partial x_1} \right)_{i-\frac{3}{4},j-1} + \left(\frac{\partial \mathscr{E}_3}{\partial x_1} \right)_{i-\frac{3}{4},j}
-\left(\frac{\partial \mathscr{E}_3}{\partial x_1} \right)_{i-\frac{1}{4},j-1} - \left(\frac{\partial \mathscr{E}_3}{\partial x_1} \right)_{i-\frac{1}{4},j}
 \right) \, .
\end{split}
\end{equation}
Using Equation~\eqref{eq:gs05-slope-diff}, the GS05 expression reduces to
\begin{equation}  \label{eq:gs05-emf-average-x1-flow-2}
\begin{split}
\mathscr{E}_{3,i-\frac{1}{2},j-\frac{1}{2}} =& \frac{1}{4} (  \mathscr{E}_{3,i,j-\frac{1}{2}} + \mathscr{E}_{3,i-1,j-\frac{1}{2}} +\mathscr{E}_{3,i-\frac{1}{2},j} +\mathscr{E}_{3,i-\frac{1}{2},j-1} ) + \frac{h}{8}\frac{2}{h} \left( 2\mathscr{E}_{3,i,j-\frac{1}{2}}  - \mathscr{E}^r_{3,i-1,j} - \mathscr{E}^r_{3,i-1,j-1}  \right) \\
+&\frac{h}{16}\frac{2}{h} \left(
(2\mathscr{E}_{3,i-\frac{1}{2},j-1}  - \mathscr{E}^r_{3,i-1,j-1} - \mathscr{E}^r_{3,i,j-1} )
+ (2\mathscr{E}_{3,i-\frac{1}{2},j}  - \mathscr{E}^r_{3,i-1,j} - \mathscr{E}^r_{3,i,j} )
\right) \, .
\end{split}
\end{equation}
After collecting the terms in the expression, we observe that the GS05 upwinding scheme has essentially acted as a switch between the two intercell \(x_2\) face HLL fluxes, with
\begin{equation}\label{eq:gs05-emf-average-x1-flow-3}
\begin{split}
\mathscr{E}_{3,i-\frac{1}{2},j-\frac{1}{2}} =&   \frac{3}{4}\mathscr{E}_{3,i-1,j-\frac{1}{2}} + \frac{1}{4}\mathscr{E}_{3,i,j-\frac{1}{2}} + \frac{1}{2}(\mathscr{E}_{3,i-\frac{1}{2},j} +\mathscr{E}_{3,i-\frac{1}{2},j-1} )  \\
-& \frac{3}{8}\left(\mathscr{E}^r_{3,i-1,j} +\mathscr{E}^r_{3,i-1,j-1} \right) -\frac{1}{8}\left(\mathscr{E}^r_{3,i,j-1} +\mathscr{E}^r_{3,i,j} \right) \, ,
\end{split}
\end{equation}
where the shared upwind \(x_1\) direction at both \(i-\frac{1}{2},j\) and \(i-\frac{1}{2},j-1\) interfaces biases the terms above with larger \(\frac{3}{4},-\frac{3}{8}\) coefficients relative to the downstream quantities with \(\frac{1}{4},-\frac{1}{8}\) coefficients.
The interface quantities in the previous equation have been upwinded in 1D by a Riemann solver. Therefore, in general they cannot be described exclusively in terms of the reconstructed quantities of a single cell. We now insert the expressions for the 1D HLL fluxes in Equations~\eqref{eq:hll-emf-x1-x1-flow} and~\eqref{eq:hll-emf-x2-x1-flow} in order to separate out the HLL explicit dissipation terms as
\begin{equation}\label{eq:gs05-emf-average-x1-flow-4}
\begin{split}
\mathscr{E}_{3,i-\frac{1}{2},j-\frac{1}{2}} =& \frac{3}{8}\left(\left(\mathscr{E}_{3,i-1,j-\frac{1}{2}}^{L_2} + \mathscr{E}_{3,i-1,j-\frac{1}{2}}^{R_2} \right)
- \alpha_2(\langle B_1^{R_2} \rangle_{i-1,j-\frac{1}{2}}  - \langle B_1^{L_2} \rangle_{i-1,j-\frac{1}{2}}) \right)  \\
+& \frac{1}{8} \left(\left(\mathscr{E}_{3,i,j-\frac{1}{2}}^{L_2} +  \mathscr{E}_{3,i,j-\frac{1}{2}}^{R_2} \right)
- \alpha_2(\langle B_1^{R_2} \rangle_{i,j-\frac{1}{2}}  - \langle B_1^{L_2} \rangle_{i,j-\frac{1}{2}}) \right)  \\
+& \frac{\alpha^+_1 \mathscr{E}_{3,i-\frac{1}{2},j}^{L_1} + \alpha^-_1 \mathscr{E}_{3,i-\frac{1}{2},j}^{R_1} + \alpha^+_1 \mathscr{E}_{3,i-\frac{1}{2},j-1}^{L_1} + \alpha^-_1 \mathscr{E}_{3,i-\frac{1}{2},j-1}^{R_1}}{2(\alpha^+_1 + \alpha^-_1)}  \\
+& \frac{\alpha^+_1 \alpha^-_1}{2(\alpha^+_1 + \alpha^-_1)} (\langle B_2^{R_1} \rangle_{i-\frac{1}{2},j}  - \langle B_2^{L_1} \rangle_{i-\frac{1}{2},j} + \langle B_2^{R_1} \rangle_{i-\frac{1}{2},j-1}  - \langle B_2^{L_1} \rangle_{i-\frac{1}{2},j-1}) \\
-& \frac{3}{8}\left(\mathscr{E}^r_{3,i-1,j} +\mathscr{E}^r_{3,i-1,j-1} \right) -\frac{1}{8}\left(\mathscr{E}^r_{3,i,j-1} +\mathscr{E}^r_{3,i,j} \right) \, .
\end{split}
\end{equation}
These terms include Riemann states of \(\mathbf{B}\) that are reconstructed in adjacent cells. We then use Equation~\eqref{eq:gs05-corner-b} to combine the \(B_2\) terms from the \(x_1\) interface Riemann solutions, resulting in
\begin{equation}\label{eq:gs05-emf-average-x1-flow-5}
\begin{split}
\mathscr{E}_{3,i-\frac{1}{2},j-\frac{1}{2}} =& -\frac{3\alpha_2}{8}\left(\langle B_1^{R_2} \rangle_{i-1,j-\frac{1}{2}}  - \langle B_1^{L_2} \rangle_{i-1,j-\frac{1}{2}} \right) - \frac{\alpha_2}{8} \left(\langle B_1^{R_2} \rangle_{i,j-\frac{1}{2}}  - \langle B_1^{L_2} \rangle_{i,j-\frac{1}{2}} \right) \\
+& \frac{\alpha^+_1 \left( \mathscr{E}_{3,i-\frac{1}{2},j}^{L_1} + \mathscr{E}_{3,i-\frac{1}{2},j-1}^{L_1}  \right) + \alpha^-_1 \left( \mathscr{E}_{3,i-\frac{1}{2},j}^{R_1} + \mathscr{E}_{3,i-\frac{1}{2},j-1}^{R_1}\right)}{2(\alpha^+_1 + \alpha^-_1)} \\
+& \frac{\alpha^+_1 \alpha^-_1}{(\alpha^+_1 + \alpha^-_1)} \left(\langle {B}_2^{R_1} \rangle_{i-\frac{1}{2},j-\frac{1}{2}}  - \langle {B}_2^{L_1} \rangle_{i-\frac{1}{2},j-\frac{1}{2}}  \right) \\
+& \frac{3}{8}\left(\left(\mathscr{E}_{3,i-1,j-\frac{1}{2}}^{L_2} + \mathscr{E}_{3,i-1,j-\frac{1}{2}}^{R_2} \right) - \left(\mathscr{E}^r_{3,i-1,j} +\mathscr{E}^r_{3,i-1,j-1} \right) \right)  \\
+& \frac{1}{8}\left( \left(\mathscr{E}_{3,i,j-\frac{1}{2}}^{L_2} +  \mathscr{E}_{3,i,j-\frac{1}{2}}^{R_2} \right) - \left(\mathscr{E}^r_{3,i,j-1} +\mathscr{E}^r_{3,i,j} \right) \right) \, .
\end{split}
\end{equation}
Next, we consider only the \(B_1\) terms
\begin{equation}\label{eq:b1-continuous-bias}
\begin{split}
&-\frac{\alpha_2}{2}\left(
\left(\frac{3}{4}\langle B_1^{R_2} \rangle_{i-1,j-\frac{1}{2}}  + \frac{1}{4}\langle B_1^{R_2} \rangle_{i,j-\frac{1}{2}}  \right)
- \left(\frac{3}{4}\langle B_1^{L_2} \rangle_{i-1,j-\frac{1}{2}} + \frac{1}{4}\langle B_1^{L_2} \rangle_{i,j-\frac{1}{2}} \right)
\right)  \\
& \approx -\frac{\alpha_2}{2}\left(\langle B_1^{R_2} \rangle_{i-\frac{3}{4},j-\frac{1}{2}}  - \langle B_1^{L_2} \rangle_{i-\frac{3}{4},j-\frac{1}{2}} \right) + \mathcal{O}(\Delta x^2)  \\
& \approx -\frac{\alpha_2}{2}\left(\langle B_1^{R_2} \rangle_{i-\frac{1}{2},j-\frac{1}{2}}  - \langle B_1^{L_2} \rangle_{i-\frac{1}{2},j-\frac{1}{2}} \right) + \mathcal{O}(\Delta x^2) \, ,
\end{split}
\end{equation}
which as in the previous proof, are guaranteed to provide a continuous approximation to an intermediate linear reconstructed value since \(B_1\) is continuous in \(x_1\). Unlike Equation~\eqref{eq:gs05-corner-b}, this linear interpolation is biased to the left of the \(i-\frac{1}{2},j-\frac{1}{2}\) corner position, which corresponds to the upstream direction. Nevertheless, it is a consistent approximation at second-order accuracy, so Equation~\eqref{eq:gs05-emf-average-x1-flow-5} can be written as
\begin{equation} \label{eq:gs05-approx-uct-x1-flow}
\begin{split}
\mathscr{E}_{3,i-\frac{1}{2},j-\frac{1}{2}} \approx & -\frac{\alpha_2}{2}\left(\langle B_1^{R_2} \rangle_{i-\frac{1}{2},j-\frac{1}{2}}  - \langle B_1^{L_2} \rangle_{i-\frac{1}{2},j-\frac{1}{2}} \right)  \\
+& \frac{\alpha^+_1 \left( \mathscr{E}_{3,i-\frac{1}{2},j}^{L_1} + \mathscr{E}_{3,i-\frac{1}{2},j-1}^{L_1}  \right) + \alpha^-_1 \left( \mathscr{E}_{3,i-\frac{1}{2},j}^{R_1} + \mathscr{E}_{3,i-\frac{1}{2},j-1}^{R_1}\right)}{2(\alpha^+_1 + \alpha^-_1)}  \\
+& \frac{\alpha^+_1 \alpha^-_1}{(\alpha^+_1 + \alpha^-_1)} \left(\langle {B}_2^{R_1} \rangle_{i-\frac{1}{2},j-\frac{1}{2}}  - \langle {B}_2^{L_1} \rangle_{i-\frac{1}{2},j-\frac{1}{2}}  \right)  \\
+& \frac{3}{8}\left(\left(\mathscr{E}_{3,i-1,j-\frac{1}{2}}^{L_2} + \mathscr{E}_{3,i-1,j-\frac{1}{2}}^{R_2} \right) - \left(\mathscr{E}^r_{3,i-1,j} +\mathscr{E}^r_{3,i-1,j-1} \right) \right)  \\
+& \frac{1}{8}\left( \left(\mathscr{E}_{3,i,j-\frac{1}{2}}^{L_2} +  \mathscr{E}_{3,i,j-\frac{1}{2}}^{R_2} \right) - \left(\mathscr{E}^r_{3,i,j-1} +\mathscr{E}^r_{3,i,j} \right) \right) \, .
\end{split}
\end{equation}
It is impossible for the general expression in Equation~\eqref{eq:gs05-approx-uct-x1-flow} to exactly match the UCT solution in Equation~\eqref{eq:uct-hll-x1-flow} due to the variable factors of \(\alpha_1^\pm\) in the \(x_1\) face-centered emf flux states that are absent for the other terms. However, we can show second-order accurate agreement under further simplifying assumptions.

We first assume usage of the simple HLL wavespeed estimates of Davis \cite{Davis1988}, which are defined as
\begin{subequations}  \label{eq:davis-hll-mhd}
\begin{align}
S^{L_1} =& \min \left(\lambda^-_1(\mathbf{W}^{L_1}, B_1), \lambda^-_1(\mathbf{W}^{R_1}, B_1)\right) \, ,  \\
S^{R_1} =& \max \left(\lambda^+_1(\mathbf{W}^{L_1}, B_1), \lambda^+_1(\mathbf{W}^{R_1}, B_1) \right) \, ,
\end{align}
\end{subequations}
where \( \lambda^\pm_1 \) are the largest and smallest eigenvalues of the system in the \(x_1\) direction. For the MHD system, these are related to the fast magnetosonic wavespeeds \(\lambda_1^\pm = v_1 \pm c_1^f\). We refer the reader to Equation 55 of LD2004 for comparison.

Next, we consider the limiting case of \(v_1 \gg 0\) such that \(\alpha_1^- \equiv -\min(0,S^{L_1}) = 0\). For the above two-speed flux approximation applied to the MHD system, the supersonic limiting case occurs when \(v_1 \geq c_1^f\). In such a case, the \(x_1\) explicit dissipation term goes to zero in Equation~\eqref{eq:gs05-approx-uct-x1-flow}, and the expression becomes
\begin{equation} \label{eq:gs05-supersonic-x1-flow}
\begin{split}
\mathscr{E}_{3,i-\frac{1}{2},j-\frac{1}{2}} \approx & -\frac{\alpha_2}{2}\left(\langle B_1^{R_2} \rangle_{i-\frac{1}{2},j-\frac{1}{2}}  - \langle B_1^{L_2} \rangle_{i-\frac{1}{2},j-\frac{1}{2}} \right)
+ \frac{\mathscr{E}_{3,i-\frac{1}{2},j}^{L_1} + \mathscr{E}_{3,i-\frac{1}{2},j-1}^{L_1}  }{2}  \\
+& \frac{3}{8}\left(\left(\mathscr{E}_{3,i-1,j-\frac{1}{2}}^{L_2} + \mathscr{E}_{3,i-1,j-\frac{1}{2}}^{R_2} \right) - \left(\mathscr{E}^r_{3,i-1,j} +\mathscr{E}^r_{3,i-1,j-1} \right) \right)  \\
+& \frac{1}{8}\left( \left(\mathscr{E}_{3,i,j-\frac{1}{2}}^{L_2} +  \mathscr{E}_{3,i,j-\frac{1}{2}}^{R_2} \right) - \left(\mathscr{E}^r_{3,i,j-1} +\mathscr{E}^r_{3,i,j} \right) \right) \, .
\end{split}
\end{equation}
Now, recall that the reconstructed emf is simplified in this 1D flow case. The continuity of \(B_2\) across the \(x_2\) interface further simplifies the reconstructed emf states
\begin{subequations}  \label{eq:emf-x2-interface-x1-flow}
\begin{align}
\mathscr{E}_{3,i,j-\frac{1}{2}}^{L_2} =& -v_{1,i,j-\frac{1}{2}}^{L_2} B_{2,i,j-\frac{1}{2}} \, ,  \\
\mathscr{E}_{3,i,j-\frac{1}{2}}^{R_2} =&-v_{1,i,j-\frac{1}{2}}^{R_2} B_{2,i,j-\frac{1}{2}} \, ,
\end{align}
\end{subequations}
since they only depend on the \(v_1\) reconstruction. Rearranging the emf terms in Equation~\eqref{eq:gs05-supersonic-x1-flow} and using the continuity of \(B_2\) in \(x_2\) results in
\begin{equation} \label{eq:gs05-supersonic-x1-flow-2}
\begin{split}
\mathscr{E}_{3,i-\frac{1}{2},j-\frac{1}{2}} \approx & -\frac{\alpha_2}{2}\left(\langle B_1^{R_2} \rangle_{i-\frac{1}{2},j-\frac{1}{2}}  - \langle B_1^{L_2} \rangle_{i-\frac{1}{2},j-\frac{1}{2}} \right)
+ \frac{\mathscr{E}_{3,i-\frac{1}{2},j}^{L_1} + \mathscr{E}_{3,i-\frac{1}{2},j-1}^{L_1}  }{2}  \\
+& \frac{1}{2}\left(\frac{3}{4}\left(-v_{1,i-1,j-\frac{1}{2}}^{L_2} B_{2,i-1,j-\frac{1}{2}} - \mathscr{E}^r_{3,i-1,j} \right) + \frac{1}{4} \left(-v_{1,i,j-\frac{1}{2}}^{L_2} B_{2,i,j-\frac{1}{2}}- \mathscr{E}^r_{3,i,j}\right) \right)  \\
+& \frac{1}{2}\left( \frac{3}{4}\left(-v_{1,i-1,j-\frac{1}{2}}^{R_2} B_{2,i-1,j-\frac{1}{2}}- \mathscr{E}^r_{3,i-1,j-1} \right)
+ \frac{1}{4}\left(-v_{1,i,j-\frac{1}{2}}^{R_2} B_{2,i,j-\frac{1}{2}} - \mathscr{E}^r_{3,i,j-1}  \right) \right) \, .
\end{split}
\end{equation}
In contrast, no such simplification can be made for the emf across the \(x_1\) interfaces, which satisfy
\begin{subequations}  \label{eq:emf-x1-interface-x1-flow}
\begin{align}
\mathscr{E}_{3,i-\frac{1}{2},j}^{L_1} =& -v_{1,i-\frac{1}{2},j}^{L_1} B_{2,i-\frac{1}{2},j}^{L_1} \, ,  \\
\mathscr{E}_{3,i-\frac{1}{2},j}^{R_1} =&-v_{1,i-\frac{1}{2},j}^{R_1} B_{2,i-\frac{1}{2},j}^{R_1} \, .
\end{align}
\end{subequations}
Because \(B_2\) may be discontinuous in \(x_1\), combining these terms in a similar fashion as the \(B_1\) terms in Equation~\eqref{eq:b1-continuous-bias} is impossible without making further simplifying assumptions about the field. The general expression in terms of the velocity and magnetic field is
\begin{equation} \label{eq:gs05-supersonic-x1-flow-3}
\begin{split}
\mathscr{E}_{3,i-\frac{1}{2},j-\frac{1}{2}} \approx & -\frac{\alpha_2}{2}\left(\langle B_1^{R_2} \rangle_{i-\frac{1}{2},j-\frac{1}{2}}  - \langle B_1^{L_2} \rangle_{i-\frac{1}{2},j-\frac{1}{2}} \right)
- \frac{1}{2} \left(v_{1,i-\frac{1}{2},j}^{L_1} B_{2,i-\frac{1}{2},j}^{L_1} + v_{1,i-\frac{1}{2},j-1}^{L_1} B_{2,i-\frac{1}{2},j-1}^{L_1}  \right) \\
-& \frac{1}{2}\left(\frac{3}{4}\left(v_{1,i-1,j-\frac{1}{2}}^{L_2} B_{2,i-1,j-\frac{1}{2}} - v_{1,i-1,j}B_{2,i-1,j} \right) + \frac{1}{4} \left(v_{1,i,j-\frac{1}{2}}^{L_2} B_{2,i,j-\frac{1}{2}}- v_{1,i,j}B_{2,i,j}\right) \right)  \\
-& \frac{1}{2}\left( \frac{3}{4}\left( v_{1,i-1,j-\frac{1}{2}}^{R_2} B_{2,i-1,j-\frac{1}{2}}- v_{1,i-1,j-1}B_{2,i-1,j-1} \right)
+ \frac{1}{4}\left(v_{1,i,j-\frac{1}{2}}^{R_2} B_{2,i,j-\frac{1}{2}} - v_{1,i,j-1}B_{2,i,j-1}  \right) \right) \, .
\end{split}
\end{equation}
If the \(x_1\)-reconstruction step produces identical Riemann states \(B_{2,i-\frac{1}{2},j}^{L_1}, B_{2,i-\frac{1}{2},j-1}^{L_1} \), then the GS05 approximation to the emf at the cell corner finally reduces to
\begin{multline}
\mathscr{E}_{3,i-\frac{1}{2},j-\frac{1}{2}} \approx  -\frac{\alpha_2}{2}\left(\langle B_1^{R_2} \rangle_{i-\frac{1}{2},j-\frac{1}{2}}  - \langle B_1^{L_2} \rangle_{i-\frac{1}{2},j-\frac{1}{2}} \right)  \\
- \frac{\left( \langle v^{L_1L_2}_{1}\rangle_{i-\frac{1}{2},j-\frac{1}{2}} + \langle v_{1}^{L_1R_2} \rangle_{i-\frac{1}{2},j-\frac{1}{2}} \right)  \langle {B}_2^{L_1} \rangle_{i-\frac{1}{2},j-\frac{1}{2}}}{2} \, . \label{eq:gs05-supersonic-x1-flow-4}
\end{multline}

Having considered the GS05 algorithm's behavior in this limiting case, we turn our attention to the upwind constrained transport approach. Under the assumption of supersonic wavespeed estimates, the UCT formula in Equation~\eqref{eq:uct-hll-x1-flow} becomes
\begin{equation} \label{eq:uct-hll-supersonic-x1-flow}
\langle \mathscr{E}^U_{3} \rangle_{i-\frac{1}{2},j-\frac{1}{2}} = \frac{
\langle \mathscr{E}^{L_1L_2}_{3}\rangle_{i-\frac{1}{2},j-\frac{1}{2}} + \langle \mathscr{E}_{3}^{L_1R_2} \rangle_{i-\frac{1}{2},j-\frac{1}{2}}  }{2}
 - \frac{\alpha_2}{2}( \langle {B}_1^{R_2} \rangle_{i-\frac{1}{2},j-\frac{1}{2}}  - \langle {B}_1^{L_2} \rangle_{i-\frac{1}{2},j-\frac{1}{2}} ) \, .
\end{equation}
Since \(v_2=0\) in this case,
\begin{multline}
\langle \mathscr{E}^U_{3} \rangle_{i-\frac{1}{2},j-\frac{1}{2}} = \frac{
- \left( \langle v^{L_1L_2}_{1}\rangle_{i-\frac{1}{2},j-\frac{1}{2}} + \langle v_{1}^{L_1R_2} \rangle_{i-\frac{1}{2},j-\frac{1}{2}} \right)  \langle {B}_2^{L_1} \rangle_{i-\frac{1}{2},j-\frac{1}{2}}}{2}  \\
  - \frac{\alpha_2}{2}( \langle {B}_1^{R_2} \rangle_{i-\frac{1}{2},j-\frac{1}{2}}  - \langle {B}_1^{L_2} \rangle_{i-\frac{1}{2},j-\frac{1}{2}} ) \, ,  \label{eq:uct-hll-supersonic-x1-flow-2}
\end{multline}
which matches Equation~\eqref{eq:gs05-supersonic-x1-flow-4}.

\bibliographystyle{plainnat}\bibliography{FelkerStone2017}

\begin{thebibliography}{64}
\providecommand{\natexlab}[1]{#1}
\providecommand{\url}[1]{\texttt{#1}}
\expandafter\ifx\csname urlstyle\endcsname\relax
  \providecommand{\doi}[1]{doi: #1}\else
  \providecommand{\doi}{doi: \begingroup \urlstyle{rm}\Url}\fi

\bibitem[Amano(2015)]{Amano2015}
Takanobu Amano.
\newblock {Divergence-free approximate Riemann solver for the quasi-neutral
  two-fluid plasma model}.
\newblock \emph{Journal of Computational Physics}, 299:\penalty0 863--886,
  2015.
\newblock \doi{10.1016/j.jcp.2015.07.035}.

\bibitem[Balsara(2010)]{Balsara2010}
Dinshaw~S. Balsara.
\newblock {Multidimensional HLLE Riemann solver: Application to Euler and
  magnetohydrodynamic flows}.
\newblock \emph{Journal of Computational Physics}, 229\penalty0 (6):\penalty0
  1970--1993, 2010.
\newblock \doi{10.1016/j.jcp.2009.11.018}.

\bibitem[Balsara(2012)]{Balsara2012}
Dinshaw~S. Balsara.
\newblock {A two-dimensional HLLC Riemann solver for conservation laws:
  Application to Euler and magnetohydrodynamic flows}.
\newblock \emph{Journal of Computational Physics}, 231\penalty0 (22):\penalty0
  7476--7503, 2012.
\newblock \doi{10.1016/j.jcp.2011.12.025}.

\bibitem[Balsara(2014)]{Balsara2014a}
Dinshaw~S. Balsara.
\newblock {Multidimensional Riemann problem with self-similar internal
  structure. Part I - Application to hyperbolic conservation laws on structured
  meshes}.
\newblock \emph{Journal of Computational Physics}, 277:\penalty0 163--200,
  2014.
\newblock \doi{10.1016/j.jcp.2014.07.053}.

\bibitem[Balsara and Kim(2004)]{BalsaraKim2004}
Dinshaw~S. Balsara and Jongsoo Kim.
\newblock A comparison between divergence-cleaning and staggered-mesh
  formulations for numerical magnetohydrodynamics.
\newblock \emph{The Astrophysical Journal}, 602\penalty0 (2):\penalty0
  1079--1090, 2004.
\newblock \doi{10.1086/381051}.

\bibitem[Balsara and Shu(2000)]{BalsaraShu2000}
Dinshaw~S. Balsara and Chi-Wang Shu.
\newblock Monotonicity preserving weighted essentially non-oscillatory schemes
  with increasingly high order of accuracy.
\newblock \emph{Journal of Computational Physics}, 160\penalty0 (2):\penalty0
  405--452, 2000.
\newblock \doi{10.1006/jcph.2000.6443}.

\bibitem[Balsara and Spicer(1999)]{BalsaraSpicer1999}
Dinshaw~S. Balsara and Daniel~S. Spicer.
\newblock A staggered mesh algorithm using high order {Godunov} fluxes to
  ensure solenoidal magnetic fields in magnetohydrodynamic simulations.
\newblock \emph{Journal of Computational Physics}, 149\penalty0 (2):\penalty0
  270--292, 1999.
\newblock \doi{10.1006/jcph.1998.6153}.

\bibitem[Barad and Colella(2005)]{BaradColella2005}
Michael Barad and Phillip Colella.
\newblock {A fourth-order accurate local refinement method for Poisson's
  equation}.
\newblock \emph{Journal of Computational Physics}, 209\penalty0 (1):\penalty0
  1--18, 2005.
\newblock \doi{10.1016/j.jcp.2005.02.027}.

\bibitem[Brackbill and Barnes(1980)]{BrackbillBarnes1980}
J.U Brackbill and D.C Barnes.
\newblock The effect of nonzero {$\nabla \cdot B$} on the numerical solution of
  the magnetohydrodynamic equations.
\newblock \emph{Journal of Computational Physics}, 35\penalty0 (3):\penalty0
  426--430, 1980.
\newblock \doi{10.1016/0021-9991(80)90079-0}.

\bibitem[Brio and Wu(1988)]{BrioWu1988}
M.~Brio and C.C. Wu.
\newblock {An upwind differencing scheme for the equations of ideal
  magnetohydrodynamics}.
\newblock \emph{Journal of Computational Physics}, 75\penalty0 (2):\penalty0
  400--422, 1988.
\newblock \doi{10.1016/0021-9991(88)90120-9}.

\bibitem[Cargo and Gallice(1997)]{CargoGalice1997}
P.~Cargo and G.~Gallice.
\newblock Roe matrices for ideal {MHD} and systematic construction of {Roe}
  matrices for systems of conservation laws.
\newblock \emph{Journal of Computational Physics}, 136\penalty0 (2):\penalty0
  446--466, 1997.
\newblock \doi{10.1006/jcph.1997.5773}.

\bibitem[Colella et~al.(2009)Colella, Dorr, Hittinger, McCorquodale, and
  Martin]{ColellaDorrHittingerMartin2009}
P.~Colella, M.~Dorr, J.~Hittinger, P.~McCorquodale, and D.~F. Martin.
\newblock High-order finite-volume methods on locally-structured grids.
\newblock \emph{Numerical Modeling of Space Plasma Flows: ASTRONUM-2008},
  406:\penalty0 1--9, 2009.

\bibitem[Colella et~al.(2011)Colella, Dorr, Hittinger, and
  Martin]{ColellaDorrHittingerMartin2011}
P.~Colella, M.~R. Dorr, J.~A~F Hittinger, and D.~F. Martin.
\newblock {High-order, finite-volume methods in mapped coordinates}.
\newblock \emph{Journal of Computational Physics}, 230\penalty0 (8):\penalty0
  2952--2976, 2011.
\newblock \doi{10.1016/j.jcp.2010.12.044}.

\bibitem[Colella and Sekora(2008)]{ColellaSekora2008}
Phillip Colella and Michael~D. Sekora.
\newblock {A limiter for PPM that preserves accuracy at smooth extrema}.
\newblock \emph{Journal of Computational Physics}, 227\penalty0 (15):\penalty0
  7069--7076, 2008.
\newblock \doi{10.1016/j.jcp.2008.03.034}.

\bibitem[Colella and Woodward(1984)]{ColellaWoodward1984}
Phillip Colella and Paul~R. Woodward.
\newblock {The Piecewise Parabolic Method (PPM) for gas-dynamical simulations}.
\newblock \emph{Journal of Computational Physics}, 54\penalty0 (1):\penalty0
  174--201, 1984.
\newblock \doi{10.1016/0021-9991(84)90143-8}.

\bibitem[Crockett et~al.(2005)Crockett, Colella, Fisher, Klein, and
  McKee]{Crockett2005}
Robert~K. Crockett, Phillip Colella, Robert~T. Fisher, Richard~I. Klein, and
  Christopher~F. McKee.
\newblock {An unsplit, cell-centered Godunov method for ideal MHD}.
\newblock \emph{Journal of Computational Physics}, 203\penalty0 (2):\penalty0
  422--448, 2005.
\newblock \doi{10.1016/j.jcp.2004.08.021}.

\bibitem[Dai and Woodward(1998)]{DaiWoodward1998b}
Wenlong Dai and Paul~R. Woodward.
\newblock On the divergence-free condition and conservation laws in numerical
  simulations for supersonic magnetohydrodynamical flows.
\newblock \emph{The Astrophysical Journal}, 494\penalty0 (1):\penalty0
  317--335, 1998.
\newblock \doi{10.1086/305176}.

\bibitem[Davis(1988)]{Davis1988}
S.~F. Davis.
\newblock Simplified second-order {Godunov}-type methods.
\newblock \emph{SIAM Journal on Scientific and Statistical Computing},
  9\penalty0 (3):\penalty0 445--473, 1988.
\newblock \doi{10.1137/0909030}.

\bibitem[de~la Rosa and Munz(2016)]{NunezMunz2016-I}
Jonatan~N{\'{u}}{\~{n}}ez de~la Rosa and Claus-Dieter Munz.
\newblock xtroem-fv: a new code for computational astrophysics based on very
  high order finite-volume methods {\textendash} {I}. magnetohydrodynamics.
\newblock \emph{Monthly Notices of the Royal Astronomical Society},
  455\penalty0 (4):\penalty0 3458--3479, 2016.
\newblock \doi{10.1093/mnras/stv2531}.

\bibitem[Dedner et~al.(2002)Dedner, Kemm, Kr{\"{o}}ner, Munz, Schnitzer, and
  Wesenberg]{Dedner2002}
A.~Dedner, F.~Kemm, D.~Kr{\"{o}}ner, C.-D. Munz, T.~Schnitzer, and
  M.~Wesenberg.
\newblock Hyperbolic divergence cleaning for the {MHD} equations.
\newblock \emph{Journal of Computational Physics}, 175\penalty0 (2):\penalty0
  645--673, 2002.
\newblock \doi{10.1006/jcph.2001.6961}.

\bibitem[Dumbser et~al.(2013)Dumbser, Zanotti, Hidalgo, and
  Balsara]{DumbserZanotti2013}
Michael Dumbser, Olindo Zanotti, Arturo Hidalgo, and Dinshaw~S. Balsara.
\newblock {ADER-WENO finite volume schemes with space-time adaptive mesh
  refinement}.
\newblock \emph{Journal of Computational Physics}, 248:\penalty0 257--286,
  2013.
\newblock \doi{10.1016/j.jcp.2013.04.017}.

\bibitem[Einfeldt et~al.(1991)Einfeldt, Munz, Roe, and
  Sj{\"{o}}green]{Einfeldt1991}
B.~Einfeldt, C.~D. Munz, P.~L. Roe, and B.~Sj{\"{o}}green.
\newblock {On Godunov-type methods near low densities}.
\newblock \emph{Journal of Computational Physics}, 92\penalty0 (2):\penalty0
  273--295, 1991.
\newblock \doi{10.1016/0021-9991(91)90211-3}.

\bibitem[Einfeldt(1988)]{Einfeldt1988}
Bernd Einfeldt.
\newblock On {Godunov}-type methods for gas dynamics.
\newblock \emph{SIAM Journal on Numerical Analysis}, 25\penalty0 (2):\penalty0
  294--318, 1988.
\newblock \doi{10.1137/0725021}.

\bibitem[Evans and Hawley(1988)]{EvansHawley1988}
Charles~R. Evans and John~F. Hawley.
\newblock {Simulation of magnetohydrodynamic flows - A constrained transport
  method}.
\newblock \emph{The Astrophysical Journal}, 332\penalty0 (2):\penalty0 659,
  1988.
\newblock \doi{10.1086/166684}.

\bibitem[Falle(1991)]{Falle1991}
S.~A. E.~G. Falle.
\newblock {Self-similar jets}.
\newblock \emph{Monthly Notices of the Royal Astronomical Society},
  250\penalty0 (3):\penalty0 581--596, 1991.
\newblock \doi{10.1093/mnras/250.3.581}.

\bibitem[Fromang et~al.(2006)Fromang, Hennebelle, and Teyssier]{Fromang2006}
S.~Fromang, P.~Hennebelle, and R.~Teyssier.
\newblock {A high order Godunov scheme with constrained transport and adaptive
  mesh refinement for astrophysical magnetohydrodynamics}.
\newblock \emph{Astronomy \& Astrophysics}, 457\penalty0 (2):\penalty0
  371--384, 2006.
\newblock \doi{10.1051/0004-6361:20065371}.

\bibitem[Gardiner and Stone(2005)]{GardinerStone2005}
Thomas~A. Gardiner and James~M. Stone.
\newblock An unsplit {Godunov} method for ideal {MHD} via constrained
  transport.
\newblock \emph{Journal of Computational Physics}, 205\penalty0 (2):\penalty0
  509--539, 2005.
\newblock \doi{10.1016/j.jcp.2004.11.016}.

\bibitem[Gardiner and Stone(2008)]{GardinerStone2008}
Thomas~A. Gardiner and James~M. Stone.
\newblock An unsplit {Godunov} method for ideal {MHD} via constrained transport
  in three dimensions.
\newblock \emph{Journal of Computational Physics}, 227\penalty0 (8):\penalty0
  4123--4141, 2008.
\newblock \doi{10.1016/j.jcp.2007.12.017}.

\bibitem[Gottlieb and Shu(1998)]{GottliebShu1998}
Sigal Gottlieb and Chi-Wang Shu.
\newblock {Total variation diminishing Runge-Kutta schemes}.
\newblock \emph{Mathematics of Computation of the American Mathematical
  Society}, 67\penalty0 (221):\penalty0 73--85, 1998.
\newblock \doi{10.1090/S0025-5718-98-00913-2}.

\bibitem[Gottlieb et~al.(2001)Gottlieb, Shu, and Tadmor]{GottliebShuTadmor2001}
Sigal Gottlieb, Chi-Wang Shu, and Eitan Tadmor.
\newblock Strong stability-preserving high-order time discretization methods.
\newblock \emph{SIAM Review}, 43\penalty0 (1):\penalty0 89--112, 2001.
\newblock \doi{10.1137/S003614450036757X}.

\bibitem[Gottlieb et~al.(2009)Gottlieb, Ketcheson, and
  Shu]{GottliebKetchesonShu2009}
Sigal Gottlieb, David~I. Ketcheson, and Chi-Wang Shu.
\newblock {High order strong stability preserving time discretizations}.
\newblock \emph{Journal of Scientific Computing}, 38\penalty0 (3):\penalty0
  251--289, 2009.
\newblock \doi{10.1007/s10915-008-9239-z}.

\bibitem[Guzik et~al.(2015)Guzik, Gao, Owen, McCorquodale, and
  Colella]{Guzik2015}
Stephen~M. Guzik, Xinfeng Gao, Landon~D. Owen, Peter McCorquodale, and Phillip
  Colella.
\newblock {A freestream-preserving fourth-order finite-volume method in mapped
  coordinates with adaptive-mesh refinement}.
\newblock \emph{Computers and Fluids}, 123:\penalty0 202--217, 2015.
\newblock \doi{10.1016/j.compfluid.2015.10.001}.

\bibitem[Harten et~al.(1983)Harten, Lax, and van Leer]{HartenLaxLeer1983}
Amiram Harten, Peter~D. Lax, and Bram van Leer.
\newblock On upstream differencing and {Godunov}-type schemes for hyperbolic
  conservation laws.
\newblock \emph{SIAM Review}, 25\penalty0 (1):\penalty0 35--61, 1983.
\newblock \doi{10.1137/1025002}.

\bibitem[Jiang and Shu(1996)]{JiangShu1996}
Guang-Shan Jiang and Chi-Wang Shu.
\newblock Efficient implementation of weighted {ENO} schemes.
\newblock \emph{Journal of Computational Physics}, 126\penalty0 (1):\penalty0
  202--228, 1996.
\newblock \doi{10.1006/jcph.1996.0130}.

\bibitem[Ketcheson(2008)]{Ketcheson2008}
David~I. Ketcheson.
\newblock Highly efficient strong stability-preserving {Runge-Kutta} methods
  with low-storage implementations.
\newblock \emph{SIAM Journal on Scientific Computing}, 30\penalty0
  (4):\penalty0 2113--2136, 2008.
\newblock \doi{10.1137/07070485X}.

\bibitem[Ketcheson(2010)]{Ketcheson2010}
David~I Ketcheson.
\newblock {Runge-Kutta methods with minimum storage implementations}.
\newblock \emph{Journal of Computational Physics}, 229\penalty0 (5):\penalty0
  1763--1773, 2010.
\newblock \doi{10.1016/j.jcp.2009.11.006}.

\bibitem[Li and Shu(2005)]{LiShu2005}
Fengyan Li and Chi-Wang Shu.
\newblock {Locally divergence-free discontinuous Galerkin methods for MHD
  equations}.
\newblock \emph{Journal of Scientific Computing}, 22-23\penalty0
  (June):\penalty0 413--442, 2005.
\newblock \doi{10.1007/s10915-004-4146-4}.

\bibitem[Loffeld and Hittinger(2017)]{Loffeld2017}
J~Loffeld and JAF Hittinger.
\newblock On the arithmetic intensity of high-order finite-volume
  discretizations for hyperbolic systems of conservation laws.
\newblock \emph{The International Journal of High Performance Computing
  Applications}, 2017.
\newblock \doi{10.1177/1094342017691876}.

\bibitem[Londrillo and {Del Zanna}(2000)]{Londrillo2000}
P.~Londrillo and L.~{Del Zanna}.
\newblock High-order upwind schemes for multidimensional magnetohydrodynamics.
\newblock \emph{The Astrophysical Journal}, 530\penalty0 (1):\penalty0
  508--524, 2000.
\newblock \doi{10.1086/308344}.

\bibitem[Londrillo and {Del Zanna}(2004)]{Londrillo2004}
P.~Londrillo and L.~{Del Zanna}.
\newblock {On the divergence-free condition in Godunov-type schemes for ideal
  magnetohydrodynamics: the upwind constrained transport method}.
\newblock \emph{Journal of Computational Physics}, 195\penalty0 (1):\penalty0
  17--48, 2004.
\newblock \doi{10.1016/j.jcp.2003.09.016}.

\bibitem[Luo et~al.(2008)Luo, Baum, and L{\"{o}}hner]{Luo2008}
Hong Luo, Joseph~D. Baum, and Rainald L{\"{o}}hner.
\newblock {A discontinuous Galerkin method based on a Taylor basis for the
  compressible flows on arbitrary grids}.
\newblock \emph{Journal of Computational Physics}, 227\penalty0 (20):\penalty0
  8875--8893, 2008.
\newblock \doi{10.1016/j.jcp.2008.06.035}.

\bibitem[Matsumoto et~al.(2016)Matsumoto, Asahina, Kudoh, Kawashima, Matsumoto,
  Takahashi, Minoshima, Zenitani, Miyoshi, and Matsumoto]{Matsumoto2016}
Yosuke Matsumoto, Yuta Asahina, Yuki Kudoh, Tomohisa Kawashima, Jin Matsumoto,
  Hiroyuki~R. Takahashi, Takashi Minoshima, Seiji Zenitani, Takahiro Miyoshi,
  and Ryoji Matsumoto.
\newblock Magnetohydrodynamic simulation code {CANS+}: Assessments and
  applications.
\newblock submitted to Publ. Astron. Soc. Japan, 2016.

\bibitem[McCorquodale and Colella(2011)]{McCorquodaleColella2011}
Peter McCorquodale and Phillip Colella.
\newblock {A high-order finite-volume method for conservation laws on locally
  refined grids}.
\newblock \emph{Communications in Applied Mathematics and Computational
  Science}, 6\penalty0 (1):\penalty0 1--25, 2011.
\newblock \doi{10.2140/camcos.2011.6.1}.

\bibitem[Mignone(2014)]{Mignone2014}
Andrea Mignone.
\newblock {High-order conservative reconstruction schemes for finite volume
  methods in cylindrical and spherical coordinates}.
\newblock \emph{Journal of Computational Physics}, 270:\penalty0 784--814,
  2014.
\newblock \doi{10.1016/j.jcp.2014.04.001}.

\bibitem[Mignone et~al.(2010)Mignone, Tzeferacos, and Bodo]{Mignone2010a}
Andrea Mignone, Petros Tzeferacos, and Gianluigi Bodo.
\newblock {High-order conservative finite difference GLM-MHD schemes for
  cell-centered MHD}.
\newblock \emph{Journal of Computational Physics}, 229\penalty0 (17):\penalty0
  5896--5920, 2010.
\newblock \doi{10.1016/j.jcp.2010.04.013}.

\bibitem[Miyoshi and Kusano(2005)]{MiyoshiKusano2005}
Takahiro Miyoshi and Kanya Kusano.
\newblock A multi-state {HLL} approximate {Riemann} solver for ideal
  magnetohydrodynamics.
\newblock \emph{Journal of Computational Physics}, 208\penalty0 (1):\penalty0
  315--344, 2005.
\newblock \doi{10.1016/j.jcp.2005.02.017}.

\bibitem[Mocz et~al.(2014)Mocz, Vogelsberger, Sijacki, Pakmor, and
  Hernquist]{Mocz2014}
Philip Mocz, Mark Vogelsberger, Debora Sijacki, R{\"{u}}diger Pakmor, and Lars
  Hernquist.
\newblock {A discontinuous Galerkin method for solving the fluid and
  magnetohydrodynamic equations in astrophysical simulations}.
\newblock \emph{Monthly Notices of the Royal Astronomical Society},
  437\penalty0 (1):\penalty0 397--414, 2014.
\newblock \doi{10.1093/mnras/stt1890}.

\bibitem[Olschanowsky et~al.(2014)Olschanowsky, Strout, Guzik, Loffeld, and
  Hittinger]{Olschanowsky2014}
Catherine Olschanowsky, Michelle~Mills Strout, Stephen Guzik, John Loffeld, and
  Jeffrey Hittinger.
\newblock A study on balancing parallelism, data locality, and recomputation in
  existing {PDE} solvers.
\newblock In \emph{{SC}14: International Conference for High Performance
  Computing, Networking, Storage and Analysis}. {IEEE}, 2014.
\newblock \doi{10.1109/sc.2014.70}.

\bibitem[Orszag and Tang(1979)]{OrszagTang1979}
Steven~A. Orszag and Cha-Mei Tang.
\newblock {Small-scale structure of two-dimensional magnetohydrodynamic
  turbulence}.
\newblock \emph{Journal of Fluid Mechanics}, 90\penalty0 (01):\penalty0 129,
  1979.
\newblock \doi{10.1017/S002211207900210X}.

\bibitem[Peterson and Hammett(2013)]{PetersonHammett2013}
J.~L. Peterson and G.~W. Hammett.
\newblock Positivity preservation and advection algorithms with applications to
  edge plasma turbulence.
\newblock \emph{SIAM Journal on Scientific Computing}, 35\penalty0
  (3):\penalty0 B576--B605, 2013.
\newblock \doi{10.1137/120888053}.

\bibitem[Powell et~al.(1999)Powell, Roe, Linde, Gombosi, and {De
  Zeeuw}]{Powell1999}
Kenneth~G. Powell, Philip~L. Roe, Timur~J. Linde, Tamas~I. Gombosi, and
  Darren~L. {De Zeeuw}.
\newblock A solution-adaptive upwind scheme for ideal magnetohydrodynamics.
\newblock \emph{Journal of Computational Physics}, 154\penalty0 (2):\penalty0
  284--309, 1999.
\newblock \doi{10.1006/jcph.1999.6299}.

\bibitem[Roe(1981)]{Roe1981}
P.~L. Roe.
\newblock {Approximate Riemann solvers, parameter vectors, and difference
  schemes}.
\newblock \emph{Journal of Computational Physics}, 43\penalty0 (2):\penalty0
  357--372, 1981.
\newblock \doi{10.1016/0021-9991(81)90128-5}.

\bibitem[Ryu and Jones(1995)]{RyuJones1995}
Dongsu Ryu and T.~W. Jones.
\newblock {Numerical magetohydrodynamics in astrophysics: Algorithm and tests
  for one-dimensional flow}.
\newblock \emph{The Astrophysical Journal}, 442:\penalty0 228, 1995.
\newblock \doi{10.1086/175437}.

\bibitem[Ryu et~al.(1998)Ryu, Miniati, Jones, and
  Frank]{RyuMinatiJonesFrank1998}
Dongsu Ryu, Francesco Miniati, T.~W. Jones, and Adam Frank.
\newblock A divergence-free upwind code for multidimensional
  magnetohydrodynamic flows.
\newblock \emph{The Astrophysical Journal}, 509\penalty0 (1):\penalty0
  244--255, 1998.
\newblock \doi{10.1086/306481}.

\bibitem[Shu and Osher(1988)]{ShuOsher1988}
Chi-Wang Shu and Stanley Osher.
\newblock {Efficient implementation of essentially non-oscillatory
  shock-capturing schemes}.
\newblock \emph{Journal of Computational Physics}, 77\penalty0 (2):\penalty0
  439--471, 1988.
\newblock \doi{10.1016/0021-9991(88)90177-5}.

\bibitem[Shu and Osher(1989)]{ShuOsher1989}
Chi-Wang Shu and Stanley Osher.
\newblock {Efficient implementation of essentially non-oscillatory
  shock-capturing schemes, II}.
\newblock \emph{Journal of Computational Physics}, 83\penalty0 (1):\penalty0
  32--78, 1989.
\newblock \doi{10.1016/0021-9991(89)90222-2}.

\bibitem[Sod(1978)]{Sod1978}
Gary~A. Sod.
\newblock {A survey of several finite difference methods for systems of
  nonlinear hyperbolic conservation laws}.
\newblock \emph{Journal of Computational Physics}, 27\penalty0 (1):\penalty0
  1--31, 1978.
\newblock \doi{10.1016/0021-9991(78)90023-2}.

\bibitem[Spiteri and Ruuth(2002)]{SpiteriRuuth2002}
Raymond~J. Spiteri and Steven~J. Ruuth.
\newblock A new class of optimal high-order strong-stability-preserving time
  discretization methods.
\newblock \emph{SIAM Journal on Numerical Analysis}, 40\penalty0 (2):\penalty0
  469--491, 2002.
\newblock \doi{10.1137/S0036142901389025}.

\bibitem[Stone and Gardiner(2009)]{StoneGardiner2009}
James~M. Stone and Thomas Gardiner.
\newblock {A simple unsplit Godunov method for multidimensional MHD}.
\newblock \emph{New Astronomy}, 14\penalty0 (2):\penalty0 139--148, 2009.
\newblock \doi{10.1016/j.newast.2008.06.003}.

\bibitem[Stone et~al.(2008)Stone, Gardiner, Teuben, Hawley, and
  Simon]{Stone2008}
James~M. Stone, Thomas~A. Gardiner, Peter Teuben, J.~F. Hawley, and J.~B.
  Simon.
\newblock Athena: A new code for astrophysical {MHD}.
\newblock \emph{The Astrophysical Journal Supplement Series}, 178:\penalty0
  137--177, 2008.
\newblock \doi{10.1086/588755}.

\bibitem[Susanto et~al.(2013)Susanto, Ivan, Sterck, and Groth]{Susanto2013}
A.~Susanto, L.~Ivan, H.~De Sterck, and C.P.T. Groth.
\newblock High-order central {ENO} finite-volume scheme for ideal {MHD}.
\newblock \emph{Journal of Computational Physics}, 250:\penalty0 141--164,
  2013.
\newblock \doi{10.1016/j.jcp.2013.04.040}.

\bibitem[Toro et~al.(1994)Toro, Spruce, and Speares]{Toro1994}
E.~F. Toro, M.~Spruce, and W.~Speares.
\newblock {Restoration of the contact surface in the HLL-Riemann solver}.
\newblock \emph{Shock Waves}, 4\penalty0 (1):\penalty0 25--34, 1994.
\newblock \doi{10.1007/BF01414629}.

\bibitem[T{\'{o}}th(2000)]{Toth2000}
G{\'{a}}bor T{\'{o}}th.
\newblock The {$\nabla \cdot B=0$} constraint in shock-capturing
  magnetohydrodynamics codes.
\newblock \emph{Journal of Computational Physics}, 161\penalty0 (2):\penalty0
  605--652, 2000.
\newblock \doi{10.1006/jcph.2000.6519}.

\bibitem[Zanna et~al.(2007)Zanna, Zanotti, Bucciantini, and
  Londrillo]{DelZanna2007}
L.~Del Zanna, O.~Zanotti, N.~Bucciantini, and P.~Londrillo.
\newblock {ECHO}: a {Eulerian} conservative high-order scheme for general
  relativistic magnetohydrodynamics and magnetodynamics.
\newblock \emph{Astronomy {\&} Astrophysics}, 473\penalty0 (1):\penalty0
  11--30, 2007.
\newblock \doi{10.1051/0004-6361:20077093}.

\end{thebibliography}

\end{document}